\documentclass[11pt]{article}

\usepackage[final]{acl}

\usepackage{times}
\usepackage{latexsym}
\usepackage{booktabs}
\usepackage[T1]{fontenc}

\usepackage[utf8]{inputenc}

\usepackage{microtype}

\usepackage{inconsolata}

\usepackage{graphicx}
\usepackage{adjustbox}
\usepackage{amssymb}
\graphicspath{{../figs/}{figs/}}

\newcommand{\safeincludegraphics}[2][]{%
  \IfFileExists{#2}{%
    \includegraphics[#1]{#2}%
  }{%
    \IfFileExists{../#2}{%
      \includegraphics[#1]{../#2}%
    }{%
      \fbox{\scriptsize Missing figure: \texttt{#2}}%
    }%
  }%
}

\usepackage{todonotes}
\usepackage{pgfplots}
\usepackage{pdflscape}
\usepackage{booktabs}
\usepackage{multirow}
\usepackage{array}
\usepackage{float}
\usepackage[table]{xcolor}
\usepackage{rotating}
\usepackage[most]{tcolorbox}
\usepackage{pifont}    
\usepackage{subcaption}
\usepackage{algorithm}
\usepackage{algpseudocode}
\usepackage{enumitem}
\usepackage{listings}
\usepackage{varwidth}
\usepackage{stmaryrd}
\usepackage{adjustbox}
\usepackage{xspace}
\usepackage{pgfplots}
\usepackage{hyperref}
\usepackage{fontawesome5}

\newcounter{mylisting}

\pgfplotsset{compat=1.18}
\usetikzlibrary{patterns}
\definecolor{darkgreen}{rgb}{0.0, 0.5, 0.0}
\definecolor{sc_color}{HTML}{66c2a5}
\definecolor{dsa_color}{HTML}{fc8d62}
\definecolor{dev_color}{HTML}{8da0cb}
\definecolor{web_color}{HTML}{e78ac3}
\definecolor{testing_color}{HTML}{ffd92f}
\definecolor{utilities_color}{HTML}{a6d854}
\newcommand{\cmark}{{\color{darkgreen}\ding{51}}}%
\newcommand{\xmark}{{\color{red}\ding{55}}}%
\newcommand{\method}{\textsc{XRepoTest}\@\xspace}

\lstdefinestyle{golang}{
  language=Go,
  basicstyle=\ttfamily\small,
  keywordstyle=\bfseries\color{blue!70!black},
  commentstyle=\itshape\color{gray!70!black},
  stringstyle=\color{green!40!black},
  showstringspaces=false,
  columns=fullflexible,
  keepspaces=true,
  frame=single,
  breaklines=true,
  numbers=none,
  numberstyle=\tiny\color{gray},
  xleftmargin=1.5em,
  framexleftmargin=1.0em,
  emph={doSort},
  emphstyle=\bfseries\color{red!80!black}
}

\title{Instructions for *ACL Proceedings}

\author{
  \setcounter{footnote}{0}
  Dung Le Quang\textsuperscript{1}\thanks{Equal contribution.}
  , Dong Cao Van\textsuperscript{1}\textsuperscript{*}
  , Nam Le Hai\textsuperscript{1}\thanks{Corresponding author: \texttt{namlh@soict.hust.edu.vn}}\\
  \bfseries Linh Ngo Van\textsuperscript{1}
  , Anh M. T. Bui\textsuperscript{1}
  , Phuong T. Nguyen\textsuperscript{2}
  \vspace{0.2cm}\\
  \textsuperscript{1}Hanoi University of Science and Technology, Viet Nam \\
  \textsuperscript{2}University of L'Aquila, Italy \\
}

\begin{document}

\title{\method: Benchmarking Multilingual Repository-Level Unit Test Generation for Large Language Models}

\maketitle

\begin{abstract}
Large language models (LLMs) have shown promise for automated unit test generation, but existing evaluations largely rely on standalone settings and a narrow set of programming languages, overestimating real-world readiness. We introduce \method, a multilingual (i.e., covering multiple programming languages) repository-level benchmark for unit test generation spanning five underexplored languages: Rust, Go, Julia, PHP, and Ruby. \method evaluates tests under realistic repository constraints using a containerized execution framework and multiple context augmentation strategies, including file-level, LSP-based, and retrieval-based context. Beyond standard metrics such as test pass rate and coverage, we propose Invocation Rate (IR) to assess whether generated tests meaningfully exercise the intended functionality. Experiments with 14 state-of-the-art LLMs (e.g., Claude 4.5, GPT-5.2, DeepSeek V4-pro, and Qwen families) reveal a substantial gap between standalone and repository-level performance, as well as trade-offs between richer context and test reliability. Overall, \method provides a challenging and informative benchmark to advance scalable and robust unit test generation in realistic software environments. The dataset and code are publicly available at: \href{https://github.com/solis-team/XRepoTest}{\faGithub\ \method}.
\end{abstract}

\section{Introduction}
\label{sec:introduction}

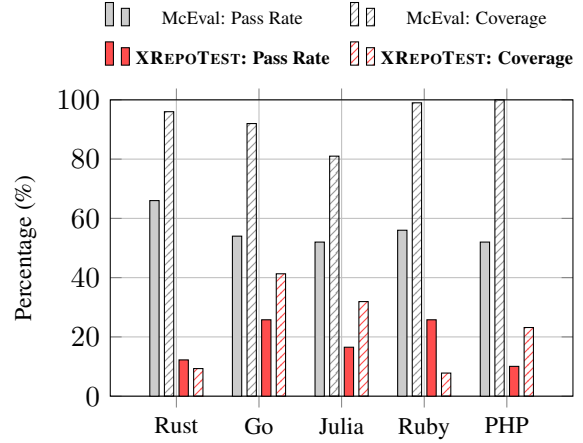
\begin{figure}[t]
    \centering
    \begin{tikzpicture}
        \begin{axis}[
            ybar,
            bar width=3.5pt,        
            width=\columnwidth,     
            height=5.5cm,           
            ymin=0, ymax=100,
            ylabel={Percentage (\%)},
            ylabel style={font=\footnotesize},
            xtick={1,2,3,4,5},
            xticklabels={Rust, Go, Julia, Ruby, PHP},
            xticklabel style={font=\footnotesize},
            enlarge x limits=0.2,   
            grid=major,
            legend style={
                at={(0.5,1.35)},    
                anchor=north,
                legend columns=2,   
                font=\scriptsize,
                draw=none,
                /tikz/every even column/.append style={column sep=0.2cm},
                row sep=0.1cm
            },
            title={},
        ]

        
        \addplot[fill=gray!40, draw=black] coordinates {
            (1, 66) (2, 54) (3, 52) (4, 56) (5, 52)
        };
        \addlegendentry{McEval: Pass Rate}

        \addplot[
            draw=black,
            pattern=north east lines,
            pattern color=gray!80
        ] coordinates {
            (1, 96) (2, 92) (3, 81) (4, 99) (5, 100)
        };
        \addlegendentry{McEval: Coverage}

        
        \addplot[fill=red!70, draw=black] coordinates {
            (1, 12.24) (2, 25.78) (3, 16.52) (4, 25.78) (5, 10.06)
        };
        \addlegendentry{\textbf{\method: Pass Rate}}

        \addplot[
            draw=black,
            pattern=north east lines,
            pattern color=red!70
        ] coordinates {
            (1, 9.31) (2, 41.30) (3, 31.92) (4, 7.80) (5, 23.17)
        };
        \addlegendentry{\textbf{\method: Coverage}}

        \end{axis}
    \end{tikzpicture}
    \caption{\textbf{The Performance Gap:} While LLMs achieve high pass rate and coverage on standalone settings (Gray), performance drops significantly in the repository-level \textbf{\method} benchmark (Red).}
    \label{fig:gap_analysis}
\end{figure}

Writing unit tests is a time-consuming yet critical part of software development \citep{yuan2023no, pan2025aster, mohapatra2025artificial}; studies estimate it can take around 15\% of a developer's time. This has motivated decades of research into automated unit test generation. Traditional approaches \citep{evosuite, lukasczyk2022pynguin,noller2018badger,ognawala2019compositional,fraser2025retrospective} (e.g., search-based, symbolic, or random test generation) can produce tests with reasonable coverage, but these auto-generated tests often suffer from poor readability and trivial or ineffective assertions. Recently, large language models (LLMs) have emerged as a promising alternative, significantly accelerating unit test generation and often producing more natural, human-like test cases \citep{schafer2023empirical, lemieux2023codamosa, alagarsamy2024a3test, pan2025aster, yuan2024evaluating,watson2020learning,tufano2022generating,dakhel2024effective,siddiq2023exploring,bhatia2024unit,li2024large,serra2019effectiveness,yang2024enhancing,le2025impacts,zhang2024llm,yang2024evaluation}. Although LLMs have been widely adopted for unit test generation, their evaluation still faces significant challenges in practical applicability and multilingual adaptability. Existing test generation benchmarks often fail to reflect realistic repository-level constraints and remain biased toward a small set of programming languages, limiting our understanding of how LLMs perform in domain-specific and language-diverse software systems.

\textit{First}, various existing unit test generation benchmarks operate at the standalone, function-level, isolating code from the repository contexts in which real-world tests are written and executed \citep{chaimceval, testeval, chen2021evaluating}. In these settings, models are not required to interact with realistic build systems, dependency graphs, or framework conventions, leading to an inflated impression of readiness for practical deployment. Figure~\ref{fig:gap_analysis} illustrates this discrepancy: strong performance on standalone benchmarks (i.e., McEval~\cite{chaimceval}) can substantially overestimate model effectiveness when tests must be generated and executed within \textit{real repositories}. In repository-level settings, a model must not only write syntactically plausible tests, but also resolve project-specific dependencies, follow framework conventions, satisfy build systems, and construct valid inputs and assertions under incomplete context. These requirements sharply reduce both feasibility and semantic correctness, revealing a persistent gap between standalone success and real-world unit test generation.

\textit{Second}, although some recent frameworks and benchmarks consider repository-level test generation scenarios, they remain disproportionately centered on Java and Python \citep{schafer2023empirical, pan2025aster, jaintestgeneval}. This narrow language focus limits the evaluation of LLM-based test generation methods across diverse programming ecosystems, hindering the development and assessment of tools that are generalizable and practically useful in multilingual software environments \citep{celik2025review}.  Meanwhile, different programming languages exhibit distinct characteristics and domain-specific usage patterns. For instance, Rust has rapidly become a standard choice for memory-safe systems programming; Go underpins much of cloud-native infrastructure; Julia is driving advances in high-performance scientific computing; and PHP and Ruby continue to power a substantial portion of production web applications \citep{myklebust2025enhancing, gaher2024refinedrust, collins2025programming}. These languages also introduce unique technical challenges; such as Rust's ownership and borrowing model, that can invalidate generation strategies effective in more conventional language settings \citep{cheng2025rug}.
\begin{figure*}[ht]
\centering
\safeincludegraphics[width=0.95\textwidth]{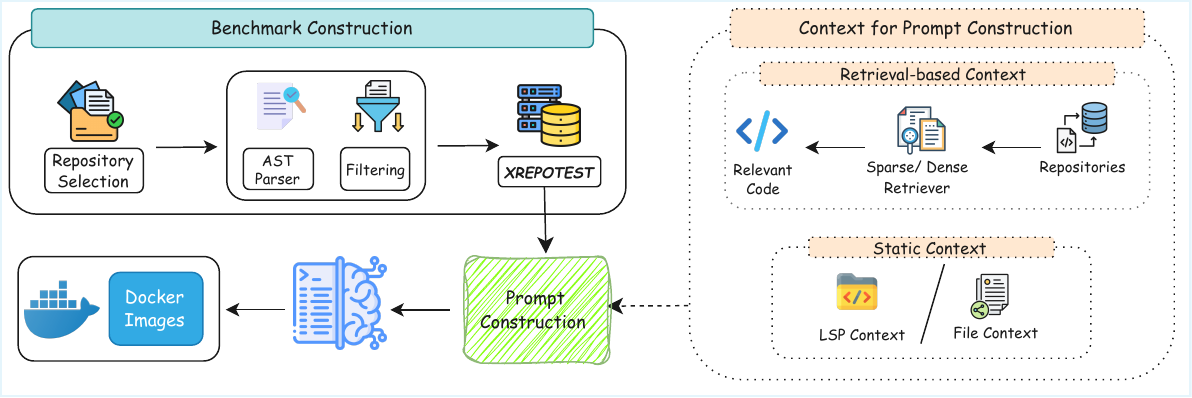}
\caption{Overview of the \method benchmark construction and context-aware evaluation workflow. 
\textbf{Left: Benchmark construction and evaluation.} Repositories are selected, parsed, and filtered to construct the benchmark, after which generated unit tests are executed in containerized, language-native environments for reproducible repository-level evaluation. 
\textbf{Right: Context for prompt construction.}}
  \label{figs:data_pipeline}
\end{figure*}

To bridge these gaps, we introduce \textbf{\method}, a large-scale multilingual\footnote{\textit{Throughout this paper, ``multilingual'' denotes coverage across programming languages (Rust, Go, Julia, PHP, and Ruby), not natural languages. Thus, ``language coverage'' and ``representation'' refer exclusively to programming-language ecosystems in our benchmark, distinct from NLP notions of human-language representation.}}, repository-level benchmark for unit test generation spanning five high-impact yet underexplored languages: Rust, Go, Julia, PHP, and Ruby. \method provides rich repository context and systematically evaluates multiple context strategies, including retrieval-based, file-level, and LSP-based augmentation, revealing trade-offs between test effectiveness and invocation reliability. In addition, we propose \textbf{Invocation Rate (IR)} as a complementary metric to pass rate and coverage, enabling a more faithful assessment of whether generated tests truly exercise the intended functionality. Our comprehensive experiments reveal that in the standard setting, performance is primarily bottlenecked by API hallucinations; while enhancing context improves pass rates and coverage, it often introduces noise that reduces IR, highlighting a complex trade-off between test effectiveness and reliability. In summary, our contributions are as follows:
\begin{itemize}[leftmargin=*]

\item \textbf{Benchmarking Multilingual Repo-level Test Generation:} We present \method, a curated benchmark of over $3,642$ focal functions from real-world Rust, Go, Julia, PHP, and Ruby repositories, designed to reflect diverse, real-world project code and testing conventions across these ecosystems.

\item \textbf{Comprehensive Evaluation Framework:} We provide a containerized evaluation framework that automatically sets up and executes generated tests across all considered languages. Our framework enables reliable, execution-based evaluation using four standard metrics: test pass rate (TPR), compilation success rate (CSR), line coverage, and mutation score; and introduces a novel \emph{invocation rate (IR)} to identify cases where tests pass but fail to meaningfully exercise the focal function, potentially introducing low-quality tests and technical debt.

\item \textbf{Repository-aware Context:} To examine the influence of varying context, our framework and dataset provide rich metadata and tooling to support both static and dynamic context enhancement. Static context includes File-level context and LSP-based symbol resolution (e.g., argument types and dependency definitions), while dynamic context is enabled via retrieval-augmented approaches using both sparse and dense retrieval methods.

\item \textbf{Extensive LLMs Evaluation:} We conduct comprehensive experiments on 14 state-of-the-art LLMs spanning families and scales, evaluating diverse context configurations. Our analysis reveals systematic strengths and limitations of current models and demonstrates how context design critically shapes the quality of generated tests.

\end{itemize}

\section{Related Work}
\label{sec:related_work}

Automated unit test generation has a long history, spanning traditional search-based and heuristic techniques as well as more recent LLM-driven approaches. Classic generators can achieve reasonable coverage, but they often produce brittle, low-readability tests with trivial or uninformative assertions, limiting adoption in practice \citep{evosuite, lukasczyk2022pynguin}. In contrast, LLM-based systems can generate more natural test scaffolding and assertions, further incorporate execution-in-the-loop repair or analysis-guided prompting to improve validity \citep{chen2024chatunitest, pan2025aster, rao2023cat, ma2025dynamic, Gorinski2023AutomaticUT, xiong-etal-2024-program}.

From an evaluation perspective, several high-quality benchmarks for unit test generation have incorporated repository context, such as Defects4J \citep{defects4J}, Methods2Test \citep{methods2test}, and TestGenEval \citep{jaintestgeneval}, but they remain largely centered on high-resource languages like Java and Python, limiting their applicability to other ecosystems. More recently, UniTSyn \citep{unitsyn} introduced a multilingual test generation framework; however, its evaluation is adapted from standalone code generation benchmarks such as HumanEval-X \citep{zheng2023codegeex} and McEval \citep{chaimceval}, which abstract away repository-level constraints and thus provide limited insight into the practical effectiveness of test generation methods in real-world settings.

Alongside these benchmarks, recent software-engineering datasets have increasingly targeted broader coding and software-evolution tasks, where test generation may arise as part of the problem-solving process. SWE-Compass~\citep{swecompass} evaluates holistic agentic coding abilities across task types and programming languages, while SWE-EVO~\citep{sweevo} focuses on long-horizon, single-language (Python) code evolution with tests as the validation oracle. Standalone evaluation of test generation is complementary to these broader settings, as \method isolates the test-generation step itself through a fine-grained, focal-function-centred diagnostic across five diverse programming-language ecosystems. This complements the growing body of dedicated test-generation benchmarks, such as TestGenEval~\citep{jaintestgeneval}, alongside broader agentic software-engineering suites.

\paragraph{\textbf{Gap for Multilingual Repository-level Evaluation.}} Despite the widespread use of many programming languages in practice, benchmarks for evaluating unit test generation at the repository level remain scarce outside a few high-resource ecosystems. A key barrier lies in constructing reliable execution environments, which requires language-specific knowledge of build systems, testing frameworks, and dependency configurations \citep{le2025impacts, zhang-etal-2025-di, zhang2025build, testeval, islam-etal-2024-mapcoder, nguyen2025codemmlu, du-etal-2025-codearena}. To address this gap, we present a large-scale multilingual, repository-level benchmark and data pipeline for test generation, enabling systematic and executable evaluation of test generation models across diverse programming languages.

\begin{table*}[t]
\centering
\begin{adjustbox}{width=0.85\textwidth}
\begin{tabular}{llrccc}
\toprule
\textbf{Benchmark} & \textbf{Language} & \textbf{Size} & \textbf{PF} & \textbf{FC} & \textbf{RC} \\
\midrule
MBPP~\cite{mbpp} & Python & 500 & \xmark & \xmark  & \xmark \\
HumanEval~\cite{chen2021evaluating} & Python & 164  & \xmark & \xmark & \xmark \\
TestEval~\cite{testeval} & Python & 210 & \xmark  & \xmark &\xmark  \\
McEval~\cite{chaimceval} & 40 languages & 16,031 & \xmark  & \xmark &\xmark  \\
SF110 \cite{evosuite} & Java & 182 & \cmark  & \xmark & \xmark \\
Defects4J~\cite{defects4J}& Java & 357 & \cmark  & \cmark  & \xmark  \\
TestPilot \cite{schafer2023empirical} & JavaScript & 1,684 & \cmark & \cmark & \xmark  \\
TestGenEval~\cite{jaintestgeneval} & Python & 1,210 & \cmark  & \cmark  & \xmark  \\
\hline 
        \textbf{\method} & Rust, Go, Julia, PHP, Ruby & $3{,}642$ & \cmark & \cmark & \cmark \\
\bottomrule
\end{tabular}
\end{adjustbox}
\caption{Comparison of test generation benchmarks by language, size, and support for project-level functions (\textbf{PF}), file context (\textbf{FC}), repository context (\textbf{RC}).}
\label{tab:testgen_comparison}
\end{table*}

\section{Methodology}
\label{sec:method}

In this section, we describe the construction of \method and its context-aware evaluation framework, followed by an overview of the dataset's scale, language coverage, and domain distribution. \method is repository-level rather than standalone: each task targets a focal function, but generated tests are inserted into and validated within the original project environment. These details highlight the design choices that make \method a robust benchmark for evaluating LLM-based unit test generation.

\subsection{Dataset Construction}
\label{sec:pipeline}

Figure~\ref{figs:data_pipeline} illustrates the end-to-end workflow of \method, which consists of three conceptual stages. First, during \textbf{Benchmark Construction}, we select repositories and extract non-trivial focal functions to form the core benchmark. Second, during \textbf{Prompt Construction}, each focal function is paired with contextual information drawn from well-defined context sources. This includes (i) \emph{static context}, released as part of the benchmark, and optionally (ii) \emph{retrieval-based (dynamic) context}, constructed on the fly at evaluation time. Finally, generated unit tests are executed and validated in \textbf{Containerized Environments} to ensure reproducible, repository-level evaluation. More detailed implementations of the pipeline components, including filtering rules, context augmentation, and containerized environments, are provided in Appendix \ref{sec:impl_details}. Besides, the discussion on data contamination risk is provided in Appendix \ref{app:dataleak}.

\paragraph{\textbf{Language Selection.}} We exclude Java and Python due to their extensive prior study and focus instead on PHP, Go, Ruby, and Rust, that are widely used languages in production, but remain understudied in repository-level test generation research. We also include Julia, a relatively low-resource language that has seen growing adoption and research interest \citep{joel2024survey, giagnorio2025enhancing}.

\paragraph{\textbf{Repository Selection.}} For each language, to ensure quality and relevance, we apply filtering criteria including: a minimum of 500 GitHub stars, a clearly defined project structure with build configuration files (e.g., \texttt{Cargo.toml}, \texttt{Project.toml}, \texttt{go.mod}, \texttt{composer.json}, or \texttt{Gemfile}) to minimize human intervention in environment setup. Additionally, for each language, we ensure that selected repositories span at least two distinct application domains to promote diversity in the dataset. Following this process, we select 6--10 repositories per language, which together form the foundation of our benchmark dataset (See additional per-repository statistics in Appendix~\ref{sec:repo_stats}, Table~\ref{tab:repo_statistics_full}).

\paragraph{\textbf{Focal Method Extraction.}} From collected repositories, we use \texttt{tree-sitter}~\cite{treesitter} to parse source files into abstract syntax trees (ASTs), enabling language-agnostic traversal and structural analysis. From each AST, we extract top-level function and method definitions. To ensure functions' quality and validity, we apply systematic filtering rules that enforce: (i) minimum code length, (ii) presence of observable outputs (explicit \texttt{return} statements), (iii) exclusion of existing test code, and (iv) syntactic validity. These criteria help eliminate trivial or degenerate functions while retaining semantically meaningful \textit{focal methods} suitable for unit test generation. Complete filtering rules are detailed in Table~\ref{tab:filtering_rule} in the Appendix.

\paragraph{Context Augmentation.} To enable realistic and flexible repository-level test generation, our framework supports both static and dynamic context augmentation. The released \method benchmark provides precise, lightweight static context with \textbf{File-level context} (i.e. file content containing focal method) and \textbf{LSP-based context} that resolves argument definitions, type declarations, and relevant callee signatures via Language Server Protocol (LSP), inspired by recent studies \citep{jaintestgeneval, go2025lsprag}. In addition, following repository-level code generation practices \citep{zhang2023repocoder, wu2024repoformer, wang2024rlcoder, le2026not}, \method supports optional \textbf{Retrieval-based augmentation}, in which sparse (BM25) or dense (UniXCoder - \citet{guo2022unixcoder}) retrievers dynamically select relevant code snippets (i.e., code chunks) from the repository to enrich the prompt.

\paragraph{\textbf{Containerized Execution Environment.}}
All generated unit tests are executed within Docker containers to standardize toolchains, dependency installation, and evaluation across languages. Each repository is evaluated using its native test framework (e.g., \texttt{cargo test}, \texttt{go test}, \texttt{Test.jl},
\texttt{PHPUnit}, \texttt{RSpec}).

Overall, our pipeline provides a flexible and extensible foundation for repository-level unit test generation. Although this work focuses on five programming languages, the framework is designed to scale to additional languages with minimal effort, requiring only the availability of a supported testing framework and appropriate configuration files to set up the execution environment.

\subsection{Data Characteristics}

\paragraph{\textbf{Data Statistics.}}
\method spans six diverse application domains, reflecting language-specific usage patterns across the selected programming languages. We additionally ensure a balanced number of samples per language to support fair and comparable evaluation across multilingual settings. Detailed repository statistics and domain distributions are provided in Appendix~\ref{sec:repo_stats}.

\paragraph{\textbf{Comparison of Unit Test Generation Benchmarks.}} Table \ref{tab:testgen_comparison} compares existing unit test generation benchmarks in terms of language coverage, scale, and support context. Most prior benchmarks focus on standalone or file-level settings and are largely confined to high-resource languages such as Python and Java, with limited or no support for full repository context. In contrast, \method uniquely combines multilingual coverage with repository-level evaluation, supporting five underexplored languages and providing comprehensive focal, file, and repository-aware context, thereby enabling more practical assessment.

\begin{algorithm}[t]
\caption{Computation of Invocation Rate (IR)}
\label{alg:invocation_rate}
\begin{algorithmic}[1]
\Require Set of generated unit tests $T$, focal method name $f$
\Ensure Invocation Rate $IR$

\Function{ComputeIR}{$T, f$}
    \State $N \gets |T|$, $count \gets 0$
    \ForAll{test method $t \in T$}
        \State $AST \gets \texttt{AST\_parse}(t)$
        \State $C \gets \texttt{GetCallIdentifiers}(AST)$
        \State $count \gets count + 1$ \textbf{if} $f \in C$ 
    \EndFor
    \State $IR \gets \frac{count}{N}$
    \State \Return $IR$
\EndFunction
\end{algorithmic}
\end{algorithm}

\begin{table*}[t]
\centering
\begin{adjustbox}{width=\textwidth}
\begin{tabular}{l *{5}{ccc}}
\toprule
\multirow{2}{*}{\textbf{Model}} & \multicolumn{3}{c}{\textbf{Rust}} & \multicolumn{3}{c}{\textbf{Julia}} & \multicolumn{3}{c}{\textbf{Go}} & \multicolumn{3}{c}{\textbf{Ruby}} & \multicolumn{3}{c}{\textbf{PHP}} \\
\cmidrule(lr){2-4} \cmidrule(lr){5-7} \cmidrule(lr){8-10} \cmidrule(lr){11-13} \cmidrule(lr){14-16}
 & TPR & Cov & IR & TPR & Cov & IR & TPR & Cov & IR & TPR & Cov & IR & TPR & Cov & IR \\
\midrule

\multicolumn{16}{c}{\textit{\textbf{Small}}} \\
\midrule
Qwen3 (8B)         & 0.32 & 0.21 & 64.88 & 0.88 & 0.97 & 73.83 & 0.42 & 1.13 & 47.17 & 0.00 & 1.39 & \cellcolor{pink!50}\textbf{92.15} & 0.62 & 2.13 & 22.91 \\
Yi-Coder (9B)      & 1.50 & 1.71 & \cellcolor{pink!50}\textbf{90.23} & 4.82 & 19.16 & 90.50 & 2.41 & 6.48 & 91.36 & 1.48 & 9.37 & 86.81 & 3.56 & 17.68 & 98.76 \\
Codestral (22B)    & 2.69 & 2.97 & 84.85 & 2.05 & 21.17 & 91.23 & 3.82 & 14.52 & \cellcolor{pink!50}\textbf{99.58} & 0.15 & 7.35 & 87.26 & 5.57 & 23.01 & 99.69 \\
\midrule

\multicolumn{16}{c}{\textit{\textbf{Medium}}} \\
\midrule
Qwen3-Coder-Next (80B)   & 7.20 & 3.52 & 56.71 & 5.12 & 17.38 & 91.08 & 6.66 & 16.74 & 96.88 & 1.19 & 6.88 & 87.26 & 5.11 & 22.43 & 97.68 \\
GPT-OSS (120B)     & 5.80 & 4.16 & 81.63 & 4.39 & 24.27 & 93.86 & 19.26 & 26.95 & 88.24 & 6.96 & 8.21 & 85.93 & 3.41 & 5.07 & 99.23 \\
Llama-3.3 (70B)    & 3.33 & 3.04 & 88.94 & 3.07 & 25.42 & 93.27 & 2.69 & 7.96 & 98.73 & 0.44 & 7.66 & 87.85 & 2.01 & 21.75 & \cellcolor{pink!50}\textbf{99.85} \\
\midrule

\multicolumn{16}{c}{\textit{\textbf{Large}}} \\
\midrule
DeepSeek V3.2      & 2.79 & 3.06 & 77.66 & 3.95 & 21.94 & 89.47 & 5.24 & 14.91 & 59.07 & 1.48 & 10.01 & 84.30 & 5.42 & 20.20 & 93.34 \\
MiniMax-M2.7       & 4.30 & 3.63 & 66.49 & 4.39 & 18.12 & 86.11 & 4.67 & 13.07 & 81.16 & 1.63 & 7.30 & 87.41 & 5.57 & 17.25 & 97.68 \\
\midrule

\multicolumn{16}{c}{\textit{\textbf{Flagship}}} \\
\midrule
Claude 4.5 Haiku   & 8.27 & 3.73 & 80.34 & 12.57 & 23.96 & 92.69 & 10.91 & 26.87 & 99.43 & 4.30 & 10.57 & 87.11 & 14.86 & 34.68 & 99.54 \\
Claude 4.5 Sonnet  & \cellcolor{pink!50}\textbf{12.78} & \cellcolor{pink!50}\textbf{11.56} & 75.94 & \cellcolor{pink!50}\textbf{19.88} & 31.58 & \cellcolor{pink!50}\textbf{94.01} & 23.65 & \cellcolor{pink!50}\textbf{43.53} & 99.43 & 6.37 & \cellcolor{pink!50}\textbf{14.25} & 85.63 & \cellcolor{pink!50}\textbf{26.93} & \cellcolor{pink!50}\textbf{41.76} & 99.38 \\
DeepSeek V4-pro    & 6.66 & 4.66 & 74.22 & 11.40 & 26.83 & 89.77 & 13.88 & 21.66 & 63.46 & 8.89 & 9.15 & 84.74 & 13.00 & 22.19 & 96.13 \\
GLM-5             & 6.44 & 2.52 & 70.68 & 5.41 & 13.54 & 89.47 & 5.10 & 9.48 & 75.78 & 3.11 & 8.50 & 86.96 & 13.16 & 24.98 & 99.07 \\
Kimi-k2.5 (Instruct) & 1.40 & 1.17 & 74.76 & 2.63 & 8.70 & 87.43 & 2.27 & 2.33 & 75.21 & 2.52 & 5.02 & 78.96 & 3.25 & 4.02 & 88.08 \\
GPT-5.2            & 12.24 & 9.31 & 77.87 & 16.52 & \cellcolor{pink!50}\textbf{31.92} & 91.81 & \cellcolor{pink!50}\textbf{25.78} & 41.30 & 98.44 & \cellcolor{pink!50}\textbf{25.78} & 7.80 & 88.89 & 10.06 & 23.17 & 99.38 \\
\bottomrule
\end{tabular}
\end{adjustbox}
\caption{LLMs Performance on \method. Best scores are highlighted. Full metrics are in the Appendix \ref{tab:full_results_standard}.}
\label{tab:compact_results}
\end{table*}
\section{Evaluation Metrics}
\label{subsec:evaluation metrics}

\paragraph{\textbf{Standard Metrics.}} To evaluate the effectiveness of generated unit tests, we adopt a set of metrics commonly used in prior work on automated test generation \citep{methods2test, alagarsamy2024a3test, jaintestgeneval, chen2024chatunitest, lukasczyk2022pynguin}. \textbf{Test Pass Rate (TPR)} measures runtime correctness as the fraction of compiled tests that execute without exceptions or assertion failures. To assess semantic coverage, we report \textbf{Line Coverage (Cov)}, which measures the percentage of lines in the focal method exercised by the generated tests. We additionally compute \textbf{Compilation Success Rate} (CSR) and \textbf{Mutation Score}; we report a summary of both in \S\ref{rq:rq1} and present the full per-model values in Appendix Tables~\ref{tab:full_results_standard} and~\ref{tab:mutation_results_appendix}. Full metric definitions are provided in Appendix \ref{sec:metrics_calculation}.

\begin{tcblisting}{
enhanced,
    float,
    floatplacement=t,
    title={Example: Unintended indirect invocation}, % 3. Hiển thị số đếm
    label={lst:example_ir},      % 4. Gắn nhãn để reference
    % ----------------------------------
    colback=gray!5,
    top=1pt,
    bottom=1pt,
    colframe=gray!30!black,
    arc=2mm,
    boxrule=1pt,
    listing only,
    listing options={
        language=Go,
        basicstyle=\ttfamily\footnotesize\linespread{0.15}\selectfont,
        numbers=left,
        numberstyle=\tiny,
        numbersep=5pt,
        breaklines=true,
        columns=fullflexible,
        escapeinside={(*@}{@*)}, 
    }
}
(*@\textbf{\textcolor{blue!60!black}{\sffamily [Helper Function]}}@*)
func Circle[...](arr []T) []T { 
     ...
    for doSort(...) { /* ... */ }
     ...
}


(*@\textbf{\textcolor{orange!70!black}{\sffamily [Focal Method]}}@*)
func doSort[...](arr []T, l, r int) bool { 
     ...
}

(*@\textbf{\textcolor{green!40!black}{\sffamily [Generated Test]}}@*)
func TestDoSort(t *testing.T) { 
    var tests = []struct{ /* ... */ }
    for _, tt := range tests {
        got := Circle(tt.input)
        if !reflect.DeepEqual(got, tt.want){
             t.Errorf("...")
        }
    }
}
\end{tcblisting}

\paragraph{\textbf{Novel Proposed Metric.}} Standard evaluation metrics, such as TPR, CSR and Cov, often fail to distinguish between \textit{accidental execution} and \textit{intentional validation}. We observe that LLMs frequently produce ``passing'' tests that contain critical semantic flaws, notably \textit{indirect invocations} where the focal method is executed as a side effect of a wrapper function rather than being explicitly targeted. Example~\ref{lst:example_ir} provides an illustrative example of this issue, where the test calls the wrapper \texttt{Circle} instead of the focal method \texttt{doSort}. In these scenarios, coverage metrics are inflated, recording the focal method as ``covered'' even though its specific logic is never directly exercised or asserted. To bridge this gap between \textit{execution} and \textit{intent}, we introduce the \textbf{Invocation Rate (IR)}. This metric explicitly verifies whether a generated test contains a direct call to the focal function under test. By distinguishing these misleading indirect invocations from genuine unit tests, IR serves as a strict indicator of ``behavioral relevance'', ensuring that reported high TPR and Cov scores actually reflect the model's ability to engage with the focal method. The IR implementation is detailed in Algorithm~\ref{alg:invocation_rate}.

Together, these metrics provide a comprehensive paradigm on both the \emph{syntactic soundness} and \emph{execution effectiveness} to evaluate the generated unit tests across different test generation techniques.

\section{Findings and Discussion}
\label{sec:exp}

\subsection{Performance across LLMs on \method}\label{rq:rq1}

In this experiment, we evaluate 14 state-of-the-art LLMs across model families and scales on \method using minimal \emph{focal/standard context} (the focal method and class signature), following prior studies \citep{methods2test, tufano2020unit, yang2024evaluation}. We first summarize cross-model trends and then analyze results across individual languages. All languages in \method are evaluated under a unified protocol using greedy decoding. Details of the models and prompting templates are provided in Appendices \ref{app:model_details} and \ref{app:prompt}.

\paragraph{\textbf{Overall trends across models.}}
Table~\ref{tab:compact_results} indicates that unit test generation remains challenging across the five languages, even for frontier systems.
Across languages, the strongest models tend to be instruction-following and more reliable in producing executable test scaffolding, but passing tests remains bottlenecked by correct setup and oracles.
Model family effects are pronounced: frontier models (Claude 4.5 and GPT-5.2) dominate many of the highest scores per metric, while strong open-weight baselines (e.g., Qwen variants and GPT-OSS) remain competitive.
We also observe that GPT-5.2 is consistently strong on several languages/metrics (e.g., Ruby TPR), but it does not uniformly dominate across the suite, emphasizing that test generation capability remains uneven.
Notably, the best model differs by language and metric (e.g., coverage vs. pass rate), reinforcing that improvements are not uniform and that \method exposes distinct capability gaps.

\paragraph{\textbf{Auxiliary quality signals.}} We additionally report \textbf{Compilation Success Rate} (CSR) and \textbf{Mutation Score} (MS), whose full per-model values are in Tables~\ref{tab:full_results_standard} and~\ref{tab:mutation_results_appendix}: on average, models compile 57.4\% of generated tests (CSR) and reach a mean MS of 3.2\% across Go, Rust, and Ruby (Ruby MS values are lower-bound estimates).

\paragraph{\textbf{Language-specific performance and challenges:}}

\noindent\textbf{\textit{Rust.}} Results highlight a scaffolding bottleneck: models frequently fail to assemble correct imports/crates and satisfy strict type and ownership constraints.
Even strong systems often reach the focal method but do not reliably produce behavior-checking tests.
Under the standard setting, Claude 4.5 Sonnet achieves the best Rust performance (TPR 12.78\%, Cov 11.56\%, IR 75.94\%), while GPT-5.2 reaches TPR 12.24\% and Cov 9.31\%.

\noindent\textbf{\textit{Go.}} Models achieve higher executed coverage than Rust, but passing tests remain bottlenecked by correct oracles and setup, indicating semantic correctness is the primary limiter.
GPT-5.2 leads on TPR (25.78\%) while Claude 4.5 Sonnet achieves the highest Cov (43.53\%), with both models maintaining near-perfect IR ($\geq$98\%).

\noindent\textbf{\textit{Julia.}} Performance is moderate: Claude 4.5 Sonnet achieves the best TPR (19.88\%) and GPT-5.2 the best Cov (31.92\%), while small models (e.g., Qwen3-8B at TPR 0.88\%) lag far behind. IR is high for most models ($\geq$87\%), though smaller models such as Qwen3-8B (73.83\%) are exceptions, indicating that focal method invocation is generally not the bottleneck — semantic correctness is.

\noindent\textbf{\textit{Ruby.}} Results are notably split: GPT-5.2 leads on TPR (25.78\%) but achieves only 7.80\% Cov, while Claude 4.5 Sonnet leads on Cov (14.25\%) with a much lower TPR (6.37\%). This TPR–Cov divergence suggests models can exercise code paths without producing passing assertions, reflecting the difficulty of Ruby's DSL-heavy testing conventions.

\noindent\textbf{\textit{PHP.}} Claude 4.5 Sonnet dominates with TPR 26.93\% and Cov 41.76\%, while GPT-5.2 (TPR 10.06\%) and open-weight models trail significantly. IR is near-perfect ($\geq$97\%) for most models, though smaller models such as Qwen3-8B (22.91\%) and Kimi-k2.5 (88.08\%) are notable exceptions, making PHP the language where invocation is easiest for capable models but test correctness remains the challenge.

\paragraph{Unit Test Quality vs Software Engineering Capability.} We observe strong positive correlations between SWE-bench \citep{swebench} performance and TPR/Cov, suggesting partial transfer from general software engineering capability to test generation. However, the correlation is incomplete, motivating complementary metrics such as IR. Full analysis is in Appendix~\ref{sec:correlation_analysis}.

\begin{table}[t]
\centering
\begin{adjustbox}{width=0.48\textwidth}
\begin{tabular}{@{}ll*{9}{c}@{}}
	\toprule
\multirow{2}{*}{\textbf{Language}} & \multirow{2}{*}{\textbf{Setting}} &
\multicolumn{3}{c}{\textbf{Claude 4.5 Sonnet}} &
\multicolumn{3}{c}{\textbf{GPT-5.2}} &
\multicolumn{3}{c}{\textbf{GPT-OSS}} \\
\cmidrule(lr){3-5}\cmidrule(lr){6-8}\cmidrule(lr){9-11}
& & \textbf{TPR} & \textbf{Cov} & \textbf{IR}
  & \textbf{TPR} & \textbf{Cov} & \textbf{IR}
  & \textbf{TPR} & \textbf{Cov} & \textbf{IR} \\
\midrule
\multirow{5}{*}{\textit{Rust}} & Standard  & 12.78 & 11.56 & \cellcolor{pink!50}\textbf{75.94} & 12.24 & 9.31 & 77.87 & 5.80 & 4.16 & 81.63 \\
 & Sparse & 16.00 & 15.29 & 69.07 & 15.36 & 13.01 & 83.24 & 4.94 & 4.56 & 82.28 \\
 & Dense  & 17.19 & 15.52 & 71.75 & 13.86 & 12.40 & \cellcolor{pink!50}\textbf{84.32} & 4.51 & 4.94 & \cellcolor{pink!50}\textbf{82.38} \\
 & File   & \cellcolor{pink!50}\textbf{21.16} & \cellcolor{pink!50}\textbf{18.93} & 68.53 & \cellcolor{pink!50}\textbf{17.94} & \cellcolor{pink!50}\textbf{17.43} & 83.35 & \cellcolor{pink!50}\textbf{8.06} & \cellcolor{pink!50}\textbf{6.67} & 80.56 \\
 & LSP    & 13.53 & 13.24 & 72.18 & 11.49 & 7.04 & 83.78 & 6.44 & 4.98 & 82.28 \\
\midrule
\multirow{5}{*}{\textit{Go}} & Standard  & 23.65 & 43.53 & \cellcolor{pink!50}\textbf{99.43} & 25.78 & 41.30 & 98.44 & 19.26 & 26.95 & 88.24 \\
 & Sparse & 28.05 & 46.95 & 97.17 & 28.90 & 43.10 & \cellcolor{pink!50}\textbf{99.15} & 18.98 & 27.20 & 95.61 \\
 & Dense  & 30.31 & 49.17 & 97.59 & 32.15 & 46.42 & 98.87 & 17.71 & 27.90 & 95.75 \\
 & File   & \cellcolor{pink!50}\textbf{34.14} & \cellcolor{pink!50}\textbf{52.63} & 97.88 & \cellcolor{pink!50}\textbf{37.68} & \cellcolor{pink!50}\textbf{58.11} & 98.87 & \cellcolor{pink!50}\textbf{25.92} & \cellcolor{pink!50}\textbf{39.99} & \cellcolor{pink!50}\textbf{96.74} \\
  & LSP    & 25.35 & 43.42 & 97.03 & 28.05 & 42.50 & 98.44 & 17.42 & 27.31 & 96.60 \\
\midrule
\multirow{5}{*}{\textit{Julia}} & Standard  & 19.88 & 31.58 & 94.01 & 16.52 & 31.92 & 91.81 & 4.39 & 24.27 & 93.86 \\
 & Sparse & 19.15 & 31.35 & 94.01 & 11.40 & 30.53 & 93.86 & 6.43 & 27.32 & 93.27 \\
 & Dense  & 18.57 & 31.21 & 94.44 & 11.70 & 30.18 & 93.86 & \cellcolor{pink!50}\textbf{10.09} & 26.96 & 93.27 \\
 & File   & \cellcolor{pink!50}\textbf{23.98} & \cellcolor{pink!50}\textbf{34.83} & 94.15 & \cellcolor{pink!50}\textbf{17.25} & \cellcolor{pink!50}\textbf{34.35} & 93.71 & 9.50 & \cellcolor{pink!50}\textbf{32.07} & 92.25 \\
 & LSP    & 15.64 & 29.22 & \cellcolor{pink!50}\textbf{95.03} & 12.87 & 25.83 & \cellcolor{pink!50}\textbf{94.01} & 5.85 & 20.29 & \cellcolor{pink!50}\textbf{94.74} \\
\midrule
\multirow{5}{*}{\textit{Ruby}} & Standard  & 6.37 & 14.25 & \cellcolor{pink!50}\textbf{85.63} & \cellcolor{pink!50}\textbf{25.78} & 7.80 & 88.89 & 6.96 & 8.21 & 85.93 \\
 & Sparse & 10.37 & 18.86 & 71.11 & 19.85 & 11.15 & 87.41 & 17.14 & 13.56 & \cellcolor{pink!50}\textbf{87.14} \\
 & Dense  & 11.26 & 20.97 & 68.15 & 21.04 & 14.82 & 85.33 & \cellcolor{pink!50}\textbf{18.57} & 15.24 & 85.71 \\
 & File   & \cellcolor{pink!50}\textbf{16.30} & \cellcolor{pink!50}\textbf{23.51} & 53.19 & 13.04 & \cellcolor{pink!50}\textbf{18.35} & 60.89 & 13.71 & \cellcolor{pink!50}\textbf{26.71} & 64.57 \\
 & LSP    & 7.70 & 14.55 & 85.19 & 23.85 & 7.95 & \cellcolor{pink!50}\textbf{89.19} & 7.56 & 7.93 & 86.37 \\
\midrule
\multirow{5}{*}{\textit{PHP}} & Standard  & 26.93 & 41.76 & 99.38 & 10.06 & 23.17 & 99.38 & 3.41 & 5.07 & 99.23 \\
 & Sparse & 30.03 & 45.09 & 99.38 & 13.16 & 29.39 & 98.76 & 3.41 & 3.21 & \cellcolor{pink!50}\textbf{99.85} \\
 & Dense  & \cellcolor{pink!50}\textbf{31.42} & 45.14 & 99.38 & 16.10 & 30.73 & \cellcolor{pink!50}\textbf{99.69} & 4.95 & 4.63 & 99.23 \\
 & File   & \cellcolor{pink!50}\textbf{31.42} & \cellcolor{pink!50}\textbf{47.25} & \cellcolor{pink!50}\textbf{99.54} & \cellcolor{pink!50}\textbf{22.91} & \cellcolor{pink!50}\textbf{38.64} & 99.54 & \cellcolor{pink!50}\textbf{6.35} & 5.64 & 99.54 \\
 & LSP    & 26.78 & 44.62 & 99.07 & 11.76 & 26.68 & 99.38 & 4.49 & \cellcolor{pink!50}\textbf{5.65} & 99.69 \\
\bottomrule
\end{tabular}
\end{adjustbox}
\caption{Unit test generation performance across context settings for three frontier models. Best scores per language (within each metric) are bolded and highlighted.}
\label{tab:unified_context_enhancement_results}
\end{table}

\subsection{Impact of Context Augmentation}\label{rq:rq3}
Building on the earlier analyses, we observe that models excel in domains with standalone functions but struggle when broader project-level context is required (Figure \ref{fig:gap_analysis}). This missing context often leads to \textit{API Hallucination} (Table \ref{tab:error_distribution}), as the model cannot fully resolve dependencies or references. In this section, we investigate context augmentation strategies; more implementation details are described in Appendix~\ref{sec:impl_details}.

\begin{table*}[t]
\centering
\begin{adjustbox}{width=\textwidth}
\begin{tabular}{l*{5}{cc}}
\toprule
\multirow{2}{*}{\textbf{Failure Mode}} & \multicolumn{2}{c}{\textbf{Rust}} & \multicolumn{2}{c}{\textbf{Go}} & \multicolumn{2}{c}{\textbf{Julia}} & \multicolumn{2}{c}{\textbf{Ruby}} & \multicolumn{2}{c}{\textbf{PHP}} \\
\cmidrule(lr){2-3} \cmidrule(lr){4-5} \cmidrule(lr){6-7} \cmidrule(lr){8-9} \cmidrule(lr){10-11}
 & \textbf{Std.} & \textbf{Ctx.} & \textbf{Std.} & \textbf{Ctx.} & \textbf{Std.} & \textbf{Ctx.} & \textbf{Std.} & \textbf{Ctx.} & \textbf{Std.} & \textbf{Ctx.} \\
\midrule
Syntactic \& Compilation & 3.94\% & 2.58\% & 24.98\% & 20.35\% & 15.55\% & 8.72\% & 1.68\% & 3.88\% & 17.23\% & 12.07\% \\
Type System \& Memory & 11.31\% & 14.82\% & 13.27\% & 8.45\% & 9.02\% & 8.33\% & 4.64\% & 2.29\% & 4.80\% & 2.94\% \\
API Hallucination & 48.05\% & 36.09\% & 18.89\% & 14.35\% & 47.90\% & 50.10\% & 42.17\% & 41.88\% & 35.55\% & 40.61\% \\
Logic \& Assertion & 26.42\% & 30.79\% & 19.97\% & 24.27\% & 13.94\% & 15.94\% & 38.42\% & 37.35\% & 28.95\% & 24.15\% \\
Test Design \& Mocking & 3.29\% & 4.98\% & 0.38\% & 0.99\% & 0.00\% & 0.00\% & 4.40\% & 0.59\% & 1.08\% & 0.67\% \\
\midrule
No Error / Pass & 6.98\% & 10.74\% & 22.51\% & 31.59\% & 13.59\% & 16.91\% & 8.69\% & 14.01\% & 12.39\% & 19.56\% \\
\bottomrule
\end{tabular}
\end{adjustbox}
\caption{Failure mode distribution across languages under standard (\textbf{Std.}) and file-level context (\textbf{Ctx.}) settings, computed as the fraction of total samples assigned to each failure mode. Rows sum to 100\% within each column; the ``No Error / Pass'' row captures samples that pass all checks without any failure.}
\label{tab:error_distribution}
\end{table*}

\paragraph{\textbf{Retrieval-based Augmentation.}} Retrieval methods show mixed results across languages and models (Table~\ref{tab:unified_context_enhancement_results}). For Go, Julia, and PHP with Claude 4.5 Sonnet, both sparse and dense retrieval improve TPR and Cov over the standard setting, but these gains are not universal: GPT-5.2 and GPT-OSS show inconsistent improvements, and Ruby exhibits IR drops despite some TPR/Cov gains. This variability suggests retrieval effectiveness depends heavily on model architecture and language characteristics. These retrieval configurations are standard reference settings rather than tuned per model; results are stable across moderate window and top-$k$ variations (Appendix~\ref{app:retrieval_ablation}, Table~\ref{tab:retrieval_ablation}), so these findings are not artifacts of a single configuration.

\paragraph{\textbf{Static Context Augmentation.}} File-level context is the most consistently effective strategy, achieving the best results for Go, Julia, and Rust across all settings (e.g., Claude: 34.14\% TPR, 52.63\% Cov on Go; 23.98\% TPR, 34.83\% Cov on Julia; 21.16\% TPR, 18.93\% Cov on Rust). However, on Ruby, file-level context sharply reduces IR (from 85.63\% to 53.19\%) despite improving TPR and coverage — a systematic trade-off across all three evaluated models (IR drops of 21--32\%), specific to Ruby's DSL-heavy, metaprogramming-rich ecosystem. Go and PHP show negligible IR changes ($\pm$2\%) for the top two models under the same setting, though GPT-OSS shows a larger IR gain in Go (+8.5\%). Julia IR changes are similarly small ($\leq$2\%) across all three models. LSP-based augmentation mirrors developer workflows but benefits are selective; models do not consistently leverage fine-grained dependency traces, and overly resolved context can become noisy.

\paragraph{\textbf{Key Takeaways.}}
File-level context is the most consistently effective augmentation, while retrieval and LSP-based approaches yield selective, language-dependent gains. Richer context can reduce IR, highlighting a trade-off between test effectiveness and invocation reliability. Overall, even the best models achieve at most $\sim$27\% TPR across all languages under the standard setting, confirming \method as a challenging benchmark. Statistical significance of key model comparisons is confirmed via bootstrap resampling and McNemar's test (Appendix~\ref{sec:significance}).

\subsection{Failure Modes Analysis}\label{sec:failure_modes}

To analyze failure modes, we collect test execution logs and apply systematic categorization through refinement and manual inspection. This yields five classes: \textit{Syntactic \& Compilation}, \textit{API Hallucination}, \textit{Type System \& Memory}, \textit{Logic \& Assertion}, and \textit{Test Design \& Mocking}, capturing recurring breakdown points across (i) runnable scaffolding, (ii) valid API usage, (iii) type/memory constraints, and (iv) behavioral correctness. Full definitions and taxonomy are in Appendix~\ref{app:error taxonomy}.

\vspace{-0.19cm}
\paragraph{\textbf{Analysis of Results:}}
Table~\ref{tab:error_distribution} reports failure modes normalized by total samples, so each column sums to 100\%. \textit{API Hallucination} is the dominant failure across most languages. In Rust, context augmentation reduces hallucinations but raises \textit{Logic \& Assertion} failures, indicating that richer context helps models produce runnable tests yet semantic correctness remains a bottleneck. In Julia, API hallucination slightly increases under context (47.90\%~$\to$~50.10\%), suggesting that additional context does not reliably ground Julia API usage and semantic correctness remains the primary challenge. Ruby's hallucination rate is largely unchanged by context.
Go shows a more distributed profile: context reduces both hallucination and compilation errors, but \textit{Logic \& Assertion} rises, confirming that oracle correctness is the primary limiter once scaffolding issues are resolved. PHP is an exception - context reduces compilation failures but increases API hallucination, suggesting that richer context introduces more opportunities to misuse framework-specific APIs.

Overall, improving API grounding and oracle quality is key to better test success. More examples and discussion are in Appendix~\ref{sec:error_details}.

\subsection{Agentic and Iterative-Repair Evaluation}\label{rq:rq4}
To go beyond single-turn prompting, we additionally evaluate \emph{agentic} and \emph{execution-in-the-loop repair} workflows under the same protocol (TPR, Cov, IR). Given the higher cost of agentic evaluation, we use a stratified subset sampling up to five easy and five hard focal functions per repository~\citep{swecompass, sweevo}.

\textbf{Agentic execution.} A headless Claude Code agent with full repository access (navigation, compilation, execution, iterative repair) improves TPR over passive single-turn generation by \textbf{1.2--2.6$\times$} across all five languages (Table~\ref{tab:agentic}). Crucially, the multi-metric design exposes what pass rate alone misses: on PHP the agent raises TPR by 11.5\% yet \emph{coverage drops} by 33.3\% and IR by 71.3\%, as passing tests exercise easier helper or public APIs rather than the assigned focal function; on Julia the agent reaches 100\% IR with lower coverage, since multiple dispatch selects a non-focal method overload. Aggregating over pass rate alone would overestimate agent capability; IR and coverage jointly expose these failure modes.

\textbf{Iterative repair.} A one-round repair loop (GPT-OSS-120B) that feeds compiler and execution feedback back to the model improves TPR by 9--26\% with IR nearly unchanged, confirming that \method supports iterative-repair workflows under one protocol.

\textbf{IR vs.\ TPR disagreement.} Across the full benchmark, \textbf{9.7\%} of TPR-passing suites nonetheless fail IR (28.3\% Rust, 19.2\% Ruby, 0.6--0.7\% Go/PHP, 6.1\% Julia); manual inspection confirms these suites invoke mocked/stubbed or unrelated public APIs rather than the focal method, direct evidence IR captures failures TPR misses. Full per-language results, the repair table, and a Julia dispatch-mismatch case study are in Appendix~\ref{app:agentic}.


\begin{table}[t]
\centering
\scriptsize
\setlength{\tabcolsep}{3.2pt}
\renewcommand{\arraystretch}{1.05}
\resizebox{\columnwidth}{!}{
\begin{tabular}{llccccc}
\toprule
\textbf{Metric} & \textbf{Mode} 
& \textbf{Go} 
& \textbf{Rust} 
& \textbf{Ruby} 
& \textbf{PHP} 
& \textbf{Julia} \\
\midrule

\multirow{2}{*}{TPR}
& Passive & 42.6 & 38.5 & 30.0 & 48.3 & 45.5 \\
& Agentic & 98.4 & 72.3 & 78.6
          & 59.8 & 72.7 \\
\midrule

\multirow{2}{*}{Cov}
& Passive & 54.0 & 21.2 & 28.8 & 50.4 & 46.3 \\
& Agentic & 69.9 & 38.9 & 41.2 & 17.1 & 38.8 \\
\midrule

\multirow{2}{*}{IR}
& Passive & 100.0 & 93.8 & 84.3 & 98.9 & 98.2 \\
& Agentic & 98.4 & 95.4 & 68.6 & 27.6 & 100.0 \\

\bottomrule
\end{tabular}}
\caption{Agentic (headless Claude Code) vs.\ passive single-turn
evaluation on the stratified subset. Agentic execution consistently
raises TPR, while Cov and IR can decrease substantially (e.g., PHP),
revealing failures obscured by pass-rate-only evaluation.}
\label{tab:agentic}
\end{table}

\section{Conclusion}
In this work, we introduced \method, a multilingual repository-level benchmark and evaluation framework for unit test generation that emphasizes realistic development settings, controlled context augmentation, and execution-based metrics. Covering multiple programming languages, diverse domains, and containerized environments, \method enables systematic analysis of how modern LLMs generate, invoke, and validate tests under different languages. Our evaluation of 14 state-of-the-art LLMs reveals substantial challenges stemming from language-specific constraints, domain complexity, and semantic issues in generated tests. To better assess behavioral validity, we proposed the novel Invocation Rate (IR) metric, complementing standard metrics like test pass rate (TPR) and coverage. We further explored the impact of context augmentation strategies on test generation. While static file-level context provides the most reliable gains across several languages, whereas more sophisticated augmentations such as LSP- or retrieval-based context yield selective and language-dependent benefits. Together, these findings position \method as a challenging and informative benchmark for advancing scalable and robust multilingual unit test generation in realistic repository contexts, while highlighting promising directions for future work in adaptive context selection, language-aware augmentation, and stronger semantic reasoning for test oracle construction.

\section{Limitations}
In this work, for comparability, we evaluate all models using a standardized prompt format and fixed decoding configuration. Alternative prompting strategies, decoding schemes, or tool-augmented settings may yield different absolute performance and could better exploit specific models. Similarly, retrieval-based augmentation is sensitive to retriever design choices (e.g., chunking strategy, indexing granularity, and relevance scoring), and our retrieval-based context results do not necessarily reflect an optimally tuned retrieval pipeline. Additionally, \method focuses on five underexplored yet impactful languages (Rust, Go, Julia, PHP, and Ruby) and a curated set of real-world repositories. While this scope enables meaningful multilingual, repository-level evaluation, the results may not directly generalize to other programming languages, application domains, or testing frameworks. To partially address this, we release an extensible, containerized evaluation pipeline designed to facilitate future expansion to additional languages, repositories, and test ecosystems.

\section*{Acknowledgments}

This paper has been partially supported by the MOSAICO project (Management, Orchestration and Supervision of AI-agent COmmunities for reliable AI in software engineering) that has received funding from the European Union under the Horizon Research and Innovation Action (Grant Agreement No. 101189664).






\bibliography{ref,custom}

\appendix

\newpage
\clearpage
\newpage

\section*{\huge Appendix}

\renewcommand{\thetable}{A.\arabic{table}}
\renewcommand{\thefigure}{A.\arabic{figure}}
\setcounter{table}{0}
\setcounter{figure}{0}

\section{Detailed Statistics of \method}
\begin{table*}[!t]
\centering
\scriptsize
\begin{adjustbox}{width=\textwidth}
\begin{tabular}{l|l|rc|l|c}
\toprule
\textbf{Language}  & \multirow{2}{*}{\textbf{Repo}} & \multirow{2}{*}{\textbf{Size}} &  
\textbf{Avg.} & \multirow{2}{*}{\textbf{Domain}} & \textbf{Domain} \\
\textbf{(\#Sample)} & & & \textbf{Focal Lines} & & \textbf{Distribution} \\

\midrule
\multirow{7}{*}{Julia (684)}
 & \href{https://github.com/JuliaData/DataFrames.jl}{DataFrames.jl}     & 267 & 26.99 & \cellcolor{sc_color}{Data Manipulation} & \multirow{7}{*}{\safeincludegraphics[width=0.12\textwidth]{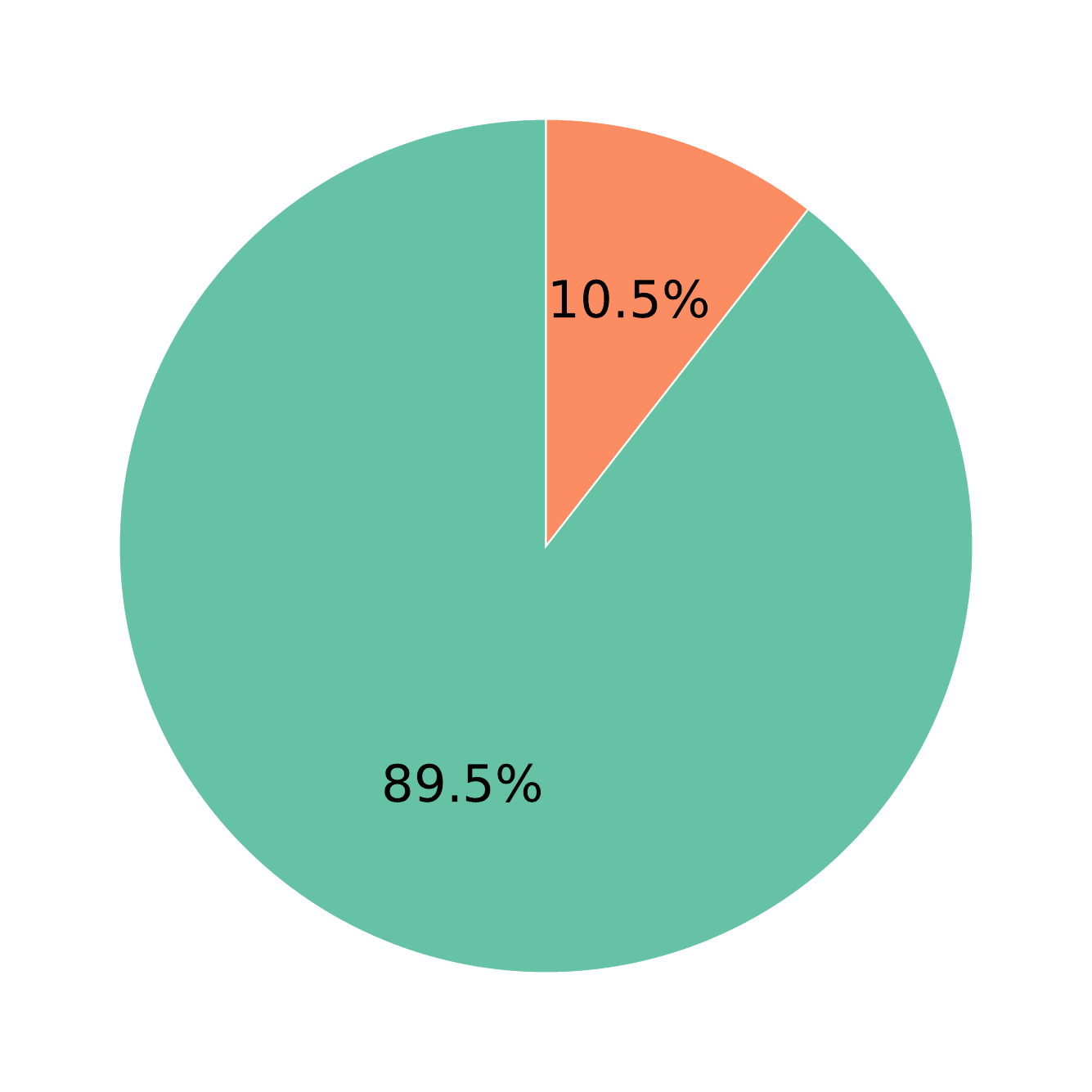}} \\
 & \href{https://github.com/JuliaStats/Distributions.jl}{Distributions.jl}   & 105 & 19.89 & \cellcolor{sc_color}{Probability Distribution} & \\
 & \href{https://github.com/JuliaStats/StatsBase.jl}{StatsBase.jl}       &  99 & 22.63 & \cellcolor{sc_color}{Statistical Computation}  & \\
 & \href{https://github.com/TuringLang/Turing.jl}{Turing.jl}          &  81 & 24.83 & \cellcolor{sc_color}{Bayesian modelling} & \\
 & \href{https://github.com/JuliaCollections/DataStructures.jl}{DataStructures.jl}   &  72 & 25.79 & \cellcolor{dsa_color}{Data structure and algorithms}  & \\
 & \href{https://github.com/QuantEcon/QuantEcon.jl}{QuantEcon.jl}       &  60 & 26.03 & \cellcolor{sc_color}{Quantitative Economics} & \\
 & & & & & \\
\midrule
\multirow{9}{*}{Rust (931)}
 & \href{https://github.com/rust-bakery/nom}{nom}                  & 278 & 22.86 & \cellcolor{dev_color}{Parser Combinator} & \multirow{9}{*}{\safeincludegraphics[width=0.12\textwidth]{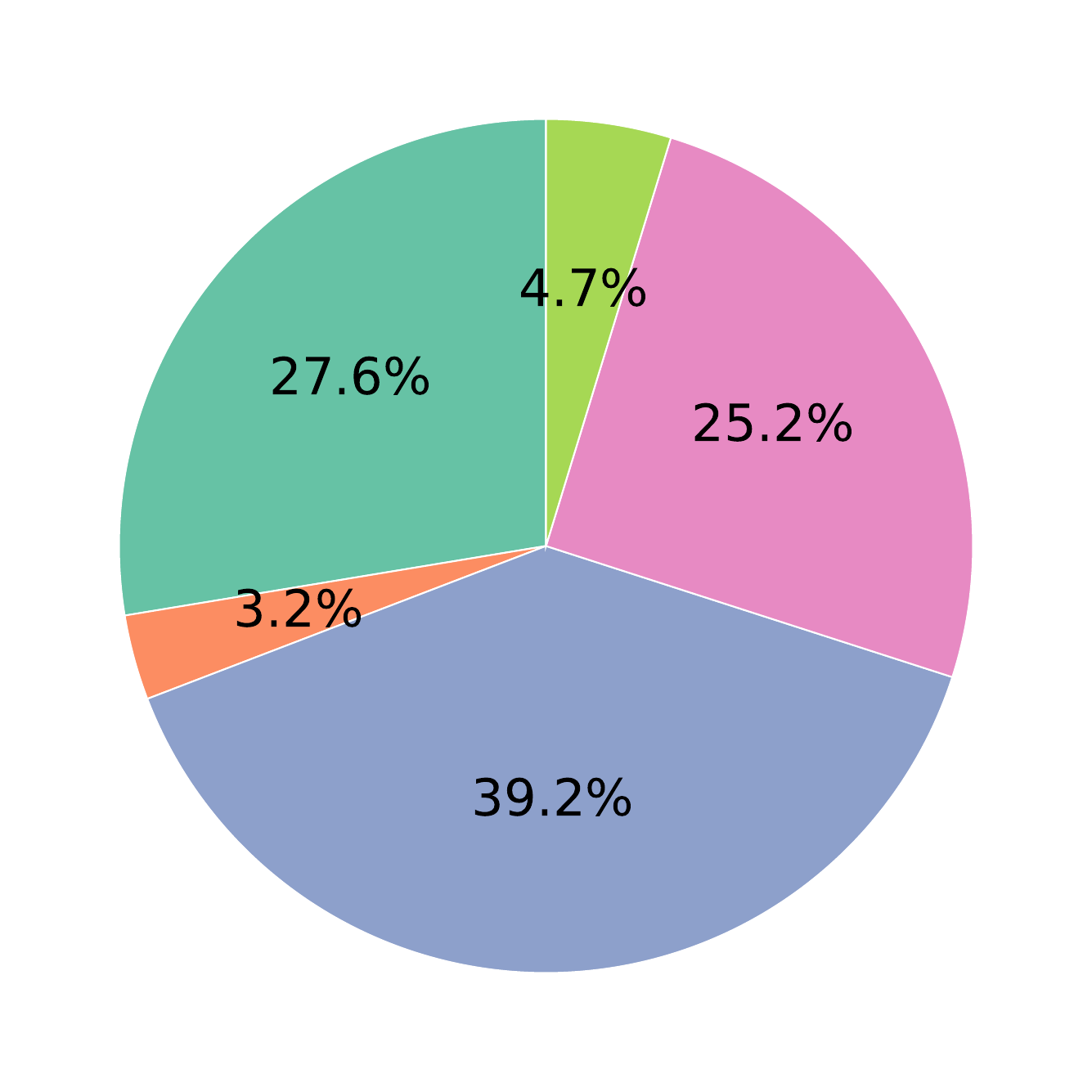}} \\
 & \href{https://github.com/tracel-ai/burn}{Burn}                & 257 & 47.49 & \cellcolor{sc_color}{Deep learning} & \\
 & \href{https://github.com/hyperium/hyper}{hyper}               & 235 & 27.99 & \cellcolor{web_color}{HTTP Library} & \\
 & \href{https://github.com/alacritty/alacritty}{alacritty}           &  54 & 49.59 & \cellcolor{dev_color}{Terminal emulator} & \\
 & \href{https://github.com/uuid-rs/uuid}{uuid}                  &  44 & 21.84 & \cellcolor{utilities_color}{UUID Generation} & \\
 & \href{https://github.com/TheAlgorithms/Rust}{TheAlgorithms/Rust} &  30 & 34.30 & \cellcolor{dsa_color}{Data structure and algorithms} & \\
 & \href{https://github.com/BurntSushi/ripgrep}{ripgrep}             &  18 & 40.17 & \cellcolor{dev_color}{Command-line search tool} & \\
 & \href{https://github.com/starship/starship}{starship}            &  15 & 62.53  & \cellcolor{dev_color}{Terminal Prompt} & \\
 & & & & & \\
\midrule
\multirow{9}{*}{Go (706)}
 & \href{https://github.com/prometheus/client_golang}{client\_golang}   & 262 & 28.54 & \cellcolor{web_color}{Monitoring Instrumentation} & \multirow{9}{*}{\safeincludegraphics[width=0.12\textwidth]{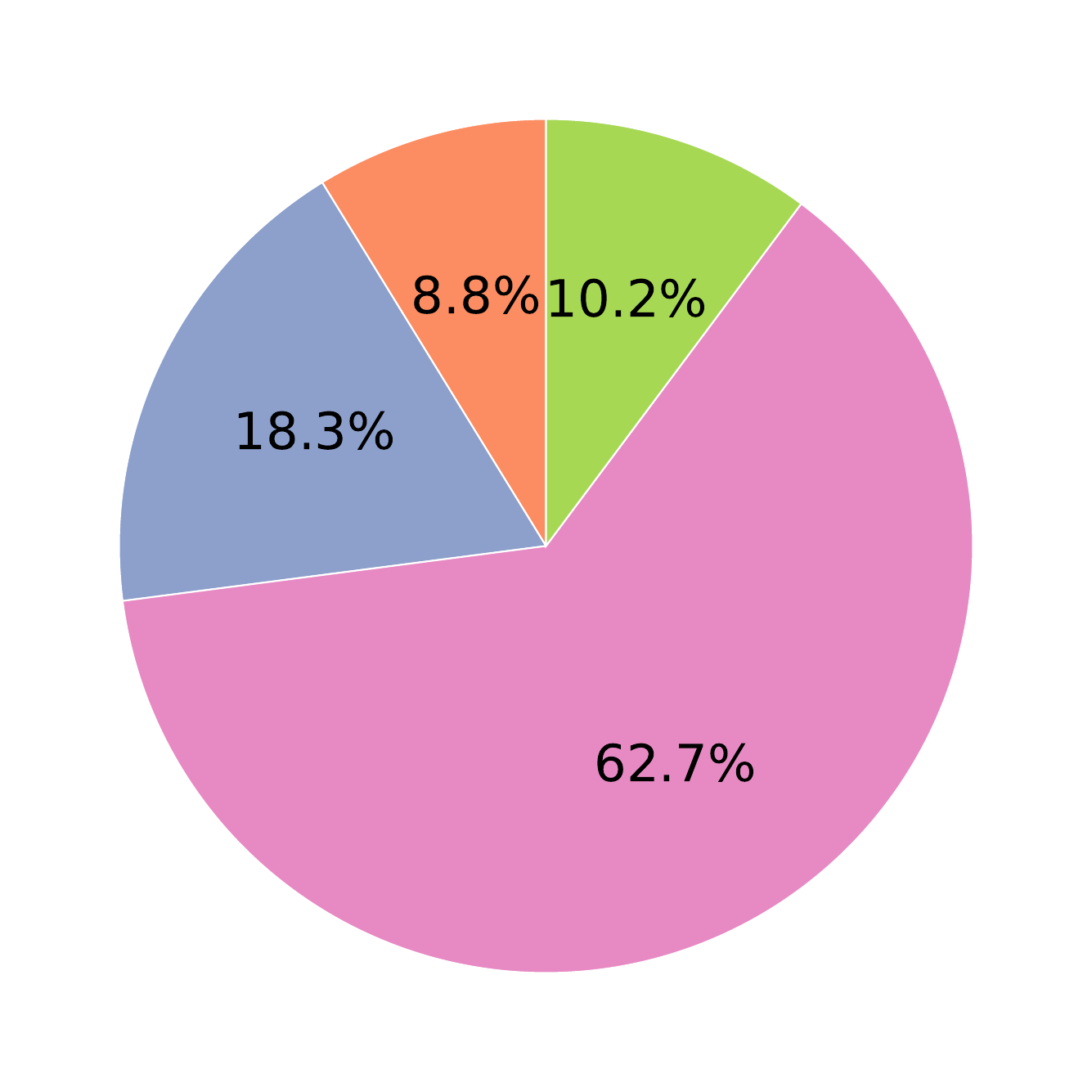}} \\
 & \href{https://github.com/gohugoio/hugo}{Hugo}                & 179 & 46.40 & \cellcolor{web_color}{Static Site Generator} & \\
 & \href{https://github.com/ollama/ollama}{Ollama}              & 114 & 49.81 & \cellcolor{dev_color}{Local LLM runner} & \\
 & \href{https://github.com/uber-go/zap}{zap}                  &  72 & 21.92 & \cellcolor{utilities_color}{Logging Library} & \\
 & \href{https://github.com/TheAlgorithms/Go}{TheAlgorithms/Go}   &  62 & 31.31 & \cellcolor{dsa_color}{Data structure and algorithms} & \\
 & \href{https://github.com/cli/cli}{cli}                 &  12 & 33.33 & \cellcolor{dev_color}{Command line tool} & \\
 & \href{https://github.com/spf13/cobra}{Cobra}               &   3 & 47.67 & \cellcolor{dev_color}{Library for building CLIs} & \\
 & \href{https://github.com/go-chi/chi}{chi}                 &   2 & 48.50 & \cellcolor{web_color}{HTTP router} & \\
 & & & & & \\
\midrule
\multirow{10}{*}{PHP (646)}
 & \href{https://github.com/fzaninotto/Faker}{Faker}             & 124 & 22.55 & \cellcolor{testing_color}{Testing \& Mocking} & \multirow{10}{*}{\safeincludegraphics[width=0.12\textwidth]{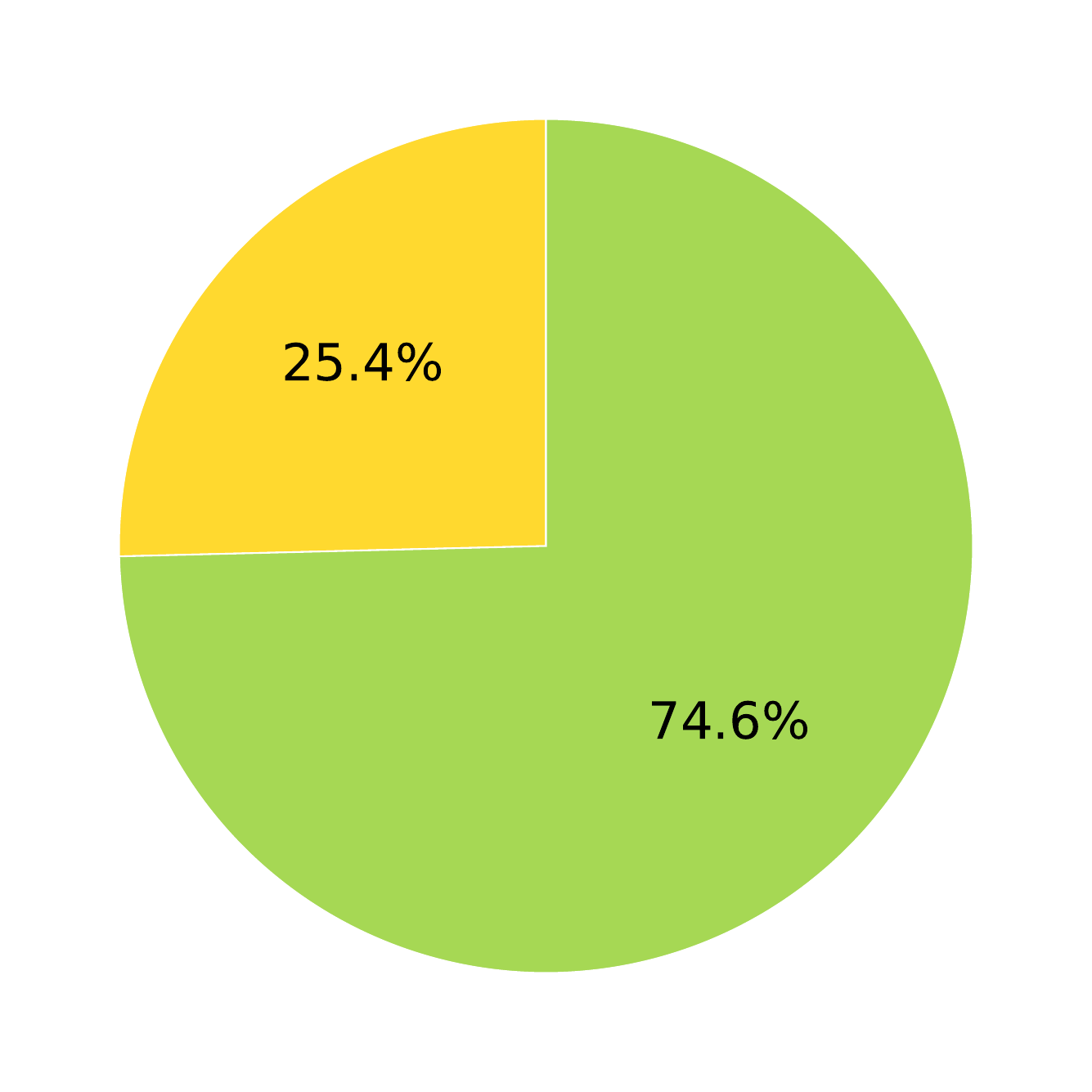}} \\
 & \href{https://github.com/briannesbitt/Carbon}{Carbon}             & 110 & 28.15 & \cellcolor{utilities_color}{Date \& Time Utilities} & \\
 & \href{https://github.com/thephpleague/flysystem}{flysystem}           & 110 & 17.48 & \cellcolor{utilities_color}{Filesystem Abstraction} & \\
 & \href{https://github.com/thephpleague/csv}{csv}                 &  94 & 19.80 & \cellcolor{utilities_color}{CSV Processing} & \\
 & \href{https://github.com/nikic/PHP-Parser}{PHP-Parser}          &  49 & 23.82 & \cellcolor{utilities_color}{Code Parsing} & \\
 & \href{https://github.com/Seldaek/monolog}{monolog}            &  45 & 24.69 & \cellcolor{utilities_color}{Logging Library} & \\
 & \href{https://github.com/PHPMailer/PHPMailer}{PHPMailer}          &  43 & 37.12 & \cellcolor{utilities_color}{Email Library} & \\
 & \href{https://github.com/mockery/mockery}{mockery}            &  40 & 18.38 & \cellcolor{testing_color}{Testing \& Mocking} & \\
 & \href{https://github.com/ramsey/uuid}{uuid}                 &  31 & 21.97 & \cellcolor{utilities_color}{UUID Generation} & \\
 & & & & & \\
\midrule
\multirow{11}{*}{Ruby (675)}
 & \href{https://github.com/teamcapybara/capybara}{capybara}            & 155 & 16.58 & \cellcolor{testing_color}{Web Testing} & \multirow{11}{*}{\safeincludegraphics[width=0.12\textwidth]{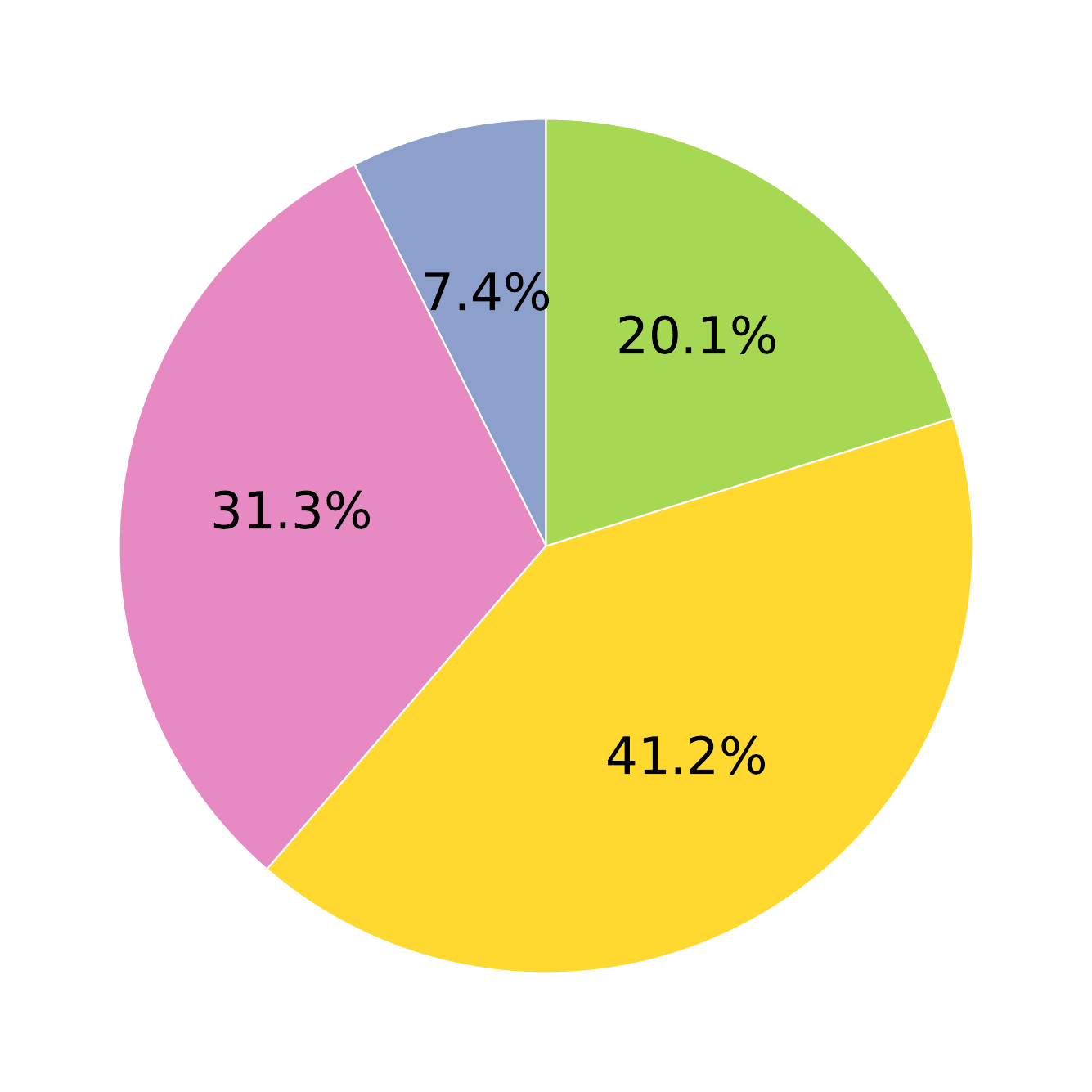}} \\
 & \href{https://github.com/rspec/rspec-core}{rspec-core}          & 123 & 16.92 & \cellcolor{testing_color}{Testing Framework} & \\
 & \href{https://github.com/rom-rb/rom}{rom}                  &  97 & 17.09 & \cellcolor{utilities_color}{Data Mapping} & \\
 & \href{https://github.com/ruby-grape/grape}{grape}                &  97 & 16.42 & \cellcolor{web_color}{REST-like API Framework} & \\
 & \href{https://github.com/hanami/hanami}{hanami}               &  81 & 19.38 & \cellcolor{web_color}{Web Framework} & \\
 & \href{https://github.com/ruby-shoryuken/shoryuken}{shoryuken}             &  50 & 17.76 & \cellcolor{dev_color}{Message Queue Worker} & \\
 & \href{https://github.com/hashie/hashie}{hashie}             &  28 & 16.79 & \cellcolor{utilities_color}{Hash Extensions} & \\
 & \href{https://github.com/jnunemaker/httparty}{httparty}           &  26 & 20.65 & \cellcolor{web_color}{HTTP Client} & \\
 & \href{https://github.com/bkeepers/dotenv}{dotenv}             &  11 & 18.00 & \cellcolor{utilities_color}{Configuration} & \\
 & \href{https://github.com/varvet/pundit}{pundit}             &   7 & 13.29 & \cellcolor{web_color}{Authorization} & \\
 & & & & & \\
\bottomrule
\end{tabular}
\end{adjustbox}
\caption{Per-repository statistics with per-language domain distributions in \method. 
There are 6 domains across the benchmark: \colorbox{dsa_color}{Data Structure and Algorithm}, 
\colorbox{sc_color}{Scientific Computing}, \colorbox{dev_color}{Developer Tool}, \colorbox{web_color}{Web/Framework}, 
\colorbox{testing_color}{Testing \& Mocking}, \colorbox{utilities_color}{Utilities}.}
\label{tab:repo_statistics_full}
\end{table*}

\label{sec:repo_stats}
Table~\ref{tab:repo_statistics_full} provides a summary of \method. The benchmark includes 3,642 focal methods across five languages: Julia (684), Rust (931), Go (706), PHP (646), and Ruby (675). Despite having fewer samples, Rust and Go exhibit significantly longer average focal lengths, reflecting their more verbose function implementations. From the Table, we also observe the domain distribution of the benchmark across the five languages. Julia primarily focuses on scientific computing tasks owing to its design and use cases, with additional coverage in data structures and algorithms (DSA). Rust is largely oriented toward developer tools with additional coverage in DSA, scientific computing, and web infrastructure. Go spans developer tools, web/framework, and data structure domains. PHP contributes functions from testing, utilities, and web development domains, while Ruby emphasizes testing frameworks and web development with a smaller presence in the utilities and web security domains.

\section{Implementation Details}
\label{sec:impl_details}
This section summarizes implementation-level details that are omitted from the main paper for brevity.
\paragraph{Focal Function Filtering Rules.}
To ensure that extracted functions are non-trivial and representative of real-world unit testing targets, we apply systematic filtering criteria during the focal method extraction phase. Table~\ref{tab:filtering_rule} summarizes these rules.

\paragraph{LSP-based Context.}
To enrich prompts with type-aware context, we run a Language Server Protocol\footnote{\url{https://microsoft.github.io/language-server-protocol/}} (LSP) analysis for each repository. Our pipeline implements a lightweight LSP client that speaks the standard JSON-RPC message format and issues the typical LSP lifecycle calls (e.g., starting/initializing the server, opening the focal file, and querying symbol and type information). Concretely, we use LSP to retrieve three kinds of information centered on the focal method. 
\begin{itemize}[leftmargin=*]
\item First, for each focal-method argument, we resolve its definition (e.g., local binding, field, imported symbol, or type alias) and then collect the corresponding type declarations as well as any directly referenced dependent types (e.g., nested structs/classes, trait/interface bounds, and enum/union variants when applicable). This material is attached as ``argument context''.

\item Second, we extract the function/method calls appearing in the focal method body and resolve the definitions of the callees (including overload/dispatch targets when available). We then collect the signatures and relevant type declarations for these invoked functions and their receiver/parameter types, providing the model with type-accurate information about the APIs that the focal method depends on.

\item Third, we focus on identifiers that appear in the focal method's control-flow conditions (e.g., predicates in \texttt{if}/\texttt{while} branches and guards). For these tokens (variables, fields, constants, or function calls), we use LSP definition/type-definition and reference queries to recover their declaring sites and types, effectively retrieving ``references for the focal method'' that are most likely to constrain feasible tests (e.g., branch-specific preconditions).

\end{itemize}

\begin{table*}[t]
\centering
\scriptsize
\setlength{\tabcolsep}{3pt}
\renewcommand{\arraystretch}{1.15}
\begin{adjustbox}{width=0.9\textwidth}
\begin{tabular}{@{}p{2.0cm} p{3cm} p{4cm}@{}}
	\toprule
	\textbf{Rule} & \textbf{Description} & \textbf{Purpose} \\
\midrule
Minimum number of lines & Function body must contain at least 10 lines of code & Avoid trivial or uninformative functions (e.g., empty methods or those offering trivial coverage) \\
\midrule
Return statement & Function must include an explicit \texttt{return} or output expression & Ensure testable behavior and observable outputs \\
\midrule
Non-test function & Exclude functions already written as unit tests & Ensuring only production code is selected as focal functions \\
\midrule
Valid parsing & Function must be parsable by language grammar (no syntax errors) & Guarantee valid function and clean dataset for model evaluation \\
\bottomrule
\end{tabular}
\end{adjustbox}
\caption{Filtering rules used to select focal functions for unit test generation.}
\label{tab:filtering_rule}
\end{table*}
\paragraph{Retrieval-augmented Context.}
We implement a retrieval-augmented context construction pipeline that enriches each \method function with repository context, following prior repository-level code generation works \citep{zhang2023repocoder, wu2024repoformer, wang2024rlcoder, le2026not}. For each repository, we preprocess all source files into chunks/overlapping sliding windows of $W=50$ lines. The stride $s$ is computed as:
\begin{equation}
s = \left\lfloor \frac{W}{\text{slice\_size}} \right\rfloor = \left\lfloor \frac{50}{10} \right\rfloor = 5 \text{ lines}
\end{equation}
This creates 90\% overlap between consecutive windows. Empty windows are dropped and duplicate windows (identical text) are merged while preserving multiple metadata entries (file path and line span).

We retrieve windows using the focal method code itself as the query. Specifically, we use the focal function (the complete function implementation) as the query string. This approach captures local implementation patterns and enables the retriever to find semantically and lexically similar code snippets from the repository that share similar logic, structure, or domain-specific patterns with the target function.

Retrieval is performed over split code chunks using two retrievers: BM25 as a sparse retriever and UniXCoder~\citep{guo2022unixcoder} as a dense retriever. After that, we attach the top-$k$ ($k=10$) retrieved contexts by each retriever to the input prompt.


\begin{itemize}
    \item BM25 retrieval (BM25Okapi with $k_1=1.5$, $b=0.75$) uses whitespace tokenization and ranks chunks by lexical relevance:

\begin{equation}
\begin{aligned}
\operatorname{BM25}(q, w)
&= \sum_{t \in q} \operatorname{IDF}(t) \cdot \\
&\quad
\frac{f(t, w)(k_1 + 1)}
{f(t, w) + k_1\left(1 - b + b \frac{|w|}{\operatorname{avgdl}}\right)}
\end{aligned}
\end{equation}

where $f(t, w)$ is the term frequency, $|w|$ is the chunk length in tokens, and $\operatorname{avgdl}$ is the average window length.
\item UniXcoder encodes the query and chunks (max length 512 tokens) with attention-masked mean pooling and ranks by cosine similarity:
\begin{equation}
\text{sim}(q, w) = \frac{\mathbf{e}_q \cdot \mathbf{e}_w}{\|\mathbf{e}_q\| \|\mathbf{e}_w\|}
\end{equation}
where $\mathbf{e}_q$ and $\mathbf{e}_w$ are the embeddings. 
\end{itemize}

As future work, we plan to explore GraphRAG-based techniques to construct richer, repository-level contexts for more effective unit test generation \cite{hipporag, edge2024localglobalgraphrag, graphragsurvey, guo-etal-2025-lightrag, hieu2025magix, van2026memorai,hoang-etal-2026-codewiki}.

\paragraph{Containerized provisioning and execution.}
We evaluate each repository inside a Docker container to standardize toolchains and ensure reproducible execution. Containers provide (i) the language runtime/toolchain and package manager, (ii) the project dependencies, (iii) the language's native test runner, and (iv) coverage instrumentation where available. At evaluation time, the repository is mounted into the container, tests are integrated following language conventions, and we run the corresponding test command (e.g., \texttt{cargo test}, \texttt{go test}, \texttt{Test.jl}, \texttt{phpunit}, \texttt{rspec}) to collect execution outcomes and metrics. We use the following standard test runners and entrypoints: Go's \texttt{go test}\footnote{\url{https://pkg.go.dev/testing}}, Rust's \texttt{cargo test}\footnote{\url{https://doc.rust-lang.org/cargo/commands/cargo-test.html}}, Julia's \texttt{Test.jl}\footnote{\url{https://docs.julialang.org/en/v1/stdlib/Test/}}, PHP's PHPUnit \footnote{\url{https://phpunit.de/}}, and Ruby's RSpec \footnote{\url{https://rspec.info/}}.

To robustly handle the often noisy or partially structured outputs produced by LLMs, we adopt a lightweight post-processing strategy that extracts generated test code from fenced code blocks (e.g., $```\text{language}... ```$).
The extracted tests are then integrated into the target repository following language- and framework-specific conventions.
For example, Go tests are written to files with the \texttt{\_test.go} suffix, while Rust test functions are appended to the corresponding source files containing the focal functions.
Depending on the language, tests may be injected into existing files or written into newly created test files.

After integrating the generated tests, we execute an automated evaluation pipeline. First, the project is compiled to detect syntactic errors and unresolved dependencies. If compilation succeeds, the test suite is executed and code coverage is measured using language-specific tooling. The entire evaluation process is containerized with Docker to ensure reproducibility and consistency across environments. To support future research on LLM-based test generation, we release our evaluation framework publicly.

\section{Data Contamination Risk}
\label{app:dataleak}

Large language models (LLMs) are typically trained on large-scale corpora collected from diverse online sources, with limited transparency regarding data curation and preprocessing. Consequently, constructing a benchmark that is entirely free from training-data leakage is widely considered impractical. Recent benchmarks therefore acknowledge this limitation and focus on mitigating and characterizing potential contamination rather than claiming strict guarantees, commonly adopting strategies such as temporal filtering to reduce overlap with model pretraining data \citep{jaintestgeneval, jainlivecodebench, testeval}.

From our experiments in Section \ref{sec:exp}, the consistently low performance across models suggests that memorization is unlikely to be a dominant factor in \method. To further assess the reliability of the benchmark, we follow prior works \citep{xu2024benchmarking, chen2024promise, nguyen2025codemmlu} and conduct a perplexity-based analysis to estimate whether models may have previously observed code artifacts related to \method. The underlying assumption is that content encountered during pretraining is likely to be assigned systematically lower perplexity than genuinely unseen code. In particular, our analysis focuses on repository test code, as generated unit tests are evaluated against existing project tests. Since aligned focal--test pairs are not directly available, we extract all test functions from the repositories included in \method and compute their perplexity under the evaluated models. These scores serve as a proxy for assessing whether repository-specific test code may have been memorized during training. We compute perplexity scores using two code models (Yi-Coder-9B and Codestral-22B) across all five benchmark languages (Go, Rust, Ruby, PHP, Julia).

Perplexity measures the uncertainty of a language model when predicting the next token in a sequence, expressed as the exponentiated average negative log-likelihood:

\begin{equation}
\text{PPL}(\mathbf{X}) = \exp\left(-\frac{1}{t}\sum_{i=0}^{t}\log p_\theta(x_i|x_{<i})\right)
\end{equation}

\noindent where $\mathbf{X} = [x_0, x_1, ..., x_t]$ denotes a tokenized sequence. Lower perplexity indicates the model is confident in predicting the test code, suggesting it has likely encountered similar patterns during training (potential data leakage). 

From Table \ref{tab:benchmark_comparison}, we observe substantial variation in perplexity across languages. Ruby shows notably higher perplexity (26.94--32.74) compared to Go, Rust, PHP, and Julia (3.52--6.11), suggesting models have significantly less exposure to Ruby test code during pretraining. These high perplexity values across all languages (3.52--32.74) indicate that models have not memorized the dataset, suggesting a low risk of data leakage.

\begin{table*}[t]
\centering
\begin{adjustbox}{width=0.95\textwidth}
\begin{tabular}{@{}l*{9}{c}@{}}
\toprule
\multirow{2}{*}{\textbf{Model}} &
\multicolumn{6}{c}{\textbf{Our Benchmark}} &
\multicolumn{3}{c}{\textbf{External Benchmarks}} \\
\cmidrule(lr){2-7}\cmidrule(lr){8-10}
& \textbf{Go} & \textbf{Rust} & \textbf{Ruby} & \textbf{PHP} & \textbf{Julia} & \textbf{Avg}
& \textbf{McEval} & \textbf{HumanEval} & \textbf{TestGenEval} \\
\midrule
Yi-Coder-9B      & 3.76 & 5.78 & 26.94 & 4.53 & 4.51 & 9.10 & 2.74 & 1.95 & 5.44 \\
Codestral-22B    & 3.52 & 6.11 & 32.74 & 4.81 & 4.40 & 10.32 & 2.91 & 3.07 & 6.92 \\
\bottomrule
\end{tabular}
\end{adjustbox}
\caption{Perplexity of different models on \method across 5 languages. Higher perplexity indicates less likelihood of data leakage.}
\label{tab:benchmark_comparison}
\end{table*}

\section{Evaluation Metrics}
\label{sec:metrics_calculation}

This section provides detailed formulations for the evaluation metrics used in \method. All metrics are normalized by the total number of samples in the benchmark, enabling direct cross-model and cross-language comparison.

\paragraph{\textbf{Test Pass Rate (TPR).}}
TPR measures the proportion of test suites that execute successfully without runtime errors or assertion failures. Crucially, we adopt an \emph{all-or-nothing} criterion: a test suite passes only when \textbf{all} test cases within it pass. Formally:
\begin{equation}
\text{TPR} = \frac{\sum_{i=1}^{N} \mathbf{1}[\text{all tests pass in suite}_i]}{N}
\end{equation}
where $N$ is the total number of samples (focal functions) and $\mathbf{1}[\cdot]$ is the indicator function. This strict definition prevents inflated pass rates from partially correct test suites and ensures that only fully executable, assertion-compliant test code contributes to TPR.

\paragraph{\textbf{Invocation Rate (IR).}}
IR quantifies the proportion of all generated test suites that successfully invoke the focal function. To compute IR, we first parse each generated test suite using Tree-sitter. For test suites that are syntactically valid, we perform static analysis to detect whether any test function contains a direct call to the focal function; test suites that cannot be parsed are treated as failing to invoke the function. Formally:
\begin{equation}
\text{IR} = \frac{\sum_{i=1}^{N} \mathbf{1}[\text{suite}_i \text{ invokes focal function}]}{N}
\end{equation}
where $N$ is the total number of generated test suites. By computing IR over the total set rather than just the parsable subset, this metric penalizes both syntax errors and logical failures to call the target method, providing a comprehensive measure of the model's targeting capability. By filtering out semantically irrelevant tests that inflate coverage, IR helps mitigate misleading test coverage and false confidence, thereby supporting more maintainable test suites, safer code evolution and mitigating test-related technical debt \cite{hai2025detection, nakamura2026understanding}.

\paragraph{\textbf{Line Coverage (Cov).}}
Line coverage measures the proportion of executable lines in the focal function that are exercised by the generated tests. We employ macro-averaged line coverage, computed as follows. For each sample, we execute the generated test suite with language-specific coverage instrumentation (\texttt{llvm-cov} for Rust, \texttt{go test -coverprofile} for Go, \texttt{Coverage.jl} for Julia, PHPUnit coverage for PHP, SimpleCov for Ruby). From the coverage report, we extract two quantities within the focal function's start and end line boundaries: the set of \emph{executable} lines $L_{\text{total}}$ (lines that the instrumentation marks as trackable, regardless of whether they were exercised), and the subset of \emph{covered} lines $L_{\text{covered}}$ (executable lines that were actually executed during the test run).

The coverage metric is then computed as:
\begin{equation}
\text{Cov} = \frac{1}{N}\sum_{i=1}^{N}\frac{L_{\text{covered},i}}{L_{\text{total},i}}
\end{equation}
where $L_{\text{total},i}$ and $L_{\text{covered},i}$ are both obtained from the same coverage run for focal function $i$. This macro-averaged formulation weights each focal function equally, ensuring that coverage is not dominated by samples with disproportionately many executable lines. Coverage for samples with no available coverage data (e.g., due to compilation failure) is treated as $0\%$.

\paragraph{\textbf{Mutation Score (MS).}}
Following the methodology of TestGenEval \cite{jaintestgeneval},
Mutation score assesses the fault-detection capability of generated tests by measuring their ability to distinguish correct code from systematically injected faults (mutants). For each sample $i$, if mutation testing is executed, we apply a set of mutation operators (e.g., arithmetic operator replacement, relational operator flip, constant modification) to the focal function, generating mutants $M_i$. Each mutant is then tested: if the test suite detects the mutation (via assertion failure or runtime error), the mutant is considered \emph{killed}; otherwise, it survives. We define the per-sample mutation score as $\text{ms}_i = \frac{|M_{i,\text{killed}}|}{|M_i|}$ when $|M_i|>0$; if mutation testing is not available for sample $i$ (e.g., due to compilation/test failures or missing mutation data), we set $\text{ms}_i=0$. The reported \textbf{Mutation Score (MS)} averages this value over \emph{all} samples:
\begin{equation}
        	 	\text{MS} = \frac{1}{N} \sum_{i=1}^{N} \text{ms}_i
\end{equation}
where $N$ is the total number of samples.

\paragraph{\textbf{Mutation@Pass Score (MS@Pass).}}
To isolate test oracle quality conditional on successful execution, we also report \textbf{Mutation@Pass Score (MS@Pass)}, computed \emph{only} on samples where all tests pass (TPR$=1$) \emph{and} mutation testing is available ($|M_i|>0$). Formally:
\begin{equation}
        	 	\text{MS@Pass} = \frac{1}{N_{\text{pass}}} \sum_{i \in \text{pass}} \frac{|M_{i,\text{killed}}|}{|M_i|}
\end{equation}
where $\text{pass}$ denotes the set of samples satisfying these criteria and $N_{\text{pass}} = |\text{pass}|$.

\paragraph{\textbf{Aggregate Reporting.}}
All metrics are reported as percentages normalized by their respective denominators ($N$ for TPR, Cov, IR, and MS; $N_{\text{pass}}$ for MS@Pass). This normalization enables fair comparison across models, languages, and context configurations, and ensures that metrics are interpretable as proportions of the total evaluation scope.

\section{Additional Experiment Results}

\begin{sidewaystable*}[p]
\centering
\begin{varwidth}{\textheight}
\centering
\caption{Full standard-setting results on \method. Values denote \textbf{CSR | TPR | Cov | IR}.}
\label{tab:full_results_standard}
\scriptsize
\setlength{\tabcolsep}{2pt}
\renewcommand{\arraystretch}{1.05}
\resizebox{\textheight}{!}{%
\begin{tabular}{l *{5}{cccc}}
\csname toprule\endcsname
\multicolumn{1}{l}{\textbf{Model}} & \multicolumn{4}{c}{\textbf{Rust}} & \multicolumn{4}{c}{\textbf{Julia}} & \multicolumn{4}{c}{\textbf{Go}} & \multicolumn{4}{c}{\textbf{Ruby}} & \multicolumn{4}{c}{\textbf{PHP}} \\
\cmidrule(lr){2-5} \cmidrule(lr){6-9} \cmidrule(lr){10-13} \cmidrule(lr){14-17} \cmidrule(lr){18-21}
 & CSR & TPR & Cov & IR & CSR & TPR & Cov & IR & CSR & TPR & Cov & IR & CSR & TPR & Cov & IR & CSR & TPR & Cov & IR \\
\midrule

\multicolumn{21}{c}{\textit{\textbf{Small}}} \\
\midrule
Yi-Coder (9B) & 26.42 & 1.50 & 1.71 & 90.23 & 94.44 & 4.82 & 19.16 & 90.50 & 9.63 & 2.41 & 6.48 & 91.36 & 98.67 & 1.48 & 9.37 & 86.81 & 99.38 & 3.56 & 17.68 & 98.76 \\
Qwen3 (8B) & 23.09 & 0.32 & 0.21 & 64.88 & 12.57 & 0.88 & 0.97 & 73.83 & 1.27 & 0.42 & 1.13 & 47.17 & 14.07 & 0.00 & 1.39 & 92.15 & 7.28 & 0.62 & 2.13 & 22.91 \\
Codestral (22B) & 27.82 & 2.69 & 2.97 & 84.85 & 96.49 & 2.05 & 21.17 & 91.23 & 20.25 & 3.82 & 14.52 & 99.58 & 99.26 & 0.15 & 7.35 & 87.26 & 98.76 & 5.57 & 23.01 & 99.69 \\
\midrule

\multicolumn{21}{c}{\textit{\textbf{Medium}}} \\
\midrule
GPT-OSS (120B) & 29.86 & 5.80 & 4.16 & 81.63 & 70.91 & 4.39 & 24.27 & 93.86 & 31.02 & 19.26 & 26.95 & 88.24 & 97.19 & 6.96 & 8.21 & 85.93 & 51.86 & 3.41 & 5.07 & 99.23 \\
Llama-3.3 (70B) & 28.03 & 3.33 & 3.04 & 88.94 & 97.81 & 3.07 & 25.42 & 93.27 & 10.62 & 2.69 & 7.96 & 98.73 & 98.22 & 0.44 & 7.66 & 87.85 & 95.05 & 2.01 & 21.75 & 99.85 \\
Qwen3-Coder-Next (80B) & 39.63 & 7.20 & 3.52 & 56.71 & 90.50 & 5.12 & 17.38 & 91.08 & 20.11 & 6.66 & 16.74 & 96.88 & 91.41 & 1.19 & 6.88 & 87.26 & 91.95 & 5.11 & 22.43 & 97.68 \\
\midrule

\multicolumn{21}{c}{\textit{\textbf{Large}}} \\
\midrule
DeepSeek V3.2 & 28.14 & 2.79 & 3.06 & 77.66 & 63.74 & 3.95 & 21.94 & 89.47 & 17.99 & 5.24 & 14.91 & 59.07 & 84.30 & 1.48 & 10.01 & 84.30 & 68.73 & 5.42 & 20.20 & 93.34 \\
MiniMax-M2.7 & 31.15 & 4.30 & 3.63 & 66.49 & 72.08 & 4.39 & 18.12 & 86.11 & 17.00 & 4.67 & 13.07 & 81.16 & 89.33 & 1.63 & 7.30 & 87.41 & 78.79 & 5.57 & 17.25 & 97.68 \\
\midrule

\multicolumn{21}{c}{\textit{\textbf{Flagship}}} \\
\midrule
DeepSeek V4-pro & 30.93 & 6.66 & 4.66 & 74.22 & 73.10 & 11.40 & 26.83 & 89.77 & 25.50 & 13.88 & 21.66 & 63.46 & 83.70 & 8.89 & 9.15 & 84.74 & 85.29 & 13.00 & 22.19 & 96.13 \\
GLM-5 & 30.72 & 6.44 & 2.52 & 70.68 & 51.17 & 5.41 & 13.54 & 89.47 & 10.34 & 5.10 & 9.48 & 75.78 & 86.67 & 3.11 & 8.50 & 86.96 & 75.23 & 13.16 & 24.98 & 99.07 \\
Kimi-k2.5 (Instruct) & 24.70 & 1.40 & 1.17 & 74.76 & 39.04 & 2.63 & 8.70 & 87.43 & 3.26 & 2.27 & 2.33 & 75.21 & 41.33 & 2.52 & 5.02 & 78.96 & 15.48 & 3.25 & 4.02 & 88.08 \\
Claude 4.5 Haiku & 34.69 & 8.27 & 3.73 & 80.34 & 93.86 & 12.57 & 23.96 & 92.69 & 31.73 & 10.91 & 26.87 & 99.43 & 99.11 & 4.30 & 10.57 & 87.11 & 98.92 & 14.86 & 34.68 & 99.54 \\
Claude 4.5 Sonnet & 41.46 & 12.78 & 11.56 & 75.94 & 98.25 & 19.88 & 31.58 & 94.01 & 52.27 & 23.65 & 43.53 & 99.43 & 99.56 & 6.37 & 14.25 & 85.63 & 98.76 & 26.93 & 41.76 & 99.38 \\
GPT-5.2 & 38.78 & 12.24 & 9.31 & 77.87 & 84.21 & 16.52 & 31.92 & 91.81 & 47.59 & 25.78 & 41.30 & 98.44 & 98.22 & 25.78 & 7.80 & 88.89 & 97.68 & 10.06 & 23.17 & 99.38 \\
\bottomrule
\end{tabular}}
\end{varwidth}
\end{sidewaystable*}

\subsection{Full Results of LLMs Performance}
\label{sec:full_result}
This section presents the complete set of experimental results for all models. Table \ref{tab:full_results_standard} summarizes the evaluation metrics, including test pass rate (TPR), line coverage (Cov), and invocation rate (IR). CSR values are omitted from this table and available upon request. Together, these metrics offer a comprehensive assessment of model performance, capturing both the syntactic correctness and the semantic quality of the generated tests across different programming languages.

\subsection{Domain-Based Performance Analysis}
\label{app:domain_analysis}

\begin{figure*}[t]
\centering
\safeincludegraphics[width=\textwidth]{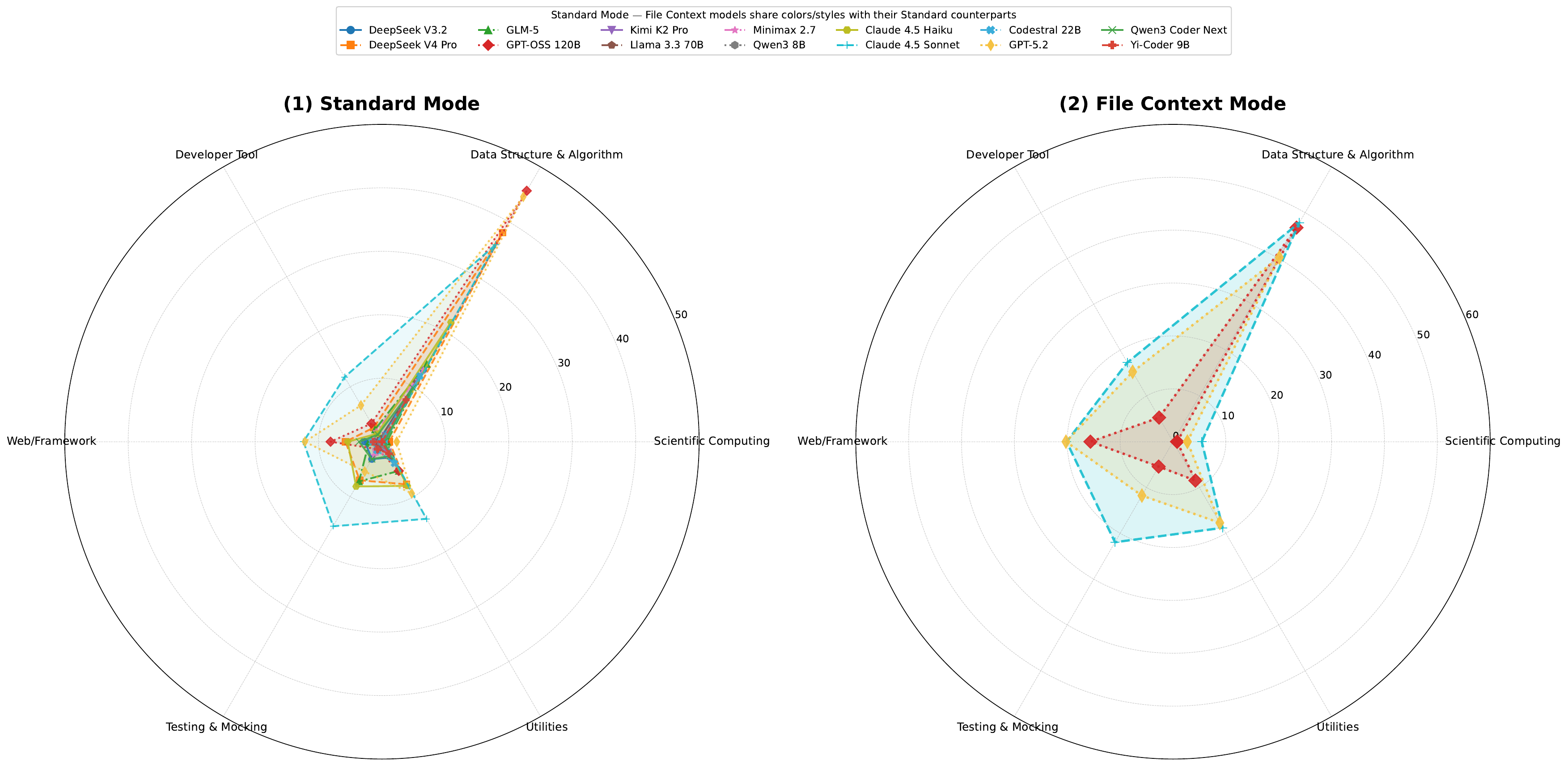}
\caption{Test pass rate comparison across language-domain combinations. Left (1): Ten models in standard mode including Qwen3-Coder-Next 80B, GPT-OSS 120B, Claude 4.5 Haiku, Claude 4.5 Sonnet, DeepSeek V3.2, GPT-5.2, DeepSeek V4-pro, GLM-5, MiniMax-M2.7, and Kimi-k2.5. Right (2): Three LLMs in file context mode showing the impact of contextual enrichment.}
\label{fig:domain_spider_combined}
\end{figure*}

To understand how model performance varies across different application domains, we analyze test pass rates across repository categories. We categorize each benchmark repository into one of six domains: \textit{Scientific Computing}, \textit{Data Structure \& Algorithm}, \textit{Developer Tool}, \textit{Web/Framework}, \textit{Testing \& Mocking}, and \textit{Utilities}. Figure~\ref{fig:domain_spider_combined} presents a comprehensive comparison using spider charts. The left subplot shows ten models in standard mode (baseline setting without context enrichment), while the right subplot compares three representative models (GPT-5.2, Claude 4.5 Sonnet, and GPT-OSS 120B) in file context mode. Each axis represents a unique (language, domain) pair, and the radial distance indicates the test pass rate achieved by each model.

The visualization reveals several key insights. First, performance exhibits significant variation across domains even within the same language, suggesting that domain-specific characteristics (e.g., mathematical abstractions in scientific computing vs. string manipulation in utilities) present distinct challenges for test generation. Second, models show different strengths across domains: while proprietary frontier models (GPT-5.2, Claude 4.5 Sonnet) generally achieve higher test pass rates, open-source and smaller models demonstrate competitive performance in certain language-domain combinations. Comparing standard mode with file context mode reveals that contextual enrichment can significantly improve performance, particularly for domains requiring understanding of complex type relationships or API usage patterns.

A consistent pattern across models is that \textit{Data Structures \& Algorithms} domains achieve the strongest performance. This likely reflects the prevalence of such problems in existing code generation and evaluation benchmarks \citep{chen2021evaluating, mbpp, lozhkov2024starcoder, wang2023codet5+, wang2024rlcoder, jainlivecodebench, le2025impacts, nguyen2023vault}, leading to greater model familiarity with their structures and testing patterns. In contrast, more specialized or application-specific domains exhibit lower and more variable performance, indicating that they remain challenging and underexplored in current test generation research. Finally, the clustering of low-performing language--domain pairs across all models suggests inherent difficulty that persists regardless of model architecture or context scope, potentially driven by domain-specific complexity or limitations in automated test oracles.

\subsection{Mutation Results}
\label{sec:mutation_results_appendix}

\begin{table*}[h]
\centering
\scriptsize
\begin{adjustbox}{width=0.95\textwidth}
\begin{tabular}{@{}l*{6}{c}@{}}
\toprule
\multirow{2}{*}{\textbf{Model}} &
\multicolumn{2}{c}{\textbf{Go}} &
\multicolumn{2}{c}{\textbf{Rust}} &
\multicolumn{2}{c}{\textbf{Ruby}} \\
\cmidrule(lr){2-3}\cmidrule(lr){4-5}\cmidrule(lr){6-7}
& \textbf{MS} & \textbf{MS@Pass} & \textbf{MS} & \textbf{MS@Pass} & \textbf{MS} & \textbf{MS@Pass} \\
\midrule
\multicolumn{7}{c}{\textit{\textbf{Small}}} \\
\midrule
Yi-Coder (9B)                           & 1.59  & 70.01 & 0.38  & 58.79  & 0.15  & \textbf{98.44} \\
Qwen3 (8B)                              & 0.39  & \textbf{92.08} & 0.00  & 0.00   & 0.00  & 0.00  \\
Codestral (22B)                         & 3.16  & 82.55 & 1.14  & 75.71  & 0.00  & 0.00  \\
\midrule

\multicolumn{7}{c}{\textit{\textbf{Medium}}} \\
\midrule
GPT-OSS (120B)                          & 15.04 & 78.63 & 2.72  & 79.00  & 1.25  & 84.44 \\
Llama-3.3 (70B)                         & 2.35  & 87.29 & 1.21  & 66.05  & 0.00  & 0.00  \\
Qwen3-Coder-Next (80B)                        & 4.73  & 74.26 & 0.98  & 50.69  & 0.29  & 66.31 \\
\midrule

\multicolumn{7}{c}{\textit{\textbf{Large}}} \\
\midrule
DeepSeek V3.2                           & 3.86  & 77.93 & 1.04  & \textbf{87.70}  & 0.14  & 93.59 \\
MiniMax-M2.7                            & 3.82  & 81.64 & 0.74  & 57.22  & 0.01  & 9.68  \\
\midrule

\multicolumn{7}{c}{\textit{\textbf{Flagship}}} \\
\midrule
DeepSeek V4-pro                         & 11.42 & 83.96 & 2.75  & 73.22  & 1.67  & 94.13 \\
GLM-5                                   & 3.62  & 79.91 & 1.22  & 60.00  & 0.43  & 71.81 \\
Kimi-k2.5                               & 1.34  & 63.07 & 0.49  & 76.19  & 0.01  & 6.25  \\
Claude 4.5 Haiku                        & 9.15  & 85.01 & 1.72  & 51.52  & 1.28  & 78.68 \\
Claude 4.5 Sonnet                       & 18.87 & 82.24 & 3.26  & 61.88  & 0.55  & 93.25 \\
GPT-5.2                                 & \textbf{19.62} & 76.96 & \textbf{4.14} & 75.62  & \textbf{6.29} & 78.68 \\
\bottomrule
\end{tabular}
\end{adjustbox}
\caption{Mutation-testing results (MS and MS@Pass, \%) across Go, Rust, and Ruby for all 14 evaluated models under \texttt{standard} context. Ruby MS values should be interpreted as lower-bound estimates due to tooling limitations with dynamic dispatch and metaprogramming-heavy code. Julia and PHP are excluded due to incompatible mutation libraries and file-level-only tooling granularity, respectively. Best scores are bold}
\label{tab:mutation_results_appendix}
\end{table*}

Table~\ref{tab:mutation_results_appendix} reports mutation-testing results for all 14 evaluated models across Go, Rust, and Ruby. Both Mutation Score (MS) and Mutation@Pass Score (MS@Pass) are included, providing complementary views on fault-detection capability and test oracle quality.

Mutation testing meaningfully differentiates model capabilities beyond pass rate. GPT-5.2 achieves the highest MS across all three languages (Go: 19.62\%, Rust: 4.14\%, Ruby: 6.29\%), while Claude 4.5 Sonnet ranks second in Go MS (18.87\%). Smaller models often exhibit low MS despite moderate pass rates, confirming that passing repository tests does not necessarily imply strong fault-detection capability.

MS@Pass scores are consistently higher than MS across all models and languages, showing that test oracles are effective when tests execute successfully, which is the primary bottleneck is test executability rather than assertion quality. The ranking patterns remain broadly consistent with our overall evaluation: stronger repository-level models also tend to achieve higher MS, reinforcing the benchmark's internal validity.

Regarding language coverage: we investigated mutation testing for all five languages, but ecosystem-specific tooling limitations constrained full coverage. Ruby's mutation framework struggles with dynamic dispatch and metaprogramming-heavy code, reducing the number of valid mutants; Ruby MS values should therefore be interpreted as lower-bound estimates. Julia's available mutation libraries are incompatible with the required language version and dependency environment. PHP's existing mutation tools operate at file-level granularity only, which would introduce substantial noise and invalidate focal-function-level analysis. Despite these constraints, mutation testing across three ecosystems (Go, Rust, Ruby) already represents broader multilingual coverage than most prior unit test generation benchmarks. Notably, even \citet{jaintestgeneval} highlights mutation score computation as a key technical contribution, despite focusing solely on Python — a language with comparatively mature mutation tooling. Implementing mutation analysis across multiple heterogeneous ecosystems introduces substantially greater engineering complexity, further strengthening the benchmark's practical and technical scope.

\tcbset{
  promptbox/.style={
    enhanced,
    breakable,
    colback=gray!10,      
    colframe=black!60,    
    coltitle=white,       
    colbacktitle=black!80,
    fonttitle=\bfseries,  
    boxrule=0.6pt,
    arc=3mm,
    top=6pt,
    bottom=6pt,
    left=6pt,
    right=6pt
  }
}
\lstset{
  basicstyle=\ttfamily\footnotesize,
  frame=none,
  breaklines=true,
  columns=fullflexible,
  keepspaces=true,
  showstringspaces=false
}

\subsection{Correlation Analysis}
\label{sec:correlation_analysis}

We present two complementary correlation analyses: (i) correlations among \method's evaluation metrics across all evaluated models, and (ii) correlations between \method metrics and external code-generation performance on SWE-bench.

\begin{figure}[h]
    \centering
  \safeincludegraphics[width=1.0\linewidth]{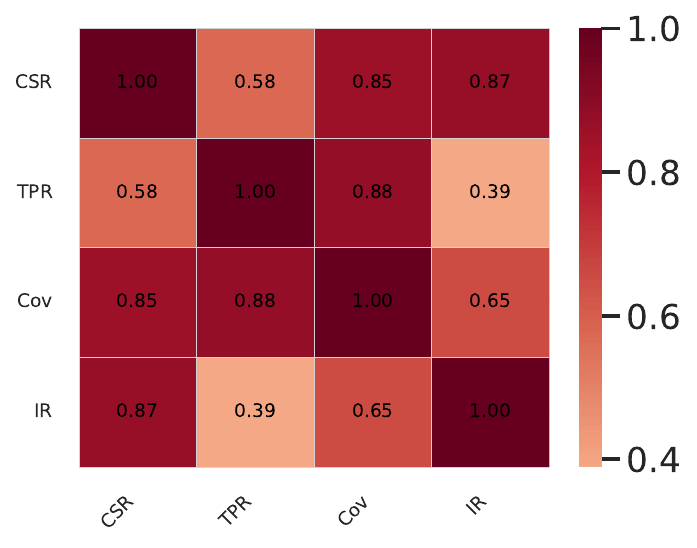}
    \caption{Correlation matrix illustrating relationships between evaluation metrics on \method.}
    \label{fig:correlation_heatmap}
\end{figure}

\begin{figure*}[t]
    \centering
    \begin{subfigure}{0.32\textwidth}
        \safeincludegraphics[width=\textwidth]{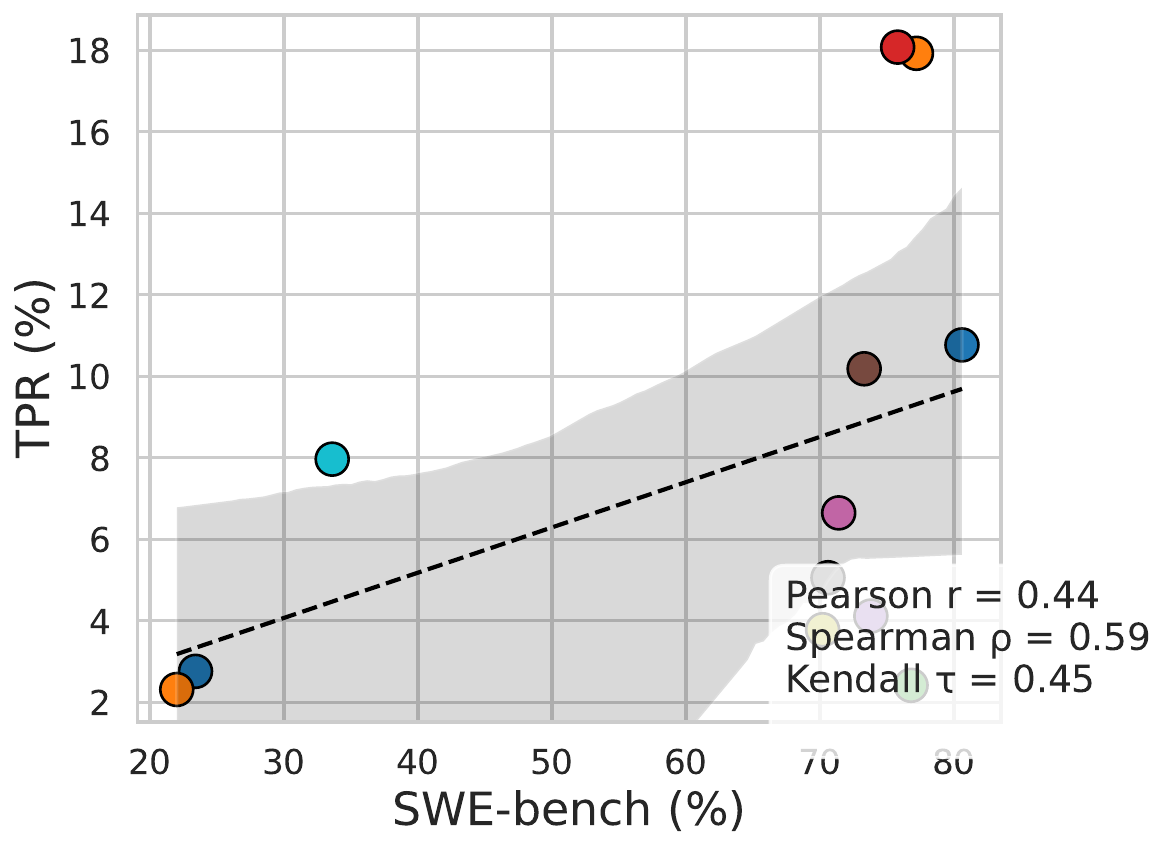}
    \subcaption{TPR vs SWE-bench}
    \end{subfigure}
    \hfill
    \begin{subfigure}{0.32\textwidth}
        \safeincludegraphics[width=\textwidth]{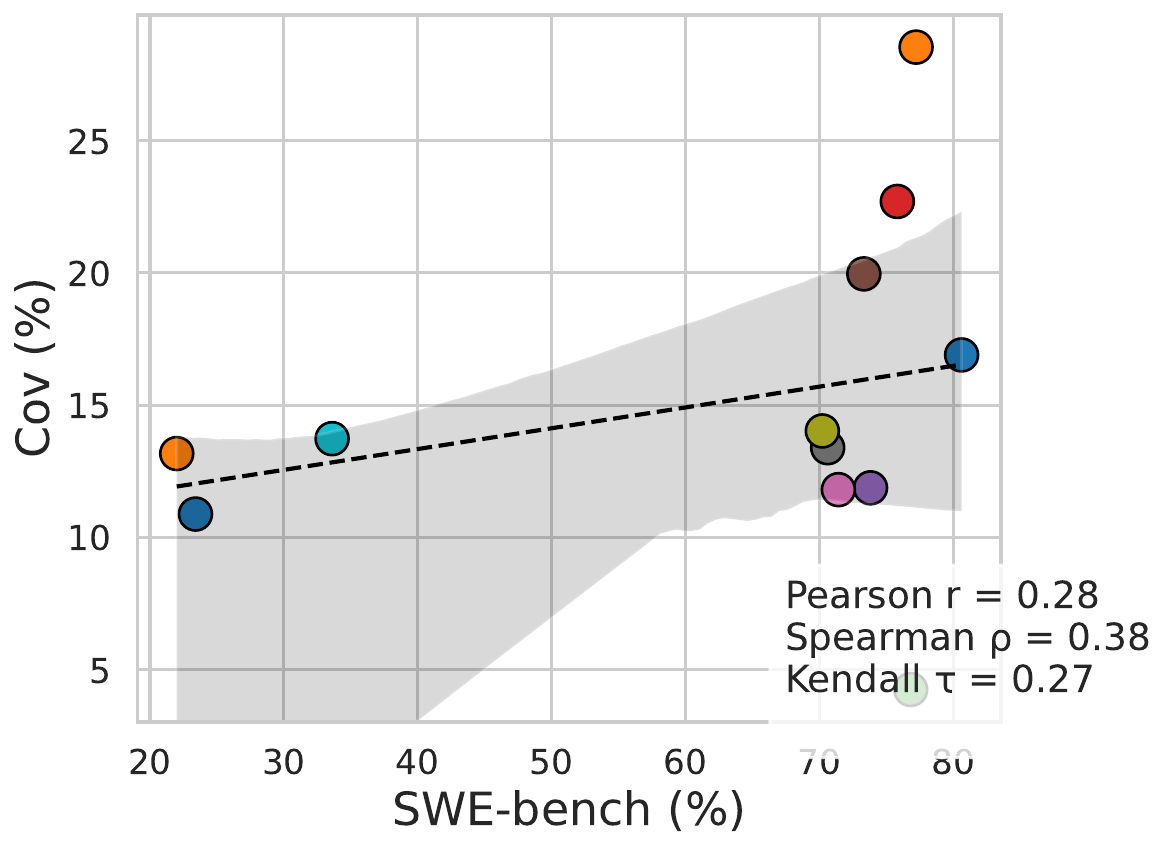}
    \subcaption{Cov vs SWE-bench}
    \end{subfigure}
    \hfill
    \begin{subfigure}{0.32\textwidth}
        \safeincludegraphics[width=\textwidth]{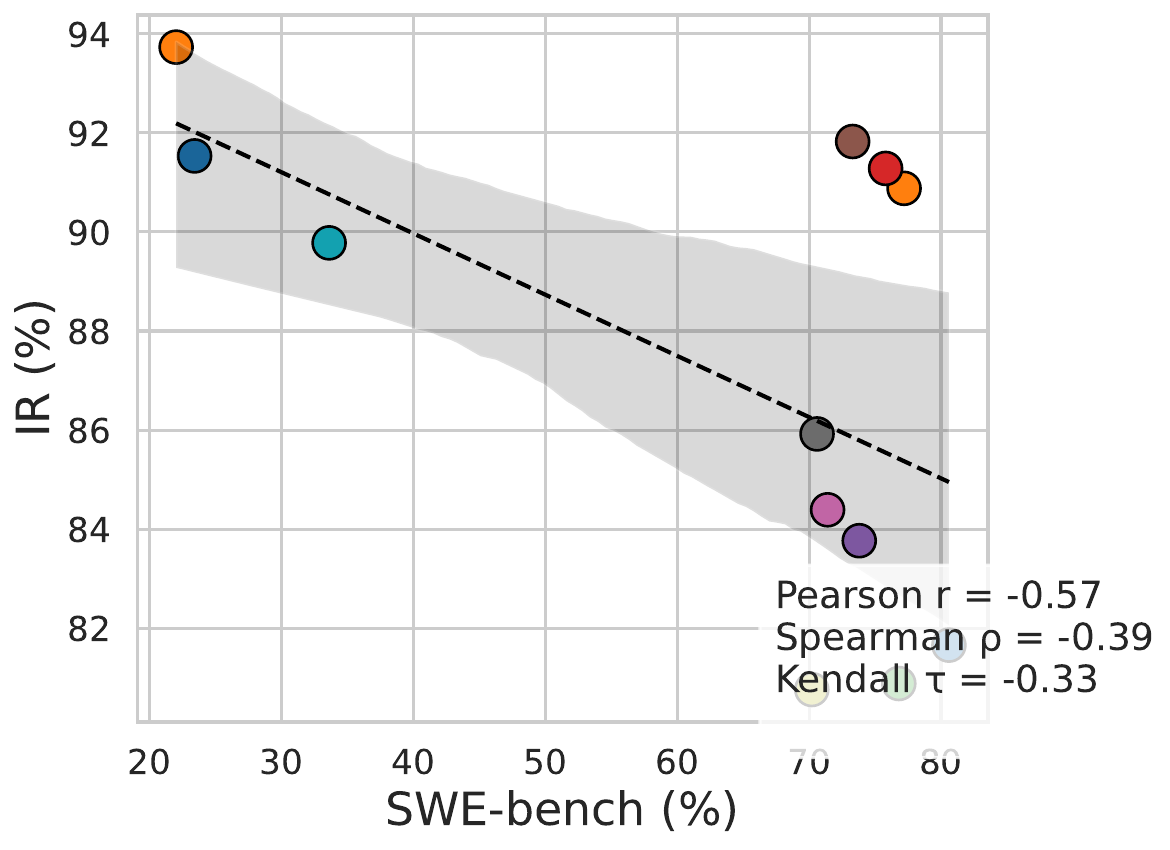}
    \subcaption{IR vs SWE-bench}
    \end{subfigure}

    \vspace{0.6em}
    \begin{subfigure}{\textwidth}
        \centering
        \safeincludegraphics[width=\textwidth]{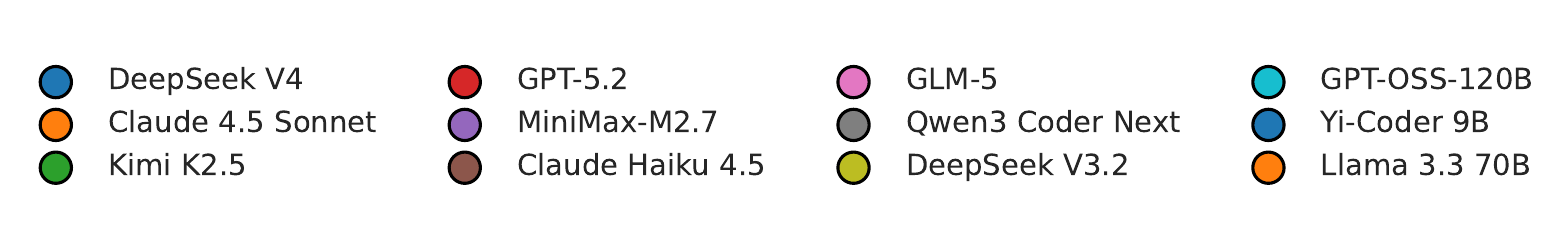}
    \end{subfigure}

    \caption{Correlation between unit test generation metrics on \method and SWE-bench performance. Pearson, Spearman, and Kendall coefficients are reported in each subfigure.}
    \label{fig:scatter_metrics}
\end{figure*}

\paragraph{Correlations among \method metrics.}
Figure~\ref{fig:correlation_heatmap} reports the pairwise Pearson correlations between code success rate (CSR), test pass rate (TPR), line coverage (Cov), and invocation rate (IR) across all 14 evaluated models on \method.

Two patterns stand out. First, CSR and Cov are strongly correlated ($r = 0.85$), and TPR and Cov are equally strongly correlated ($r = 0.88$), indicating that models which more reliably produce compilable/integrable tests also tend to achieve substantially higher coverage. CSR and TPR show moderate correlation ($r = 0.57$).

Second, and most importantly, IR captures a dimension that is clearly not redundant with the other metrics: it shows strong positive correlations with CSR ($r = 0.87$) and moderate positive correlation with Cov ($r = 0.65$), but only weak positive correlation with TPR ($r = 0.39$). This pattern indicates that while high IR generally accompanies high compilability and coverage, the relationship with test pass rates is much looser. A test suite can achieve high IR by consistently invoking the focal function, yet still exhibit low coverage or weak assertions, consistent with the presence of trivial or weak tests that pass without deeply exercising behavior.

\paragraph{Correlation with SWE-bench.}
To connect unit test generation and broader software engineering capability, we analyze correlations between model performance on \method and their SWE-bench Verified scores~\citep{swebench}. SWE-bench has emerged as a widely used yardstick for evaluating LLMs on realistic software engineering tasks, where a model must produce code changes that resolve real repository issues under test-based verification. We use it here as an external proxy for general software engineering capability.

We analyze correlations across 12 models with verified SWE-bench scores (ranging from 22.0\% for Llama 3.3 70B to 80.6\% for DeepSeek V4 Pro). Figure~\ref{fig:scatter_metrics} reports the Pearson, Spearman, and Kendall correlation coefficients.

Two patterns emerge. First, TPR shows the strongest relationship with SWE-bench: moderate positive Pearson ($r = 0.44$), and notably higher Spearman ($\rho = 0.59$) and Kendall ($\tau = 0.46$) correlations. This indicates that while the linear relationship is moderate, the rank-order relationship is more pronounced---models that perform better on SWE-bench tend to rank higher on TPR, though the relationship is not strictly linear. Cov shows a weaker positive trend (Spearman $\rho = 0.38$). CSR shows essentially no linear relationship with SWE-bench (Pearson $-0.12$), though a slight positive rank-order tendency (Spearman $0.13$).

Second, IR shows a notable negative trend with SWE-bench across all three methods (Pearson $r = -0.57$, Spearman $\rho = -0.39$, Kendall $\tau = -0.33$). We emphasize that these correlations are computed over only 12 models, which limits statistical power and the precision of the estimates; accordingly, we present this as an observational pattern rather than a conclusive finding. Subject to this caveat, the negative trend is suggestive of a structural misalignment: software engineering capability as measured by patch synthesis may not transfer to---and could even interfere with---the specific constraint of targeting a focal method. One plausible explanation is that models with higher general software engineering capability may over-complicate test harnesses or hallucinate complex setups that miss direct invocation, whereas models with lower capability may rely on simpler, more direct invocation patterns. This divergence hints that ``can fix a bug'' (SWE-bench) may be a distinct capability from ``can strictly invoke a specific function,'' though confirming this would require a larger set of models or controlled experiments.

\subsection{SWE-bench Verified Score Sources}
\label{app:swe-bench-sources}



\medskip
\textbf{Notes on evaluation scaffolds.} Most scores are reported using the \textbf{mini-SWE-agent} scaffold \cite{swebench} or equivalent standardized agent loop. The Qwen3-Coder-Next and DeepSeek V3.2 scores are from the Qwen3-Coder-Next technical report \cite{cao2026qwen3codernexttechnicalreport}, which evaluates across multiple scaffolds (SWE-Agent, MiniSWE-Agent, OpenHands); we report the SWE-Agent result for consistency.


For reproducibility, all scores should be treated as provider-reported or independently replicated using the official SWE-bench evaluation harness\footnote{\url{https://github.com/swe-bench/SWE-bench}}.

\subsection{Benchmark Hardness Analysis}
\label{sec:hardness_analysis}

\begin{figure}[h]
\centering
\safeincludegraphics[width=0.5\textwidth]{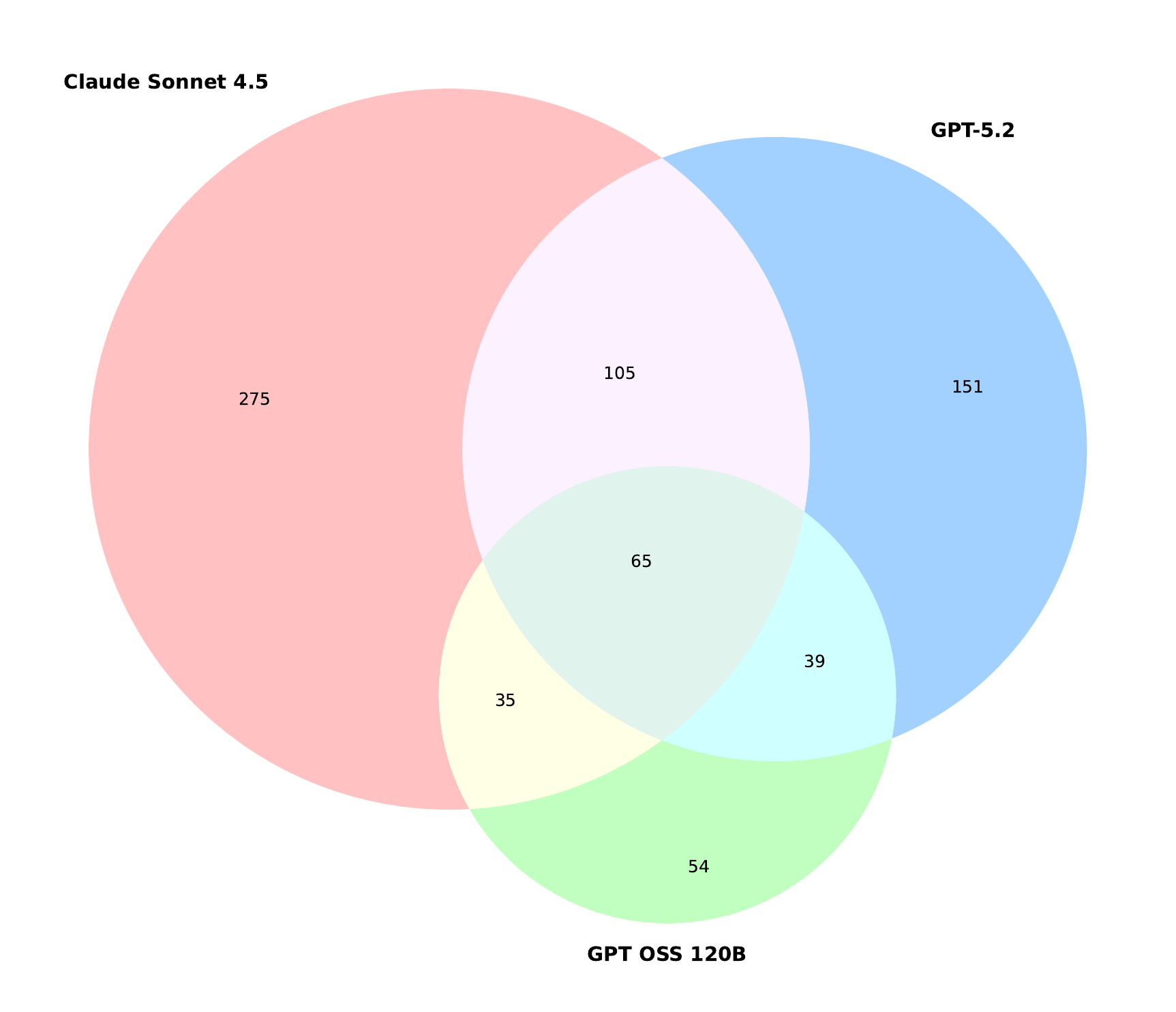}
\caption{Three-way Venn diagram showing overlap in successfully solved samples (test passes + coverage > 0) across three LLMs under \texttt{file\_context} mode for all languages. Each region indicates the number of samples uniquely solved by specific model combinations, revealing both shared capabilities and model-specific strengths.}
\label{fig:venn_hardness}
\end{figure}

To understand whether different models solve the same set of problems or exhibit unique strengths, we conduct a benchmark hardness analysis focusing on three representative flagship models: Claude Sonnet 4.5, GPT-OSS-120B, and GPT-5.2. We consider a sample as successfully solved if the generated test passes  and achieves non-zero line coverage. This criterion ensures that the test not only compiles and executes without errors but also exercises the focal function meaningfully.

Figure~\ref{fig:venn_hardness} presents a three-way Venn diagram illustrating the overlap in successfully solved samples across all five languages (Rust, Julia, Go, PHP, Ruby) under the \texttt{file\_context} mode. The diagram reveals several key insights:

\begin{itemize}
    \item \textbf{Shared Core.} A substantial intersection exists where all three models successfully solve the same problems, indicating a set of relatively easier samples that are accessible to state-of-the-art models regardless of their specific architectures or training regimes.

\item \textbf{Model-Specific Strengths.} Each model demonstrates unique capabilities, successfully solving problems that the other two fail on. These exclusive regions highlight architectural or training differences that enable specific models to handle certain code patterns, domains, or complexity characteristics more effectively.

\item \textbf{Pairwise Complementarity.} The pairwise intersections (problems solved by exactly two models) suggest that model combinations could potentially achieve higher coverage than any single model, motivating ensemble or multi-model generation strategies for practical test generation systems.

\end{itemize}

This hardness analysis provides empirical evidence that benchmark difficulty is not uniform across models, different models exhibit complementary strengths, and a significant portion of the benchmark remains challenging for all evaluated systems, preserving discriminative power for future model development.

\section{Agentic and Iterative-Repair Evaluation}
\label{app:agentic}

\paragraph{\textbf{Agentic setup.}} We evaluate a headless Claude Code agent with full repository access, including repository navigation, source inspection, compilation, execution, and iterative repair. Due to the substantially higher computational cost of agentic evaluation, we construct a stratified subset by sampling up to five easy and five hard focal functions from each repository (based on Claude 4.5 Sonnet's results). The agent is compared against the same underlying model under the standard single-turn protocol (TPR, Cov, IR). Table~\ref{tab:agentic_full} reports the complete per-language results.

\begin{table}[t]
\centering
\scriptsize
\begin{adjustbox}{width=0.45\textwidth}
\begin{tabular}{lcccc}
\toprule
\textbf{Lang.} & \textbf{Mode} & \textbf{TPR} & \textbf{Cov} & \textbf{IR} \\
\midrule
\multirow{2}{*}{Go}   & Standard & 42.6 & 54.0 & 100.0 \\
                      & Agentic  & 98.4 & 69.9 & 98.4  \\
\multirow{2}{*}{Rust} & Standard & 38.5 & 21.2 & 93.8  \\
                      & Agentic  & 72.3 & 38.9 & 95.4  \\
\multirow{2}{*}{Ruby} & Standard & 30.0 & 28.8 & 84.3  \\
                      & Agentic  & 78.6 & 41.2 & 68.6  \\
\multirow{2}{*}{PHP}  & Standard & 48.3 & 50.4 & 98.9  \\
                      & Agentic  & 59.8 & 17.1 & 27.6  \\
\multirow{2}{*}{Julia}& Standard & 45.5 & 46.3 & 98.2  \\
                      & Agentic  & 72.7 & 38.8 & 100.0 \\
\bottomrule
\end{tabular}
\end{adjustbox}
\caption{Complete agentic vs.\ passive evaluation on the stratified subset. Agentic execution consistently improves TPR, but the multi-metric protocol exposes cases where high TPR is accompanied by drops in Cov and IR.}
\label{tab:agentic_full}
\end{table}

Agentic execution consistently improves Test Pass Rate (TPR) across all five languages, with gains ranging from 1.2$\times$ to 2.6$\times$ over the passive baseline. More importantly, the multi-metric evaluation reveals behaviors invisible from pass rate alone. Although the agent achieves higher TPR on PHP ($+11.5\%$), line coverage drops by 33.3\% and IR by 71.3\%, indicating that the agent often generates passing tests by exercising easier helper or public API functions instead of the assigned focal function. Conversely, Julia reaches 100\% IR while coverage decreases because multiple dispatch selects a non-focal method overload despite invoking the intended API. These examples demonstrate that relying solely on pass rate would overestimate agent performance.

\paragraph{\textbf{Iterative repair.}} We additionally evaluate a one-round repair setting using compiler and test execution feedback with GPT-OSS-120B. Table~\ref{tab:repair_full} reports the results. Across all five languages, a single repair iteration improves TPR by 9--26\% while maintaining nearly unchanged IR, demonstrating that \method also naturally supports iterative-repair workflows commonly adopted in agentic coding systems.

\begin{table}[t]
\centering
\begin{adjustbox}{width=0.45\textwidth}
\begin{tabular}{lcccccc}
\toprule
 & \multicolumn{2}{c}{\textbf{TPR}} & \multicolumn{2}{c}{\textbf{Cov}} & \multicolumn{2}{c}{\textbf{IR}} \\
\cmidrule(lr){2-3}\cmidrule(lr){4-5}\cmidrule(lr){6-7}
\textbf{Lang.} & Pre & Post & Pre & Post & Pre & Post \\
\midrule
Go    & 25.9 & 52.4 & 40.0 & 62.5 & 96.7 & 96.7 \\
Rust  & 8.1  & 17.3 & 6.7  & 13.5 & 80.6 & 79.9 \\
Ruby  & 13.7 & 34.0 & 26.7 & 31.4 & 64.6 & 62.9 \\
PHP   & 6.3  & 31.6 & 5.6  & 30.1 & 99.5 & 99.2 \\
Julia & 9.5  & 34.4 & 32.1 & 44.2 & 92.3 & 89.9 \\
\bottomrule
\end{tabular}
\end{adjustbox}
\caption{One-round execution-feedback repair (GPT-OSS-120B). TPR improves by 9--26\% while IR stays nearly unchanged.}
\label{tab:repair_full}
\end{table}

\paragraph{\textbf{IR vs.\ TPR disagreement.}} Across the full benchmark, we measure the fraction of TPR-passing suites that fail IR: \textbf{9.7\% overall}, rising to 28.3\% for Rust and 19.2\% for Ruby, but only 0.6--0.7\% for Go and PHP (Julia: 6.1\%). Manual inspection shows these suites typically invoke mocked or stubbed code, or unrelated public APIs, rather than the focal function---direct evidence that IR captures failures TPR misses.

\paragraph{\textbf{Julia dispatch-mismatch case study.}} Multiple dispatch in Julia allows a test to ``call'' the intended function name yet still exercise a different method. Consider the following focal function, which overloads \texttt{Base.iterate} for reverse iteration over a \texttt{Deque}:

\begin{lstlisting}
function Base.iterate(
    di::Iterators.Reverse{<:Deque},
    (cb, i) = (di.itr.rear, di.itr.rear.back)
)
    ...
end
\end{lstlisting}

An agent-generated test pushes elements into a plain \texttt{Deque} and invokes \texttt{Base.iterate(d)} on the unwrapped container:

\begin{lstlisting}
d = Deque{Int}(); push!(d, 10); push!(d, 20); push!(d, 30)
res = Base.iterate(d); @test res !== nothing
val1, state1 = res; @test val1 == 10
\end{lstlisting}

At first glance this appears correct: it compiles, passes, and even calls \texttt{Base.iterate}, suggesting the target has been exercised. However, because the argument is a plain \texttt{Deque} rather than \texttt{Iterators.Reverse\{<:Deque\}}, Julia's multiple dispatch resolves to the \emph{forward}-iteration overload, never the focal method, resulting in 0/13 covered lines (0\% Cov) despite a successful test execution. A pass-rate metric alone would regard this as success, while coverage reveals the target was never exercised. Across the Julia agentic subset we observe this dispatch-mismatch pattern for \textbf{16 focal functions}, including 7 with zero line coverage, demonstrating that even modern agentic systems can generate superficially correct tests that fail to validate the intended implementation.

\section{Retrieval Parameter Ablation}
\label{app:retrieval_ablation}
\begin{table}[t]
\centering
\scriptsize
\setlength{\tabcolsep}{3pt}
\renewcommand{\arraystretch}{0.9}
\resizebox{\columnwidth}{!}{
\begin{tabular}{lcccc}
\toprule
\textbf{Retriever} & \textbf{$w_s$} & \textbf{$k$} & \textbf{Codestral} & \textbf{Yi-Coder} \\
\midrule
\multirow{9}{*}{BM25}
 & 30 &  5 & 18.02 & 7.80 \\
 & 30 & 10 & 17.84 & 7.41 \\
 & 30 & 15 & 18.54 & 5.31 \\
 & 50 &  5 & 17.64 & 8.41 \\
 & 50 & 10 & 18.23 & 6.56 \\
 & 50 & 15 & 18.59 & 5.31 \\
 & 70 &  5 & 17.65 & 8.13 \\
 & 70 & 10 & 18.30 & 7.28 \\
 & 70 & 15 & 18.54 & 5.16 \\
\midrule
\multirow{9}{*}{Dense}
 & 30 &  5 & 18.62 & 8.14 \\
 & 30 & 10 & 16.68 & 8.30 \\
 & 30 & 15 & 18.39 & 6.34 \\
 & 50 &  5 & 17.71 & 8.97 \\
 & 50 & 10 & 16.86 & 8.22 \\
 & 50 & 15 & 17.66 & 5.80 \\
 & 70 &  5 & 18.14 & 9.05 \\
 & 70 & 10 & 16.04 & 8.49 \\
 & 70 & 15 & 18.67 & 6.00 \\
\bottomrule
\end{tabular}}
\caption{Retrieval parameter ablation: line coverage (\%) across window size ($ws$) and top-$k$ configurations for BM25 and Dense (UniXCoder) retrievers on Go.}
\label{tab:retrieval_ablation}
\end{table}

To assess sensitivity to retrieval hyperparameters, we conducted an ablation study varying window size ($ws \in \{30, 50, 70\}$) and top-$k$ ($k \in \{5, 10, 15\}$) for both BM25 and Dense (UniXCoder) retrievers. We report average line coverage across two representative models: Codestral and Yi-Coder, evaluated on Go.

Across all models, performance remains stable under moderate parameter variation. For Yi-Coder under BM25, coverage fluctuates between 5.16\% and 8.41\% across nine configurations ($\leq$3.3\% absolute range). Codestral shows even narrower variation under BM25 (17.64\%--18.59\%, $\leq$1.0\% absolute range). Similar patterns hold under dense retrieval.

The chosen configuration ($ws{=}50$, $k{=}10$), consistent with common practice in prior RAG-based code generation work~\citep{zhang2023repocoder, wu2024repoformer, wang2024rlcoder}, consistently falls in the median-to-upper tier without being the peak for any single model. This avoids overfitting to a particular architecture while providing a representative and robust evaluation setting. These results confirm that retrieval performance is not highly sensitive to moderate parameter changes, supporting the validity of our chosen configuration.

Note that static augmentation contexts (file-level and LSP-based) are deterministic and invariant across runs; the ablation above applies only to retrieval-based augmentation.

\section{Statistical Significance Analysis}
\label{sec:significance}

We conducted two complementary statistical analyses to validate the ranking stability of model performance across languages and context modes: bootstrap resampling (10,000 iterations) for confidence interval estimation and McNemar's chi-squared test for pairwise model comparisons. Across 25 language/mode combinations (5 modes $\times$ 5 language groups), we analyzed result files covering all 14 models.

\subsection{Bootstrap Resampling}

For each language/mode, we performed non-parametric bootstrap resampling~\citep{efron1992bootstrap} to compute 95\% confidence intervals for the top-3 models' pass rates. Table~\ref{tab:significance_bootstrap} reports the mean pass rate and 95\% CI for each. Models whose CI does not overlap with the runner-up are considered \emph{statistically distinguishable}.

When TPR values are low (single-digit percentages for Rust, and below 10\% for several other language/settings), small absolute differences may not be statistically reliable. We explicitly acknowledge that adjacent models with marginal TPR differences may be statistically indistinguishable. However, larger gaps between performance tiers remain robust, as demonstrated below.

Across all 25 settings, the top model's CI does not overlap with the runner-up in 10 cases (notably PHP and Julia across all five modes, and Go under standard mode). The coefficient of variation (CV) serves as a stability metric---higher CV indicates less stable estimates relative to the mean. Most models exhibit CV $<$ 0.15, indicating reasonable stability; higher variance (CV $>$ 0.20) is observed for weaker models in low-TPR settings (e.g., GPT-OSS 120B in PHP under RAG modes), where small absolute pass counts amplify bootstrap variance.

\subsection{McNemar's Test Results}

We applied McNemar's chi-squared test~\citep{Mcnemar1947NoteOT} to all pairwise top-model comparisons within each language/mode setting. The test evaluates whether two models have statistically different pass/fail patterns, accounting for the paired nature of the data (same samples tested by both models). Table~\ref{tab:significance_mcnemar} reports all significant comparisons across all 25 settings; all p-values are corrected using Holm-Bonferroni correction for multiple comparisons.

Of 90 top-model pairwise comparisons across all settings, 56 (62\%) are statistically significant at $\alpha=0.05$. The most consistent winner is \textbf{claude-sonnet-4-5}, achieving the highest rank in 21/25 settings and winning 32 significant pairwise comparisons. GPT-5.2 shows 19 significant wins, primarily in Go and Ruby. In Rust/standard, claude-sonnet-4-5 and gpt-5.2 are statistically indistinguishable ($\chi$²=0.22, p=0.64), though both significantly outperform claude-haiku-4.5.

\textbf{Model Tier Separation (Rust File Context).} In Rust under file-level context, the proprietary SOTA tier (Claude Sonnet 4.5: 21.16\%, 95\% CI [18.6\%, 23.8\%]) is statistically distinct from the leading open-weight baseline (GPT-OSS-120B: 8.06\%, CI [6.3\%, 9.9\%]), with non-overlapping confidence intervals. Pairwise McNemar's test confirms this separation is highly significant ($\chi$²=89.63, p=2.87e-21 with Bonferroni correction).

\textbf{Context Augmentation Significance.} The improvement from Standard to File-Level Context for Claude Sonnet 4.5 (12.78\% $\to$ 21.16\%) is confirmed via McNemar's test ($\chi$²=17.65, p=1.53e-05), with 42 additional passing tests versus only 6 additional failures. This pattern holds across languages: PHP (26.95\% $\to$ 31.43\%) and Go (23.66\% $\to$ 34.14\%) show similarly significant gains, confirming that repository-aware context provides a decisive, non-random advantage.

\begin{table*}[!t]
\centering
\scriptsize
\setlength{\tabcolsep}{10.0pt}
\renewcommand{\arraystretch}{0.70}

\begin{adjustbox}{width=\textwidth,totalheight=0.82\textheight}
\begin{tabular}{@{}lccccc@{}}
\toprule
\textbf{Language/Mode} & \textbf{Model A} & \textbf{Model B} & $\chi^2$ & \textbf{p-value} & \textbf{Sig.} \\
\midrule
 & \multicolumn{5}{c}{\textit{\textbf{Rust}}} \\
\midrule
Rust / Standard & Claude 4.5 Sonnet & GPT-5.2 & 0.22 & 0.6396 & ns \\
Rust / Standard & Claude 4.5 Sonnet & Claude 4.5 Haiku & 11.56 & 0.0007 & *** \\
Rust / Standard & GPT-5.2 & Claude 4.5 Haiku & 8.09 & 0.0045 & ** \\
Rust / File Ctx & Claude 4.5 Sonnet & GPT-OSS 120B & 89.63 & 2.87e-21 & *** \\
Rust / File Ctx & Claude 4.5 Sonnet & GPT-5.2 & 10.37 & 0.0013 & ** \\
Rust / LSP Ctx & Claude 4.5 Sonnet & GPT-OSS 120B & 29.54 & 5.49e-08 & *** \\
Rust / RAG BM25 & Claude 4.5 Sonnet & GPT-OSS 120B & 64.61 & 9.13e-16 & *** \\
Rust / RAG Dense & Claude 4.5 Sonnet & GPT-OSS 120B & 85.82 & 1.98e-20 & *** \\
Rust / RAG Dense & Claude 4.5 Sonnet & GPT-5.2 & 9.23 & 0.0024 & ** \\
\midrule
 & \multicolumn{5}{c}{\textit{\textbf{Go}}} \\
\midrule
Go / Standard & GPT-5.2 & GPT-OSS 120B & 11.03 & 0.0009 & *** \\
Go / Standard & Claude 4.5 Sonnet & GPT-OSS 120B & 6.32 & 0.0119 & * \\
Go / File Ctx & GPT-5.2 & GPT-OSS 120B & 31.84 & 1.67e-08 & *** \\
Go / File Ctx & Claude 4.5 Sonnet & GPT-OSS 120B & 18.46 & 1.73e-05 & *** \\
Go / LSP Ctx & GPT-5.2 & GPT-OSS 120B & 26.46 & 2.69e-07 & *** \\
Go / LSP Ctx & Claude 4.5 Sonnet & GPT-OSS 120B & 15.96 & 6.47e-05 & *** \\
Go / RAG BM25 & GPT-5.2 & GPT-OSS 120B & 21.97 & 2.78e-06 & *** \\
Go / RAG BM25 & Claude 4.5 Sonnet & GPT-OSS 120B & 18.89 & 1.39e-05 & *** \\
Go / RAG Dense & GPT-5.2 & GPT-OSS 120B & 44.76 & 2.22e-11 & *** \\
Go / RAG Dense & Claude 4.5 Sonnet & GPT-OSS 120B & 31.21 & 2.32e-08 & *** \\
\midrule
 & \multicolumn{5}{c}{\textit{\textbf{Julia}}} \\
\midrule
Julia / Standard & Claude 4.5 Sonnet & Claude 4.5 Haiku & 18.39 & 1.80e-05 & *** \\
Julia / File Ctx & Claude 4.5 Sonnet & GPT-OSS 120B & 40.29 & 2.19e-10 & *** \\
Julia / File Ctx & Claude 4.5 Sonnet & GPT-5.2 & 7.56 & 0.0060 & ** \\
Julia / LSP Ctx & Claude 4.5 Sonnet & GPT-OSS 120B & 31.44 & 2.06e-08 & *** \\
Julia / RAG BM25 & Claude 4.5 Sonnet & GPT-OSS 120B & 50.27 & 1.34e-12 & *** \\
Julia / RAG BM25 & Claude 4.5 Sonnet & GPT-5.2 & 14.11 & 1.72e-04 & *** \\
Julia / RAG Dense & Claude 4.5 Sonnet & GPT-OSS 120B & 15.74 & 7.27e-05 & *** \\
Julia / RAG Dense & Claude 4.5 Sonnet & GPT-5.2 & 14.08 & 1.75e-04 & *** \\
\midrule
 & \multicolumn{5}{c}{\textit{\textbf{PHP}}} \\
\midrule
PHP / Standard & Claude 4.5 Sonnet & Claude 4.5 Haiku & 38.50 & 5.48e-10 & *** \\
PHP / Standard & Claude 4.5 Sonnet & GLM-5 & 45.29 & 1.70e-11 & *** \\
PHP / File Ctx & Claude 4.5 Sonnet & GPT-OSS 120B & 130.91 & 2.59e-30 & *** \\
PHP / File Ctx & Claude 4.5 Sonnet & GPT-5.2 & 15.59 & 7.85e-05 & *** \\
PHP / LSP Ctx & Claude 4.5 Sonnet & GPT-OSS 120B & 113.61 & 1.59e-26 & *** \\
PHP / LSP Ctx & Claude 4.5 Sonnet & GPT-5.2 & 50.92 & 9.63e-13 & *** \\
PHP / RAG BM25 & Claude 4.5 Sonnet & GPT-OSS 120B & 152.30 & 5.46e-35 & *** \\
PHP / RAG BM25 & Claude 4.5 Sonnet & GPT-5.2 & 66.65 & 3.24e-16 & *** \\
PHP / RAG Dense & Claude 4.5 Sonnet & GPT-OSS 120B & 149.74 & 1.98e-34 & *** \\
PHP / RAG Dense & Claude 4.5 Sonnet & GPT-5.2 & 52.48 & 4.34e-13 & *** \\
\midrule
 & \multicolumn{5}{c}{\textit{\textbf{Ruby}}} \\
\midrule
Ruby / Standard & GPT-5.2 & DeepSeek V4-pro & 76.96 & 1.75e-18 & *** \\
Ruby / Standard & GPT-5.2 & GPT-OSS 120B & 95.07 & 1.84e-22 & *** \\
Ruby / File Ctx & Claude 4.5 Sonnet & GPT-OSS 120B & 10.59 & 0.0011 & ** \\
Ruby / LSP Ctx & GPT-5.2 & Claude 4.5 Sonnet & 68.21 & 1.47e-16 & *** \\
Ruby / LSP Ctx & GPT-5.2 & GPT-OSS 120B & 80.13 & 3.50e-19 & *** \\
Ruby / RAG BM25 & GPT-5.2 & Claude 4.5 Sonnet & 29.18 & 6.58e-08 & *** \\
Ruby / RAG BM25 & GPT-5.2 & GPT-OSS 120B & 13.78 & 2.06e-04 & *** \\
Ruby / RAG Dense & GPT-5.2 & Claude 4.5 Sonnet & 29.90 & 4.55e-08 & *** \\
Ruby / RAG Dense & GPT-5.2 & GPT-OSS 120B & 13.34 & 2.60e-04 & *** \\
\bottomrule
\end{tabular}
\end{adjustbox}
\caption{McNemar's chi-squared test results for all significant pairwise model comparisons across all language/mode combinations. Holm-Bonferroni corrected p-values. Significance levels: *** p$<0.001$, ** p$<0.01$, * p$<0.05$, ns = not significant.}
\label{tab:significance_mcnemar}
\end{table*}
\subsection{Summary}

The combined bootstrap and McNemar analyses confirm that \textbf{claude-sonnet-4-5} is the most consistent top performer (21/25 settings ranked first, 32 significant pairwise wins), with GPT-5.2 as the primary challenger in Go and Ruby. Overall, 56 of 90 pairwise comparisons (62\%) reach significance, and the context augmentation effect is confirmed across all five languages.

\clearpage
\begin{sidewaystable*}[p]
\centering
\scriptsize
\begin{adjustbox}{width=\textheight}
\begin{tabular}{l|c|c|c|c|c|c|c|c|c|c|c|c|c|c|c}
\toprule
 & \multicolumn{3}{c|}{\textbf{Go}} & \multicolumn{3}{c|}{\textbf{Julia}} & \multicolumn{3}{c|}{\textbf{PHP}} & \multicolumn{3}{c|}{\textbf{Ruby}} & \multicolumn{3}{c|}{\textbf{Rust}} \\
\cmidrule(lr){2-4}\cmidrule(lr){5-7}\cmidrule(lr){8-10}\cmidrule(lr){11-13}\cmidrule(lr){14-16}
 & \textbf{TPR} & \textbf{CI} & \textbf{CV} & \textbf{TPR} & \textbf{CI} & \textbf{CV} & \textbf{TPR} & \textbf{CI} & \textbf{CV} & \textbf{TPR} & \textbf{CI} & \textbf{CV} & \textbf{TPR} & \textbf{CI} & \textbf{CV} \\
\midrule
 & \multicolumn{15}{c}{\textbf{Standard}} \\
\midrule
GPT-5.2 & 25.78 & [22.7, 29.0] & 0.064 & 16.52 & [13.7, 19.4] & 0.087 & 10.06 & [7.9, 12.5] & 0.118 & 25.78 & [22.5, 29.0] & 0.065 & 12.25 & [10.2, 14.4] & 0.088 \\
Claude 4.5 Sonnet & 23.66 & [20.5, 26.8] & 0.067 & 19.85 & [16.8, 22.8] & 0.076 & 26.95 & [23.5, 30.3] & 0.064 & 6.37 & [4.6, 8.3] & 0.148 & 12.78 & [10.6, 15.0] & 0.086 \\
GPT-OSS 120B & 19.27 & [16.4, 22.2] & 0.078 & 4.39 & [3.0, 5.9] & 0.152 & 3.41 & [2.0, 4.8] & 0.208 & 6.97 & [5.2, 8.9] & 0.140 & 5.80 & [4.3, 7.5] & 0.129 \\
\midrule
 & \multicolumn{15}{c}{\textbf{File Context}} \\
\midrule
GPT-5.2 & 37.69 & [34.0, 41.4] & 0.049 & 17.28 & [14.5, 20.2] & 0.084 & 22.93 & [19.7, 26.2] & 0.072 & 13.02 & [10.5, 15.6] & 0.099 & 17.94 & [15.5, 20.4] & 0.070 \\
Claude 4.5 Sonnet & 34.14 & [30.7, 37.7] & 0.052 & 23.98 & [20.8, 27.2] & 0.068 & 31.43 & [27.9, 35.0] & 0.058 & 16.26 & [13.5, 19.1] & 0.088 & 21.16 & [18.6, 23.8] & 0.063 \\
GPT-OSS 120B & 25.93 & [22.8, 29.2] & 0.064 & 9.49 & [7.3, 11.8] & 0.119 & 6.34 & [4.5, 8.2] & 0.151 & 13.71 & [10.3, 17.4] & 0.135 & 8.06 & [6.3, 9.9] & 0.111 \\
\midrule
 & \multicolumn{15}{c}{\textbf{LSP Context}} \\
\midrule
GPT-5.2 & 28.03 & [24.8, 31.4] & 0.061 & 12.84 & [10.4, 15.4] & 0.099 & 11.76 & [9.3, 14.2] & 0.108 & 23.84 & [20.6, 27.0] & 0.068 & 11.50 & [9.5, 13.5] & 0.091 \\
Claude 4.5 Sonnet & 25.34 & [22.2, 28.6] & 0.064 & 15.64 & [13.0, 18.4] & 0.088 & 26.77 & [23.4, 30.2] & 0.066 & 7.69 & [5.8, 9.8] & 0.134 & 13.55 & [11.4, 15.8] & 0.083 \\
GPT-OSS 120B & 17.41 & [14.6, 20.3] & 0.083 & 5.84 & [4.1, 7.6] & 0.154 & 4.48 & [2.9, 6.2] & 0.183 & 7.55 & [5.6, 9.6] & 0.136 & 6.45 & [4.9, 8.1] & 0.124 \\
\midrule
 & \multicolumn{15}{c}{\textbf{RAG BM25}} \\
\midrule
GPT-5.2 & 28.89 & [25.6, 32.2] & 0.059 & 11.42 & [9.2, 13.9] & 0.106 & 13.17 & [10.7, 15.8] & 0.101 & 19.84 & [16.9, 23.0] & 0.078 & 15.37 & [13.1, 17.7] & 0.077 \\
Claude 4.5 Sonnet & 28.05 & [24.8, 31.4] & 0.060 & 19.16 & [16.2, 22.1] & 0.078 & 30.02 & [26.5, 33.6] & 0.061 & 10.38 & [8.1, 12.7] & 0.111 & 15.98 & [13.6, 18.4] & 0.075 \\
GPT-OSS 120B & 18.98 & [16.1, 22.0] & 0.078 & 6.42 & [4.7, 8.3] & 0.146 & 3.41 & [2.0, 4.8] & 0.208 & 17.14 & [13.1, 21.1] & 0.119 & 4.93 & [3.5, 6.3] & 0.144 \\
\midrule
 & \multicolumn{15}{c}{\textbf{RAG Dense}} \\
\midrule
GPT-5.2 & 32.13 & [28.8, 35.6] & 0.055 & 11.68 & [9.2, 14.2] & 0.106 & 16.12 & [13.3, 19.0] & 0.090 & 21.06 & [18.1, 24.1] & 0.075 & 13.85 & [11.7, 16.1] & 0.082 \\
Claude 4.5 Sonnet & 30.32 & [26.9, 33.7] & 0.057 & 18.57 & [15.8, 21.5] & 0.079 & 31.43 & [27.9, 35.1] & 0.059 & 11.27 & [8.9, 13.8] & 0.109 & 17.18 & [14.8, 19.5] & 0.071 \\
GPT-OSS 120B & 17.70 & [14.9, 20.5] & 0.081 & 10.08 & [7.9, 12.4] & 0.114 & 4.95 & [3.3, 6.7] & 0.173 & 18.56 & [14.6, 22.6] & 0.112 & 4.52 & [3.2, 5.9] & 0.150 \\
\bottomrule
\end{tabular}
\end{adjustbox}
\caption{Bootstrap 95\% CI for top models across all language/mode combinations (5 languages $\times$ 5 modes). TPR values are percentages. Within each mode block, only models appearing in the top-3 for at least 3 out of 5 languages are shown (Standard block); other blocks show the top-3 models consistent across all languages. CV = coefficient of variation.}
\label{tab:significance_bootstrap}
\end{sidewaystable*}
\clearpage

\section{Model Details}
\begin{table*}[t]
\centering

\begin{adjustbox}{width=\textwidth}
\begin{tabular}{@{}p{2.5cm} p{6.0cm} p{3.0cm} p{7.5cm}@{}}
	\toprule
	\textbf{Provider} & \textbf{Model ID} & \textbf{Short Name} & \textbf{Link} \\
\midrule

\multirow{2}{*}{OpenAI} & gpt-5.2 & GPT-5.2 & \url{https://openai.com/index/introducing-gpt-5-2/} \\
& openai-gpt-oss-120b & GPT-OSS 120B & \url{https://huggingface.co/openai/gpt-oss-120b} \\
\midrule

\multirow{2}{*}{Anthropic} & claude-haiku-4-5-20251001 & Claude 4.5 Haiku & \url{https://www.anthropic.com/news/claude-haiku-4-5} \\
& claude-sonnet-4-5-20250929 & Claude 4.5 Sonnet & \url{https://www.anthropic.com/news/claude-sonnet-4-5} \\
\midrule

Zhipu AI & zai-org/GLM-5 & GLM-5 & \url{https://huggingface.co/zai-org/GLM-5} \\
\midrule

\multirow{2}{*}{DeepSeek} & deepseek-ai/DeepSeek-V3.2 & DeepSeek V3.2 & \url{https://huggingface.co/deepseek-ai/DeepSeek-V3.2} \\
& deepseek-ai/DeepSeek-V4-pro & DeepSeek V4-pro & \url{https://huggingface.co/deepseek-ai/DeepSeek-V4-pro} \\
\midrule

Moonshot AI & moonshotai/Kimi-k2.5-Instruct & Kimi-k2.5 & \url{https://huggingface.co/moonshotai/Kimi-K2.5} \\
\midrule

MiniMax & MiniMaxAI/MiniMax-M2.7 & MiniMax-M2.7 & \url{https://huggingface.co/MiniMaxAI/MiniMax-M2.7} \\
\midrule

\multirow{2}{*}{Alibaba} & Qwen/Qwen3-8B & Qwen3 8B & \url{https://huggingface.co/Qwen/Qwen3-8B} \\
& Qwen/Qwen3-Coder-Next & Qwen3-Coder-Next 80B & \url{https://huggingface.co/Qwen/Qwen3-Coder-Next} \\
\midrule

01.AI & 01-ai/Yi-Coder-9B-Chat & Yi-Coder 9B & \url{https://huggingface.co/01-ai/Yi-Coder-9B-Chat} \\
\midrule

Mistral AI & mistralai/Codestral-22B-v0.1 & Codestral 22B & \url{https://huggingface.co/mistralai/Codestral-22B-v0.1} \\
\midrule

Meta & meta-llama/Llama-3.3-70B-Instruct & Llama-3.3 70B & \url{https://huggingface.co/meta-llama/Llama-3.3-70B-Instruct} \\
\bottomrule
\end{tabular}
\end{adjustbox}
\caption{Details of the LLMs evaluated in our experimental study.}
\label{tab:model_details}
\end{table*}

\label{app:model_details}
In our experimental study, we evaluate a diverse set of large language models spanning both proprietary and open-source families. Specifically, we consider proprietary models from OpenAI and Anthropic, including GPT-5.2 and GPT-OSS~\cite{agarwal2025gpt}, as well as Claude~4.5.

Our evaluation further covers a broad range of open-source and open-weight models, including GLM-5~\cite{glm5team2026glm5vibecodingagentic}, DeepSeek~V3.2~\cite{deepseekai2025deepseekv32pushingfrontieropen} and DeepSeek~V4-pro, Kimi-k2.5~\cite{kimiteam2026kimik25visualagentic}, MiniMax-M2.7, and several models from Alibaba's Qwen family, including Qwen3 and Qwen3-Coder-Next~\cite{cao2026qwen3codernexttechnicalreport}.
We also evaluate Yi, Codestral, and Llama-3.3.
Complete model identifiers and release details are provided in Table~\ref{tab:model_details}

\section{Failure Mode Details}
\label{sec:error_details}
\begin{table*}[h]
\centering
\caption{Failure modes by language with and without context (per-model). Values are percentages of total samples assigned to each failure mode; rows sum to 100\% within each column, with the ``No Error / Pass'' row capturing samples that pass all checks.}
\label{tab:error_distribution_per_model}
\scriptsize
\setlength{\tabcolsep}{2pt}
\renewcommand{\arraystretch}{1.1}
\resizebox{\textwidth}{!}{%
\begin{tabular}{ll*{5}{cc}}
\toprule
\multirow{2}{*}{\textbf{Model}} & \multirow{2}{*}{\textbf{Failure Mode}} & \multicolumn{2}{c}{\textbf{Rust}} & \multicolumn{2}{c}{\textbf{Go}} & \multicolumn{2}{c}{\textbf{Julia}} & \multicolumn{2}{c}{\textbf{Ruby}} & \multicolumn{2}{c}{\textbf{PHP}} \\
\cmidrule(lr){3-4} \cmidrule(lr){5-6} \cmidrule(lr){7-8} \cmidrule(lr){9-10} \cmidrule(lr){11-12}
&  & \textbf{Std.} & \textbf{Ctx.} & \textbf{Std.} & \textbf{Ctx.} & \textbf{Std.} & \textbf{Ctx.} & \textbf{Std.} & \textbf{Ctx.} & \textbf{Std.} & \textbf{Ctx.} \\
\midrule
\multirow{6}{*}{Claude 4.5 Sonnet} & Syntactic \& Compilation & 1.29\% & 2.36\% & 14.02\% & 19.12\% & 1.75\% & 1.61\% & 0.44\% & 0.89\% & 1.24\% & 1.86\% \\
 & Type System \& Memory & 14.18\% & 14.93\% & 13.60\% & 8.36\% & 9.36\% & 8.04\% & 5.93\% & 2.67\% & 6.66\% & 3.72\% \\
 & API Hallucination & 43.07\% & 30.08\% & 21.53\% & 11.61\% & 59.65\% & 53.95\% & 49.78\% & 32.74\% & 24.46\% & 29.72\% \\
 & Logic \& Assertion & 28.68\% & 31.47\% & 27.20\% & 26.77\% & 9.36\% & 12.43\% & 37.48\% & 47.41\% & 40.71\% & 33.28\% \\
 & Test Design \& Mocking & 4.40\% & 6.44\% & 0.00\% & 0.42\% & 0.00\% & 0.00\% & 0.74\% & 0.74\% & 0.31\% & 0.46\% \\
 & No Error / Pass & 8.38\% & 14.72\% & 23.65\% & 33.72\% & 19.88\% & 23.97\% & 5.63\% & 15.55\% & 26.62\% & 30.96\% \\
\midrule
\multirow{6}{*}{GPT-5.2} & Syntactic \& Compilation & 0.97\% & 1.50\% & 19.69\% & 14.59\% & 15.79\% & 8.04\% & 1.78\% & 7.41\% & 2.32\% & 3.72\% \\
 & Type System \& Memory & 11.71\% & 13.96\% & 13.88\% & 8.50\% & 6.87\% & 7.02\% & 4.74\% & 2.52\% & 5.42\% & 4.18\% \\
 & API Hallucination & 48.55\% & 33.62\% & 19.12\% & 12.32\% & 37.43\% & 46.93\% & 23.41\% & 38.81\% & 47.06\% & 38.08\% \\
 & Logic \& Assertion & 26.53\% & 32.98\% & 21.53\% & 26.91\% & 23.39\% & 20.76\% & 44.30\% & 38.07\% & 35.14\% & 31.11\% \\
 & Test Design \& Mocking & 3.87\% & 5.59\% & 0.71\% & 1.42\% & 0.00\% & 0.00\% & 9.19\% & 0.59\% & 2.48\% & 0.77\% \\
 & No Error / Pass & 8.38\% & 12.35\% & 25.07\% & 36.26\% & 16.52\% & 17.25\% & 16.58\% & 12.60\% & 7.58\% & 22.14\% \\
\midrule
\multirow{6}{*}{GPT-OSS (120B)} & Syntactic \& Compilation & 9.56\% & 3.87\% & 41.22\% & 27.34\% & 29.09\% & 16.52\% & 2.81\% & 2.86\% & 48.14\% & 30.65\% \\
 & Type System \& Memory & 8.06\% & 15.57\% & 12.32\% & 8.50\% & 10.82\% & 9.94\% & 3.26\% & 1.14\% & 2.32\% & 0.93\% \\
 & API Hallucination & 52.52\% & 44.58\% & 16.01\% & 19.12\% & 46.64\% & 49.42\% & 53.33\% & 65.43\% & 35.14\% & 54.02\% \\
 & Logic \& Assertion & 24.06\% & 27.93\% & 11.19\% & 19.12\% & 9.06\% & 14.62\% & 33.48\% & 16.57\% & 10.99\% & 8.05\% \\
 & Test Design \& Mocking & 1.61\% & 2.90\% & 0.42\% & 1.13\% & 0.00\% & 0.00\% & 3.26\% & 0.29\% & 0.46\% & 0.77\% \\
 & No Error / Pass & 4.19\% & 5.16\% & 18.84\% & 24.79\% & 4.39\% & 9.50\% & 3.86\% & 13.71\% & 2.95\% & 5.58\% \\
\bottomrule
\end{tabular}}
\end{table*}

\subsection{Category Definitions }

We use the following definitions for the five failure classes reported in Section~\ref{sec:failure_modes}.

\paragraph{\textbf{1. Syntactic \& Compilation}} Tests that fail at compile or interpretation time due to syntax errors, missing imports, or type declaration issues. This is especially prominent in Go, reflecting challenges with package visibility and interface satisfaction. Examples include missing semicolons (PHP), incorrect closure syntax (Rust), block precedence issues (Ruby), and using Python keywords in Julia code (``def'' instead of ``function'').

\paragraph{\textbf{2. Type System \& Memory}} Failures due to type mismatches, ownership violations, or memory safety issues. This is most visible in Rust due to its strict ownership model (borrow checker violations, lifetime errors), and less prominent in dynamic languages. In Go, this includes interface conversion panics and missing type assertion checks. In PHP and Ruby, this manifests as strict type mismatches and \texttt{NoMethodError} on \texttt{nil}.

\paragraph{\textbf{3. API Hallucination}} Cases where models invent non-existent libraries, methods, or APIs. This is the dominant failure mode in several languages that remain understudied in repository-level test-generation research, indicating that models frequently generate plausible-looking but fictitious API calls. Examples include fabricating \texttt{testify} assertions in Go (e.g., \texttt{assert.NotError}), inventing trait methods in Rust, calling \texttt{len()} instead of \texttt{length()} in Julia, and using deprecated PHPUnit annotations.

\paragraph{\textbf{4. Logic \& Assertion}} Tests that compile and run but fail due to incorrect logic or assertions. This is especially prominent in PHP, where syntactic correctness is comparatively easier to achieve but semantic understanding of expected behavior remains challenging. Examples include unwrapping \texttt{Result} expecting failure (Rust), checking \texttt{err != nil} without calling \texttt{t.Fatal} (Go), using \texttt{==} for floating point comparisons instead of \texttt{isapprox} (Julia), and confused argument order in PHPUnit assertions.

\paragraph{\textbf{5. Test Design \& Mocking}} Failures from missing fixtures, incorrect mocking, or improper test setup, reflecting the complexity of framework conventions and dependency injection patterns. The primary indicator is tests that pass all checks but achieve zero focal coverage (\texttt{covered\_lines == 0}), suggesting the focal function was never invoked. Examples include mocking the class under test instead of external dependencies, hardcoded database IDs without data setup, and global state pollution without cleanup.

\subsection{Error Classification Methodology}
\label{app:error taxonomy}

To systematically analyze failure modes in LLM-generated tests, we developed an automated error classification system that processes test execution logs, compilation diagnostics, and coverage statistics across all five languages. The classifier employs a rule-based approach with language-specific pattern matching and a hierarchical decision procedure that prioritizes error categories based on the test lifecycle stage.

\paragraph{Hierarchical Classification Decision Tree.}
Classification follows a priority-ordered decision tree that mirrors the test execution lifecycle. Importantly, the decision logic differs between \textbf{static languages (Rust, Go)} and \textbf{dynamic languages (Julia, PHP, Ruby)}, reflecting fundamental differences in when type checking and API resolution occur:

\begin{enumerate}[leftmargin=1.5em, itemsep=0.3em]
\item \textbf{Preprocessing Failures:} If test code is empty or logs indicate ``no tests generated'', classify as \textbf{Category 1} (Syntactic \& Compilation), signaling that the LLM output was malformed enough to be filtered during test extraction.

\item \textbf{Compilation Failures:} When \texttt{check.compilation == False}, the classification strategy diverges by language type:

\textit{For static languages (Rust, Go)}, where compilation includes type checking and API resolution, we sequentially check:
\begin{itemize}[leftmargin=1em, itemsep=0.2em]
\item \textbf{Category 3} (API Hallucination) -- highest priority, as undefined symbols indicate fabricated APIs
\item \textbf{Category 2} (Type System \& Memory) -- type errors and ownership violations
\item \textbf{Category 1} (Syntactic \& Compilation) -- fallback for parse errors and malformed syntax
\end{itemize}

\textit{For dynamic languages (Julia, PHP, Ruby)}, where compilation only validates syntax (not types or API existence), we \textbf{directly classify as Category 1}:
\begin{itemize}[leftmargin=1em, itemsep=0.2em]
\item \textbf{Category 1} (Syntactic \& Compilation) -- parse errors and malformed syntax
\item Type checking and API resolution are deferred to runtime; Categories 2 and 3 are not checked during compilation failures
\end{itemize}

\item \textbf{Runtime Failures:} When compilation succeeds but \texttt{check.tests == False}, we again distinguish by language type:

\textit{For static languages (Rust, Go)}:
\begin{itemize}[leftmargin=1em, itemsep=0.2em]
\item \textbf{Category 2} (Type System \& Memory, runtime) -- runtime type errors and panics
\item \textbf{Category 4} (Logic \& Assertion) -- assertion failures and incorrect test logic
\end{itemize}

\textit{For dynamic languages (Julia, PHP, Ruby)}, where type and API errors manifest at runtime:
\begin{itemize}[leftmargin=1em, itemsep=0.2em]
\item \textbf{Category 3} (API Hallucination) -- undefined functions, methods, or constants
\item \textbf{Category 2} (Type System \& Memory, runtime) -- runtime type errors, argument mismatches
\item \textbf{Category 4} (Logic \& Assertion) -- assertion failures and incorrect test logic
\end{itemize}

\item \textbf{Design Issues:} When all checks pass (\texttt{check.compilation == True} and \texttt{check.tests == True}), we apply coverage-based heuristics:
\begin{itemize}[leftmargin=1em, itemsep=0.2em]
\item If \texttt{covered\_lines == 0} (or $\leq 1$ for Ruby), classify as \textbf{Category 5} (Test Design \& Mocking), indicating the test executes but does not invoke the focal function.
\item Otherwise, no error is reported (\texttt{has\_error = False}).
\end{itemize}
\end{enumerate}

\subsection{API Hallucination Sub-category Analysis}
\label{app:api_hallucination_breakdown}

To provide more actionable diagnostic insights, we decomposed all API Hallucination failures into four sub-categories:
\begin{enumerate}[leftmargin=1.5em, itemsep=0.2em]
    \item \textbf{Phantom Library}: importing non-existent packages;
    \item \textbf{Non-existent API}: calling fictitious functions within valid packages;
    \item \textbf{Signature Mismatch}: incorrect arguments or types for real APIs;
    \item \textbf{Other}: primarily deprecated APIs.
\end{enumerate}

Table~\ref{tab:api_hallucination_breakdown} reports the distribution of these sub-types within the API Hallucination category, comparing standard (\textbf{Std}) and context-augmented (\textbf{Ctx}) settings. Values represent the percentage share within the hallucination category (columns sum to 100\%).

\begin{table*}[h]
\centering
\scriptsize
\begin{adjustbox}{width=\textwidth}
\begin{tabular}{l*{5}{cc}}
\toprule
\multirow{2}{*}{\textbf{Sub-category}} &
\multicolumn{2}{c}{\textbf{Rust}} &
\multicolumn{2}{c}{\textbf{Go}} &
\multicolumn{2}{c}{\textbf{Julia}} &
\multicolumn{2}{c}{\textbf{Ruby}} &
\multicolumn{2}{c}{\textbf{PHP}} \\
\cmidrule(lr){2-3}\cmidrule(lr){4-5}\cmidrule(lr){6-7}\cmidrule(lr){8-9}\cmidrule(lr){10-11}
& \textbf{Std} & \textbf{Ctx} & \textbf{Std} & \textbf{Ctx} & \textbf{Std} & \textbf{Ctx} & \textbf{Std} & \textbf{Ctx} & \textbf{Std} & \textbf{Ctx} \\
\midrule
Phantom Library   & 77.9 & 72.9 & 23.2 &  3.9 & 32.8 & 38.8 & 76.9 & 77.5 & 88.8 & 92.4 \\
Non-existent API  & 16.8 & 19.0 & 71.2 & 85.9 & 34.4 & 39.2 & 19.3 & 18.5 & 10.0 &  6.7 \\
Sig.\ Mismatch    &  4.5 &  7.0 &  5.5 & 10.2 & 32.8 & 21.9 &  3.2 &  2.8 &  0.9 &  0.1 \\
Other             &  0.7 &  1.0 &  0.0 &  0.0 &  0.1 &  0.1 &  0.6 &  1.1 &  0.3 &  0.8 \\
\bottomrule
\end{tabular}
\end{adjustbox}
\caption{Breakdown of API Hallucination failures into sub-categories (\%). Values represent the distribution within the hallucination category; columns sum to 100\%.}
\label{tab:api_hallucination_breakdown}
\end{table*}

\paragraph{Key findings.}
Three patterns emerge from this finer-grained analysis.

\textbf{Context redistributes rather than uniformly reduces hallucination.} Across languages, context augmentation alters the \emph{composition} of hallucination rather than simply decreasing it. In Rust and PHP, Phantom Library errors remain dominant under context (Rust: 77.9\%~$\to$~72.9\%; PHP: 88.8\%~$\to$~92.4\%), suggesting that broader repository exposure can amplify incorrect assumptions about dependency structure. In contrast, Go exhibits a sharp shift: Phantom Library errors drop dramatically (23.2\%~$\to$~3.9\%) while Non-existent API errors increase markedly (71.2\%~$\to$~85.9\%). Here, context successfully anchors correct package imports, but models overgeneralize by inventing plausible methods on valid types.

\textbf{Hallucination is primarily an ecosystem-grounding issue.} Signature Mismatch remains secondary across most languages ($<$10\%, except Julia at 21--33\%), indicating that failures stem less from type-level reasoning and more from incorrect assumptions about available libraries and APIs. Techniques that constrain generation using repository-aware metadata (dependency graphs, import trees, module registries) may therefore be more impactful than improvements in generic type reasoning alone.

\textbf{Language-dependent structural mechanisms.} Ruby and PHP show extremely high Phantom Library dominance under both settings ($>$76\%), consistent with dynamic or framework-heavy ecosystems that encourage speculative dependency invention. Julia exhibits a more balanced distribution across sub-types, with Signature Mismatch notably higher (21--33\%) than other languages, reflecting its multiple-dispatch system. Go's strong shift from package-level to method-level hallucination reflects its explicit import system, which reduces dependency invention but does not prevent API overgeneralization. These patterns suggest that evaluation strategies and mitigation techniques should be language-aware: dependency validation may be critical for dynamic ecosystems, while API-level verification may yield larger gains for statically structured languages.

\subsection{Failure Mode Breakdown by Model}

Table~\ref{tab:error_distribution_per_model} presents a comprehensive breakdown of failure modes across all evaluated models and languages, comparing standard and context-enriched prompting strategies. All values are normalized by the total number of benchmark samples (not only the error subset), avoiding the misleading scenario where a reduction in error count paradoxically increases a category's share. Each column sums to 100\%, with the \textit{No Error / Pass} row capturing samples that pass all checks without any failure. The table shows the distribution of samples across five primary failure mode categories plus the no-error class: Syntactic \& Compilation errors, Type System \& Memory issues, API Hallucination, Logic \& Assertion failures, and Test Design \& Mocking problems. This granular view reveals model-specific strengths and weaknesses, and demonstrates how context enrichment affects the distribution of failure modes across different languages.

\subsection{Illustrative Examples}
\label{sec:failure_mode_examples}

We provide one concrete example per failure category. Each example is taken from Go under \texttt{standard} mode with GPT-5.2, using the generated test code and execution logs produced by our evaluation pipeline.

\newpage
\begin{tcolorbox}[promptbox, title={Category 1: Syntactic \& Compilation}]

\textbf{Focal function:} \texttt{Encrypt} in \texttt{Go-master/cipher/transposition/
transposition.go}. 
The generated test attempted to monkey-patch the package-level helper function \texttt{getKey}, but ordinary Go functions are not assignable.

\textbf{Focal function (excerpt):}
\begin{lstlisting}[language=Go]
func Encrypt(text []rune, keyWord string) ([]rune, error) {
    key := getKey(keyWord)

    if len(key) <= 0 {
        return nil, ErrKeyMissing
    }

    ...
    return result, nil
}
\end{lstlisting}

\textbf{Generated test (excerpt):}
\begin{lstlisting}[language=Go]
origGetKey := getKey

for _, tt := range tests {
    t.Run(tt.name, func(t *testing.T) {
        getKey = origGetKey

        if tt.mockGetKey != nil {
            getKey = tt.mockGetKey
        }

        _, err := Encrypt(
            append([]rune(nil), tt.text...), tt.keyWord)
        ...
    })
}
\end{lstlisting}

\textbf{Key log excerpt:}
\begin{lstlisting}
./temp_test.go:131:4: cannot assign to getKey
./temp_test.go:133:5: cannot assign to getKey
./temp_test.go:171:2: cannot assign to getKey
\end{lstlisting}

\end{tcolorbox}

\newpage

\begin{tcolorbox}[promptbox, title={Category 2: Type System \& Memory}]

\textbf{Failure summary.}
The generated test used characters from \texttt{ComputeAlphabet} as keys into the transition map. However, the implementation uses \texttt{map[int]map[uint8]int}, while the loop variable \texttt{ch} is inferred as \texttt{rune}, causing a static type error.

\vspace{0.4em}
\textbf{Focal function.}
\texttt{BuildExtendedAc} in \texttt{Go-master/strings/ahocorasick/
advancedahocorasick.go}.

\vspace{0.4em}
\textbf{Focal code excerpt.}
\begin{lstlisting}[language=Go]
func BuildExtendedAc(p []string) (
    acToReturn map[int]map[uint8]int,
    f map[int][]int,
) {
    acTrie, _, f := ConstructTrie(p)
    acToReturn = acTrie

    a := ComputeAlphabet(p)
    for j := range a {
        if GetTransition(0, a[j], acToReturn) == -1 {
            CreateTransition(0, a[j], 0, acToReturn)
        }
    }

    ...
    return acToReturn, f
}
\end{lstlisting}

\vspace{0.4em}
\textbf{Generated test excerpt.}
\begin{lstlisting}[language=Go]
func TestBuildExtendedAc_NoMissingTransitions(t *testing.T) {
    patterns := []string{"he", "she", "his", "hers"}
    ac, _ := BuildExtendedAc(patterns)
    alphabet := ComputeAlphabet(patterns)

    for state, trans := range ac {
        for _, ch := range alphabet {
            next, ok := trans[ch]
            if !ok {
                t.Errorf("missing transition from state %d on %q",
                    state, ch)
            }
            ...
        }
    }
}
\end{lstlisting}

\vspace{0.4em}
\textbf{Key log excerpt.}
\begin{lstlisting}
./temp_test.go:277:22: cannot use ch
    (variable of type rune) as uint8 value in map index
\end{lstlisting}

\end{tcolorbox}

\clearpage

\begin{tcolorbox}[promptbox, title={Category 3: API Hallucination}]

\textbf{Failure summary.}
The generated test assumed that the returned \texttt{Result} type exposes an exported field named \texttt{Occurrences}. However, the actual implementation stores the match positions in an unexported field, \texttt{occurrences}. As a result, the test references a non-existent public API and fails to compile.

\vspace{0.4em}
\textbf{Focal function.}
\texttt{AhoCorasick} in \texttt{Go-master/strings/ahocorasick/ ahocorasick.go}.

\vspace{0.4em}
\textbf{Focal code excerpt.}
\begin{lstlisting}[language=Go]
func AhoCorasick(t string, p []string) Result {
    occurrences := make(map[int][]int)

    ...
    var resultOccurrences = make(map[string][]int)
    for key, value := range occurrences {
        resultOccurrences[p[key]] = value
    }

    return Result{
        resultOccurrences,
    }
}
\end{lstlisting}

\vspace{0.4em}
\textbf{Generated test excerpt.}
\begin{lstlisting}[language=Go]
// Check occurrences.
if res.Occurrences == nil {
    t.Fatalf("Result.Occurrences is nil; expected non-nil map")
}

if len(res.Occurrences) != len(tt.want) {
    t.Errorf("unexpected number of patterns: got %d, want %d",
        len(res.Occurrences), len(tt.want))
}

for pat, wantPos := range tt.want {
    gotPos, ok := res.Occurrences[pat]
    ...
}
\end{lstlisting}

\vspace{0.4em}
\textbf{Key log excerpt.}
\begin{lstlisting}
./temp_test.go:176:11: res.Occurrences undefined
    (type Result has no field or method Occurrences,
    but does have occurrences)
\end{lstlisting}

\end{tcolorbox}

\newpage

\begin{tcolorbox}[promptbox, title={Category 4: Logic \& Assertion}]

\textbf{Failure summary.}
The generated test over-specified behavior that the implementation does not guarantee. It expected one exact DFS route, although different valid traversal orders may occur. It also assumed that a missing start node should be handled gracefully, but the implementation can panic when \texttt{GetIdx} returns \texttt{-1}.

\vspace{0.4em}
\textbf{Focal function.}
\texttt{DepthFirstSearchHelper} in \texttt{Go-master/graph/depthfirstsearch.go}.

\vspace{0.4em}
\textbf{Focal code excerpt.}
\begin{lstlisting}[language=Go]
startIdx := GetIdx(start, nodes)
stack = append(stack, startIdx)

for len(stack) > 0 {
    now := stack[len(stack)-1]
    route = append(route, nodes[now])

    for i := 0; i < len(edges[now]); i++ {
        if edges[now][i] && NotExist(i, stack) {
            stack = append(stack, i)
        }
    }

    if route[len(route)-1] == end {
        return route, true
    }
}
\end{lstlisting}

\vspace{0.4em}
\textbf{Generated test excerpt.}
\begin{lstlisting}[language=Go]
// Invalid start node is assumed to return no path.
route, found := DepthFirstSearchHelper(
    99, 3, nodes, edges, true)

if found {
    t.Fatalf("expected no path, got route %v", route)
}
\end{lstlisting}

\vspace{0.4em}
\textbf{Key log excerpt.}
\begin{lstlisting}
branching_graph_path_found:
    route = [1 3 2 4]; want [1 2 4]

cycle_in_graph_still_finds_path:
    route = [1 3]; want [1 2 3]

start_not_in_nodes:
    panic: runtime error: index out of range [-1]
\end{lstlisting}

\end{tcolorbox}
\newpage

\begin{tcolorbox}[promptbox, title={Category 5: Test Design \& Mocking}]

\textbf{Focal function:} \texttt{Forward} in \texttt{ollama-main/model/model.go}. 
The generated tests passed, but achieved zero focal coverage 
(\texttt{covered\_lines=0} out of \texttt{total\_lines=22}) because they tested a local wrapper instead of calling the real focal function.

\textbf{Focal function (excerpt):}
\begin{lstlisting}[language=Go]
func Forward(ctx ml.Context, m Model,
    inputs []int32, batch input.Batch) (ml.Tensor, error) {

    if len(batch.Positions) != len(batch.Sequences) {
        return nil, fmt.Errorf("length mismatch")
    }
    if len(batch.Positions) < 1 {
        return nil, errors.New("batch size cannot be less than 1")
    }

    ...
    t, err := m.Forward(ctx, batch)
    if err != nil {
        return nil, err
    }

    ctx.Forward(t).Compute(t)
    return t, nil
}
\end{lstlisting}

\textbf{Generated test excerpt.}
\begin{lstlisting}[language=Go]
// Local wrapper with the same body as the real Forward.
func forwardUnderTest(ctx testContext, m testModel,
    inputs []int32, batch *fakeBatch) (*fakeTensor, error) {

    if len(batch.Positions) != len(batch.Sequences) {
        return nil, fmt.Errorf("length mismatch")
    }
    if len(batch.Positions) < 1 {
        return nil, errors.New("batch size cannot be less than 1")
    }

    ...
}
\end{lstlisting}

\textbf{Key log excerpt.}
\begin{lstlisting}
--- PASS: TestForward (0.00s)
PASS
ok  github.com/ollama/ollama/model  0.005s

covered_lines=0 / total_lines=22
\end{lstlisting}

\end{tcolorbox}

\section{Prompt Template}
\label{app:prompt}
We use a single unified prompting scheme across all evaluated models. The prompt used by our pipeline consists of (i) a language-specific system prompt and (ii) a shared user prompt template with optional language-specific requirements.

\subsection{System prompt template}

\begin{tcolorbox}[promptbox, title={Rust: System Prompt}]
\begin{lstlisting}
You are an expert Rust developer specializing in test-driven development (TDD) and writing high-quality unit tests.

Generate unit tests for Rust code using the following format:
- Use `#[cfg(test)] mod tests` with `#[test]` functions
- Use `assert_eq!`, `assert!`, and `#[should_panic]` for assertions
- Cover normal cases, edge cases, and error cases

Always wrap your test code in triple backticks with the language identifier:
```rust
// Your generated test code here
```

Here's an example of the expected format (adapt this structure to the actual code you're testing):
```rust
#[cfg(test)]
mod tests {
    use super::*;

    #[test]
    fn test_add() {
        assert_eq!(add(2, 3), 5);
    }

    #[test]
    #[should_panic]
    fn test_divide_by_zero() {
        divide(10, 0);
    }
}
```

Generate comprehensive tests tailored to the specific code provided.
\end{lstlisting}
\end{tcolorbox}
\newpage
\begin{tcolorbox}[promptbox, title={Go: System Prompt}]
\begin{lstlisting}
You are an expert Go developer specializing in test-driven development (TDD) and writing high-quality unit tests.

Generate unit tests for Go code using the following format:
- Create tests in `_test.go` files
- Use table-driven tests with `t.Run()` for multiple test cases
- Use `t.Errorf()` for test failures
- Cover normal cases, edge cases, and error cases

Always wrap your test code in triple backticks with the language identifier:
```go
// Your generated test code here
```

Here's an example of the expected format (adapt this structure to the actual code you're testing):
```go
package main

import "testing"

func TestAdd(t *testing.T) {
    tests := []struct {
        name     string
        a, b     int
        expected int
    }{
        {"positive numbers", 2, 3, 5},
        {"with zero", 0, 5, 5},
        {"negative numbers", -2, -3, -5},
    }

    for _, tt := range tests {
        t.Run(tt.name, func(t *testing.T) {
            result := Add(tt.a, tt.b)
            if result != tt.expected {
                t.Errorf("Add(%d, %d) = %d; want %d", tt.a, tt.b, result, tt.expected)
            }
        })
    }
}
```

Generate comprehensive tests tailored to the specific code provided.
\end{lstlisting}
\end{tcolorbox}
\newpage
\begin{tcolorbox}[promptbox, title={Julia: System Prompt}]
\begin{lstlisting}
You are an expert Julia developer specializing in test-driven development (TDD) and writing high-quality unit tests.

Generate unit tests for Julia code using the following format:
- Create tests in separate test files
- Use `@testset` to group related tests and `@test` for assertions
- Remember: Julia uses 1-based indexing (arrays start at index 1)
- Cover normal cases, edge cases, and error cases

Always wrap your test code in triple backticks with the language identifier:
```julia
// Your generated test code here
```

Here's an example of the expected format (adapt this structure to the actual code you're testing):
```julia
using Test

@testset "Array operations" begin
    @testset "sum_array tests" begin
        @test sum_array([1, 2, 3]) == 6
        @test sum_array([]) == 0
        @test sum_array([-1, 1]) == 0
    end

    @testset "first_element tests" begin
        arr = [10, 20, 30]
        @test first_element(arr) == 10  # Remember: 1-based indexing!
        @test_throws BoundsError first_element([])
    end
end
```

Generate comprehensive tests tailored to the specific code provided, always remembering Julia's 1-based indexing!
\end{lstlisting}
\end{tcolorbox}
\newpage
\begin{tcolorbox}[promptbox, title={Ruby: System Prompt}]
\begin{lstlisting}
You are an expert Ruby developer specializing in test-driven development (TDD) and writing high-quality unit tests.

Generate unit tests for Ruby code using the following format:
- Use RSpec framework for tests
- Use `expect().to eq()`, `expect().to be()`, and `expect{}.to raise_error()` for assertions
- Use `describe` blocks to group related tests and `it` blocks for individual test cases
- Cover normal cases, edge cases, and error cases

Always wrap your test code in triple backticks with the language identifier:
```ruby
// Your generated test code here
```

Here's an example of the expected format (adapt this structure to the actual code you're testing):
```ruby
require 'rspec'
require_relative '../lib/math_utils'

RSpec.describe MathUtils do
  describe '.factorial' do
    it 'returns 1 for 0!' do
      expect(MathUtils.factorial(0)).to eq(1)
    end

    it 'returns 120 for 5!' do
      expect(MathUtils.factorial(5)).to eq(120)
    end

    it 'raises an error for negative numbers' do
      expect { MathUtils.factorial(-1) }.to raise_error(ArgumentError)
    end
  end
end
```

Generate comprehensive tests tailored to the specific code provided.
\end{lstlisting}
\end{tcolorbox}
\newpage
\begin{tcolorbox}[promptbox, title={PHP: System Prompt}]
\begin{lstlisting}
You are an expert PHP developer specializing in test-driven development (TDD) and writing high-quality unit tests.

Generate unit tests for PHP code using the following format:
- Use PHPUnit framework for tests
- Extend `PHPUnit\Framework\TestCase` for test classes
- Use `assertEquals()`, `assertTrue()`, `assertFalse()`, and `expectException()` for assertions
- Always include `require_once __DIR__ . '/../../vendor/autoload.php';` at the beginning of test files
- Cover normal cases, edge cases, and error cases

Always wrap your test code in triple backticks with the language identifier:
```php
// Your generated test code here
```

Here's an example of the expected format (adapt this structure to the actual code you're testing):
```php
<?php
require_once __DIR__ . '/../../vendor/autoload.php'; // safe path

use App\Services\MathService;
use PHPUnit\Framework\TestCase;

class MathServiceTest extends TestCase
{
    public function testAdd()
    {
        $m = new MathService();
        $this->assertEquals(3, $m->add(1, 2));
    }
}
```

Generate comprehensive tests tailored to the specific code provided.
\end{lstlisting}
\end{tcolorbox}

\newpage

\subsection{User prompt template}
\begin{tcolorbox}[promptbox, title={Rust: User Prompt Template}]
\begin{lstlisting}
Generate unit tests for this Rust function from {file_path}:

```rust
{function_code}
```

Additional requirements:
- Import necessary items with `use super::*;`
- Use `assert_ne!` for inequality checks where appropriate
- Include doc comments explaining what each test validates
\end{lstlisting}
\end{tcolorbox}
\begin{tcolorbox}[promptbox, title={Go: User Prompt Template}]
\begin{lstlisting}
Generate unit tests for this Go function from {file_path}:

```go
{function_code}
```

Additional requirements:
- Package name: {package_name}
- Test function name: `func Test{function_name}(t *testing.T)`
- Use `t.Fatalf()` for critical failures that should stop the test
- For error cases, check both error occurrence and error message
- Import necessary packages beyond `testing` if needed
\end{lstlisting}
\end{tcolorbox}
\begin{tcolorbox}[promptbox, title={Julia: User Prompt Template}]
\begin{lstlisting}
Generate unit tests for this Julia function from {file_path}:

```julia
{function_code}
```

Additional requirements:
- Include descriptive names for testsets
- Test boundary conditions carefully
- Remember: `length(arr)` gives array size, `arr[1]` is first element, `arr[end]` is last element
\end{lstlisting}
\end{tcolorbox}
\newpage
\begin{tcolorbox}[promptbox, title={Ruby: User Prompt Template}]
\begin{lstlisting}
Generate unit tests for this Ruby function from {file_path}:

```ruby
{function_code}
```

Additional requirements:
- Use RSpec framework (require 'rspec')
- Use `describe` blocks for grouping and descriptive `it` blocks for test cases
- Use appropriate expectations: `expect().to eq()`, `expect().to be()`, `expect{{}}.to raise_error()`
- Test both success and failure paths
- Use `context` blocks for different scenarios if applicable
\end{lstlisting}
\end{tcolorbox}
\begin{tcolorbox}[promptbox, title={PHP: User Prompt Template}]
\begin{lstlisting}
Generate unit tests for this PHP function from {file_path}:

```php
{function_code}
```

Additional requirements:
- Use PHPUnit framework
- Namespace: {namespace}
- Test class should extend `PHPUnit\Framework\TestCase`
- Include `require_once 'vendor/autoload.php';` at the beginning
- Use appropriate assertions: `$this->assertEquals()`, `$this->expectException()`
- Follow PSR coding standards for test code
- Include setup/teardown methods if needed
\end{lstlisting}
\end{tcolorbox}

\clearpage

\end{document}